\documentclass[aps, prb, preprint, a4paper, superscriptaddress, showpacs]{revtex4-1}

\usepackage{graphicx}

\usepackage{amsmath,amssymb,bm,amsthm}

\newcommand{\bxi}{\boldsymbol{\xi}}
\newcommand{\bS}{\boldsymbol{S}}
\newcommand{\bH}{\boldsymbol{H}}
\newcommand{\be}{\boldsymbol{e}}
\newcommand{\cH}{\mathcal{H}}
\newcommand{\udd}{\text{d}}

\newcommand{\ignore}[1]{}
\newcommand{\beq}{\begin{equation}}
\newcommand{\eeq}{\end{equation}}
\newcommand{\mbold}[1]{\mbox{\boldmath $ #1 $}}

\graphicspath{{New-figure-magnet/}}

\date{\today}

\begin{document}

\title{Temperature dependence of the threshold magnetic field for nucleation and domain wall propagation in an inhomogeneous structure with grain boundary}

\author{Sasmita Mohakud}
\affiliation{Indian Institute of Technology Kharagpur, India}
\author{Sergio Andraus}
\affiliation{Department of Physics, Graduate School of Science and Engineering, Chuo University, 1-13-27 Kasuga, Bunkyo-ku, Tokyo, 112-8551, Japan}
\email{andraus@phys.chuo-u.ac.jp}
\author{Masamichi Nishino}
\affiliation{National Institute for Materials Science, Tsukuba, Ibaraki 305-0047, Japan}
\author{Akimasa Sakuma}
\affiliation{Department of Applied Physics, Graduate School of Engineering, Tohoku University, 6-6-05 Aoba-ku, Sendai, 980-8579, Japan.}
\author{Seiji Miyashita}
\affiliation{Department of Physics, Graduate School of Science, The University of Tokyo, 7-3-1 Hongo, Bunkyo-ku, Tokyo, 113-0033, Japan}
%\affiliation{CREST, JST, K's Gobancho, 7 Gobancho, Chiyoda-ku, Tokyo 102-0076, Japan}

\begin{abstract}
In order to study the dependence of the coercive force of sintered magnets on temperature, nucleation and domain wall propagation at the grain boundary are studied
as rate-determining processes of the magnetization reversal phenomena in magnets consisting of bulk hard magnetic grains contacting via grain boundaries of a soft magnetic material. 
These systems have been studied analytically for a continuum model at zero temperature (A. Sakuma, et al. J. Mag. Mag. Mat. {\bf 84} 52 (1990)). 
In the present study, the temperature dependence is studied by making use of the stochastic Landau-Lifshitz-Gilbert equation at finite temperatures. 
In particular, the threshold fields for nucleation and domain wall propagation are obtained as functions of ratios of magnetic interactions and anisotropies of the soft and hard magnets for various temperatures. 
It was found that the threshold field for domain wall propagation is robust against thermal fluctuations, while that for nucleation is fragile. 
The microscopic mechanisms of the observed temperature dependence are discussed.
%The parameters to enhance of the threshold are proposed.  
\end{abstract}


%\keywords{Nucleation, domain wall propagation, lattice spin model, stochastic Landau-Lifshitz-Gilbert equation}

\maketitle

\section{Introduction}

The mechanisms by which the coercive force manifests itself in permanent magnets have been studied extensively.\cite{review}
It is known that a single crystal of magnetic material does not show the hysteresis phenomenon, i.e., the coercive force is absent,
and thus the structure of ensembles of fine grains plays an important role for the coercive force.
Magnetization reversal in an antiparallel field occurs as a nucleation event at some point in the system, and it propagates through the material, forming a domain.
Nucleation may occur due to intrinsic or extrinsic sources. 
Thermal fluctuations of the bulk material constitute an intrinsic source of nucleation, while extrinsic sources are due to the inhomogeneous structure of the material, e.g., misalignment of the easy axis and impurities, etc. 
Regardless of the origin, the nucleated reverse magnetization propagates throughout the material if the bulk magnetic region is connected. 

\begin{figure}[t]
  \centering
  {\includegraphics[width=0.5\textwidth]{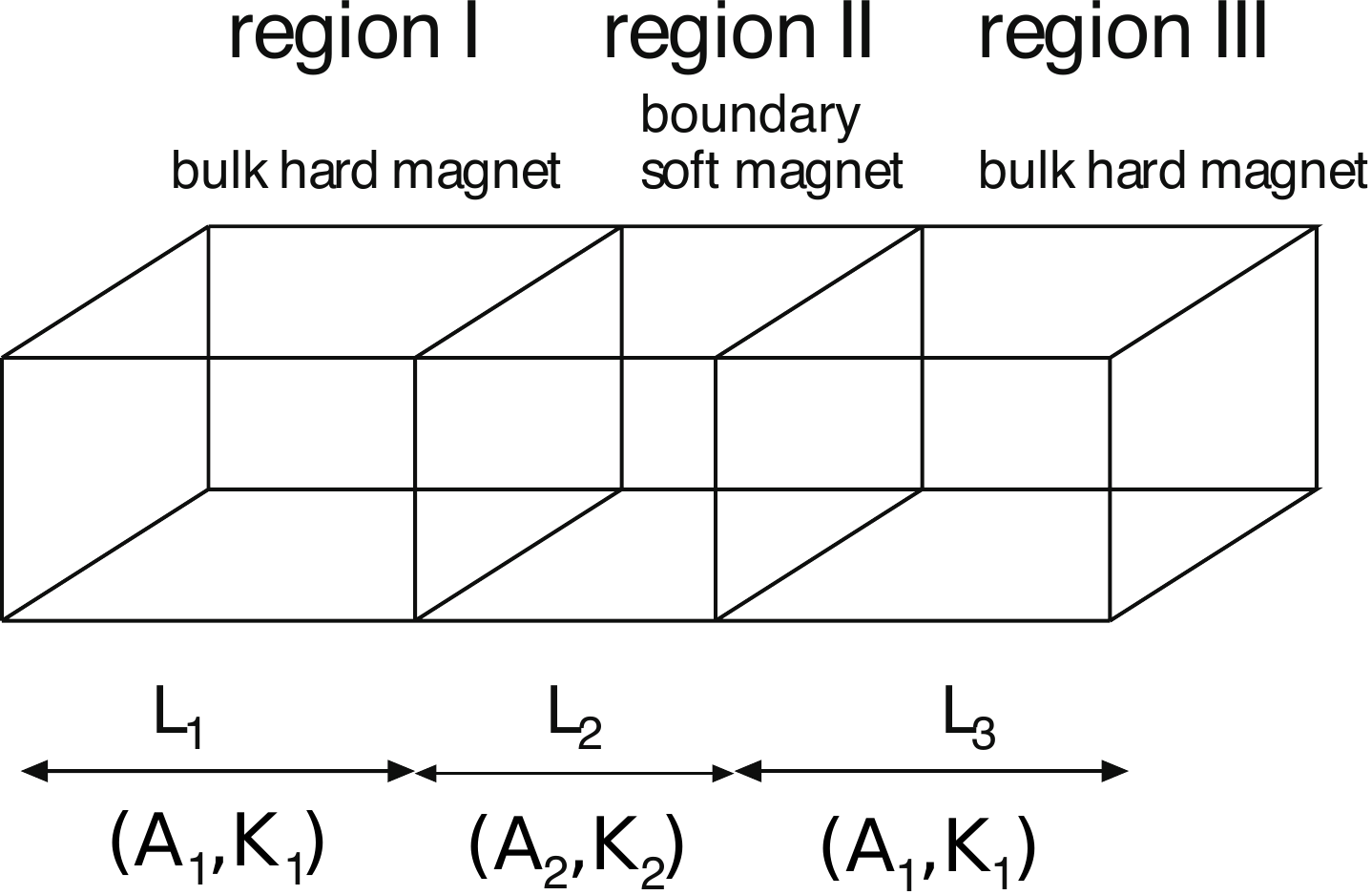}}              
\caption{Schematic picture of a system consisting of two bulk hard magnets and a boundary soft magnet. 
%In the continuum model~\cite{sakuma}, a one-dimensional system with the infinite  size is considered, where
Regions I and III are characterized by $A_1$, $K_1$ and $M_1$, while region II is characterized by $A_2$, $K_2$ and $M_2$. 
%Here, we adopt a similar notation, where $A_1$, $K_1$ and $S_1(=1)$ characterize regions I (size $L_1$) and III (size $L_3$) and $A_2$, $K_2$ and $S_2(=1)$ characterize region II (size $L_2$). 
Free boundary conditions are adopted for the lattice model. }
\label{ModelImage}
\end{figure}

To prevent the propagation of the reversed domain and maintain the coercive field, we must consider the conditions under which the pinning of the domain wall is realized.
To understand the nucleation phenomenon and also the properties of domain wall propagation at the grain boundary, one may study a system extending in one direction 
with a defect region, as depicted in Fig.~\ref{ModelImage}.
The magnetization reversal phenomena in this system have been studied at zero temperature,\cite{Friedberg} and in particular, Sakuma et al. studied the threshold fields as a function of the ratios of magnetic interactions and anisotropies of the soft and hard magnets at zero temperature analytically by solving a one-dimensional nonlinear equation, and presented the corresponding phase diagram.\cite{sakuma}
In the present paper, we study the threshold magnetic fields for nucleation and for domain wall propagation at finite temperatures by making use of simulations of the stochastic Landau-Lifshitz-Gilbert (LLG) equation\cite{review,LLG,Garcia,Evans2,NishinoLLG} for the system in Fig.~\ref{ModelImage}. 
In the analytical studies, the case in which the soft magnet grain boundary has the exchange energy $A_2<A_1$, and the anisotropy energy $K_2<K_1$, was studied.
 
The case with strong anisotropy which was studied by Hirota et al\cite{Hirota} is also interesting. However,  in the present paper we focus on the temperature dependence of the case which was considered in previous analytical studies at $T=0$. The case with strong anisotropy will be studied separately in the near future. A strong anisotropy causes the formation of narrow domain walls, which is a phenomenon that also occurs effectively in the case we studied. We discuss the narrow domain wall phenomenon in more detail in the later sections of this paper.

It has been also pointed out that the magnetic reversal of this type of systems depends on the size of the boundary region due to the so-called spring exchange effect\cite{suess, dobin}.
It is also known that this dependence is weak for large width.
With this in mind, we choose a fixed defect width larger than the domain wall width.
We consider a material similar to Nd-Fe-B, in which the correlation length ($\xi\propto \sqrt{A/K}$) is long compared to the lattice spacing, in contrast to the case of Sm-Co.
To take this fact into account, we choose the ratio $K_1/A_1=0.2$ in the hard magnet.
This value is still large when compared with real materials, but we believe that it can represent a situation where there is a long correlation length.
As for the soft magnets (grain boundary), there are various situations to
be considered.
For example, depending on the concentration ratio of Fe and Nd in the grain
boundary region of a Nd-Fe-B magnet, the parameters $A$ and $K$ would
change\cite{Hono}.
Thus we investigate the general tendency of the threshold fields in a
wide range of parameters.
The case in which the grain boundary width is narrow compared with the correlation length is interesting as well, but lies beyond the scope of this work.

%Here, we fix the width of the boundary region, which was taken to be variable in previous work.\cite{sakuma} 

In soft magnets, the magnetic order weakens with rising temperature, and the threshold field for nucleation shows a rapid decrease with increasing temperature.
However, the threshold field for the domain wall propagation is found to show a more complicated dependence due to a competition between the following mechanisms. On one hand, the reduction of ordering in the hard magnetic regions due to the rise in temperature causes a reduction of the threshold.

On the other hand, in a lattice system, the discreteness of the system becomes important when the anisotropy becomes strong, where the so-called narrow domain wall appears instead of the Bloch type wall,\cite{narrowDW} thus reducing the spring effect.
When the spring effect disappears,
the domain wall propagation is regarded as a nucleation at the surface of the hard magnet, and thus the threshold for the propagation of the domain wall has little dependence on the parameters of the soft magnet.
Because of this effect, the threshold field of domain wall depinning shows a non-monotonic dependence on the parameters under consideration, which is qualitatively different from the analytical results\cite{sakuma} even at $T=0$, and generally the temperature dependence is milder than that of the threshold field of nucleation.
As a result,
the threshold field for domain wall propagation is robust against thermal fluctuations, while that for nucleation is fragile.

%When the magnetic interaction is weak, the magnetic correlation in the boundary region does not persist at finite temperatures, and a dependence which does not exist at zero temperature is found.

The paper is organized as follows:
In Section~\ref{sectionmethod}, we explain the model under consideration and the method we used to obtain our results.
In Section~\ref{sectionnucleation}, we study the nucleation phenomenon at the boundary region and its propagation to the hard magnetic regions.
In Section~\ref{sectiondomainwallpropagation}, the domain wall propagation phenomenon is studied.
%In Section~\ref{sectioncomparison}, we compare the behavior of the nucleation and domain wall propagation phenomena with varying temperature.
In Section~\ref{sectionsummary}, we summarize and discuss our results.
In Appendix A, the temperature dependence of the anisotropy energy in the bulk magnets is given.
In Appendix B, we show how sharply the relaxation time changes with the magnetic field around the threshold values.
In Appendix C, the narrow domain wall phenomenon is explained. 

\section{Model and Method}\label{sectionmethod}

We consider a continuous magnetic system modeled by the Hamiltonian
\beq
{\cal H}=\int d\mbold{r} \left({A\over 2} (\nabla \mbold{m}(\mbold{r}))^2-Km_z(\mbold{r})^2-M\mbold{H}\cdot\mbold{m}(\mbold{r})\right),
\eeq
where $\mbold{m}$ is the unit vector of the direction of the magnetization at position $\mbold{r}$, $A$ is the exchange energy and $K$ is the anisotropy energy.
Here, it should be noted that the definition of exchange energy is different from that in
the continuum model~\cite{sakuma} by a factor of 2.
The last term is the Zeeman energy, $H$ is the magnetic field and $M$ is magnetization. 
The magnetic properties of the bulk hard magnet are specified by the exchange energy $A_1$ and anisotropy $K_1$, 
and that of the grain boundary region with width $W$ by  $A_2$ and $K_2$. 
The magnetizations in these regions are $M_1$ and $M_2$, respectively.  

A phase diagram of the threshold magnetic fields for the nucleation and domain wall depinning in a one-dimensional continuum model were obtained analytically at zero temperature for the cases $A_1>A_2$ and $K_1\ge K_2$.\cite{Friedberg,sakuma}
The threshold field for nucleation $H_{\rm NC}$ is defined to be the field above which a nucleation type solution does not exist, and that for domain wall depinning $H_{\rm DWP}$ to be the field above which a domain-wall like solution does not exist.

We adopt the following variables to parameterize the model used in the previous work\cite{sakuma}:\\
the normalized external field 
\beq h={H\over H_{\rm SW}},\quad H_{\rm SW}\equiv {2K_1\over M_1}, 
\label{hscaled}
\eeq
in which $H_{\rm SW}$ is the Stoner-Wohlfarth field of the bulk hard magnets,
the ratio of exchange energies 
\beq 
F={A_2 M_2 \over A_1 M_1},
\eeq 
and the ratio
\beq
E={A_2 K_2\over A_1 K_1}.
\eeq
Here, it should be noted that the domain wall energy is proportional to $\sqrt{AK}.$

For the nucleation process, the threshold of the normalized external field $h$ above which the nucleation occurs in the boundary region (II) for infinite width $W$ at $T=0$ is given by 
\begin{equation}
h_{\rm NCII}(0)={E\over F}.
\label{hn0}
\end{equation}
For finite width the threshold is slightly larger than this value.\cite{sakuma}

For the domain wall propagation, the threshold of the normalized external field $h$ above which the domain wall propagates from the boundary region (II) to the bulk regions (I and III) at $T=0$ is given by
\begin{equation}
h_\text{\rm DWP}(0)={1-E\over (1+\sqrt{F})^2},
\label{hp}
\end{equation}
and this quantity is known as the depinning field. 
For $h<h_\text{\rm DWP}(0)$, the domain wall is pinned at the border between the boundary and hard magnets, and does not propagate to the bulk region. 

In the case $h_{\rm NCII}(0) < h < h_{\rm DWP}(0)$, the nucleated defect region is confined.
For the magnetization reversal of the whole system the nucleated magnetization must propagate to the hard magnets (regions I and III). Thus, the threshold of the magnetization reversal in the case of nucleation, i.e., magnetization reversal from the initial configuration where all the regions are antiparallel to the applied field, is given by the largest of $h_{\rm NCII}(0)$ and $h_{\rm DWP}(0)$. Thus, the threshold field for the nucleation to propagate to the hard magnets at $T=0$ is given by
\begin{equation}
h_{\rm NC}(0)=\max\left[{E\over F},{1-E\over (1+\sqrt{F})^2}\right].
\end{equation}
%The phase diagram for $T=0$ presented by Sakuma \emph{et al.} is depicted in Fig.~\ref{NucleationPinningFigure}.

In the present paper, we study this problem in a microscopic spin system on a lattice with the shape of a long rod (Fig.~\ref{ModelImage}), modeled by the Hamiltonian
\begin{equation}\label{microscopichamiltonian}
\cH=-\sum_{\langle i,j\rangle}A_{i,j}\bS_i\cdot\bS_j-\sum_{i=1}^NK_i S_{i,z}^2-\sum_{i=1}^NH_i(t) S_{i,z},
\end{equation}
where the nearest-neighbor interaction constants $A_{i,j}$ are positive for all $i,j$, $\{K_i\}_{i=1}^N$ is a set of positive anisotropy constants, and $\bH_i(t)=H_i(t)\be_z$ is an external magnetic field pointing in the $z$-direction. 

We consider a cubic lattice of length $L_x=60$ with height $L_z=6$ and depth $L_y=6$. Each vertex of the lattice contains a spin, which we treat as a classical magnetic moment. 
 We choose units such that $g\mu_\text{B}=1$, where $g$ is the g factor and $\mu_\text{B}$ is the Bohr magneton. 
 We denote the set of spins by $\{\bS_i\}_{i=1}^N$, where $N$ is the total number of vertices in the lattice. 
 We set the magnetization of the spins to be unity, i.e., 
\beq
|\mbold{S}_i|=M_1=M_2=1.
\eeq

The time-evolution of this system is given by the Landau-Lifshitz-Gilbert equation\cite{review,LLG} for each $i=1,\ldots,N$.
\begin{equation}
\frac{\udd}{\udd t}\bS_i=-\frac{\gamma}{1+\alpha_i^2}\bS_i\times\bH_i^\text{\rm eff}-\frac{\alpha_i\gamma}{(1+\alpha_i^2)S_i}\bS_i\times\Big[\bS_i\times\bH_i^\text{\rm eff}\Big].
\label{EqLLG}
\end{equation}
The parameter $\gamma=g\mu_{\rm B}$ denotes the gyromagnetic constant and $\alpha_i$ is the damping parameter. 
The effective field $\bH_i^\text{\rm eff}$ on the $i$th spin is given by
\begin{equation}
\bH_i^\text{\rm eff}\equiv
-\frac{\partial \cH}{\partial \bS_i}=2\sum_{j:\langle i,j\rangle}A_{i,j}\bS_j+[2K_i S_{i,z}+H_i(t)] \be_z.
\end{equation}
We include thermal effects by adding a white Gaussian noise field, denoted by $\{\bxi_i(t)=(\xi_i^x,\xi_i^y,\xi_i^z)\}_{i=1}^N$, to $\bH_i^\text{\rm eff}$. Explicitly, the noise field satisfies the following properties:
\begin{equation}
\langle \xi_i^j(t)\rangle=0,\quad \langle \xi_i^j(t)\xi_k^l(s)\rangle=2 {D}_i \delta_{ik}\delta_{jl}\delta(t-s). 
\end{equation}
With the inclusion of the noise field, we treat the stochastic Landau-Lifshitz-Gilbert (SLLG) equation as a Langevin equation with the Stratonovich interpretation. 

If the following relation~\cite{Garcia,NishinoLLG} 
\begin{equation}
\frac{\alpha_i}{S_i}=\frac{\gamma {D}_i}{k_{\rm B}T},
\end{equation}
is satisfied, the system relaxes to the canonical equilibrium distribution $P_\text{eq}(\{\bS_i\}_{i=1}^N)\propto \exp [-\beta \cH(\{\bS_i\}_{i=1}^N)]$. 
Even in the case of inhomogeneous magnetic systems ($S_i \neq S_j$), 
any choice within this condition realizes the canonical equilibrium state~\cite{NishinoLLG}. However, careful thought must be given to the choice of ${D}_i$ and $\alpha_i$, which depend on $S_i$ and cause different relaxation processes.
In this study, however, we treat homogeneous magnetic moments, i.e., $S_i=1$, and we do not meet this problem.

We carried out simulations of the model by integrating Eq.~\eqref{EqLLG} numerically using a middle point method\cite{NishinoLLG} which is equivalent to the Heun method\cite{Garcia}. 

\subsection{Parameterization of the model}

The width of the domain wall is given by
\beq
\xi=\sqrt{A\over 2K}.
\eeq
In our simulations, the width of region II is 20, and we choose $\xi$ smaller than this width.

We investigate the case in which the bulk regions have the properties of a hard magnet, while region II has weaker magnetic properties. 
Therefore, we set the constants $A_{i,j}=A_1$, $K_i=K_1$ inside regions I and II, and  $A_{i,j}=A_2 < A_1$, $K_i=K_2 <K_1$ in region II.
%, with $A_1\geq A_2>0$ and $K_1\geq K_2>0$. 
%% check!!
The interaction constant $A_{i,j}$ is taken such that, if both the $i$th and $j$th particles belong to region II, $A_{i,j}=A_2$, while if any of the two particles belongs to regions I or III, $A_{i,j}=A_1$.

It is known\cite{TC} that the critical temperature of the classical Heisenberg model is about $T_{\rm c}\simeq 1.443A$ for $K=0$. 
In the model with $K=0.2A$, the critical temperature increases slightly. 
The anisotropy is defined by the anisotropy energy as $H_{\rm A}=2K/M$, 
or by the anisotropy constant given by ${\cal D}=K/M^2$. 
Hereafter we use only the anisotropy energy $K$ to avoid confusion between the anisotropy constant at each region ${\cal D}_i$ and the strength of the random field at site $i$, $D_i$.

To obtain a rough estimation of the temperature dependence of the ordering property, 
we study the temperature dependence of the square of the magnetization $\langle m_z^2(T, H=0)\rangle$,
\beq
\langle m_z^2\rangle ={\langle  \left(\sum_iS_{i,z} \right)^2 \rangle \over N^2},
\eeq
and also the temperature dependence of the anisotropy energy $K(T)$ of the bulk system
for a system of $N=20^3$ spins and $K=0.2$. 
%In Fig.~\ref{TCK}, we depict  $\langle m_z^2(T, H=0)\rangle$ and $K(T)$
In Appendix A, we depict  $\langle m_z^2(T, H=0)\rangle$ for various values of $K$.
% and $K(T)$ of systems with several values of $K(0)$. 
%for a system of $N=20^3$ spins and $D=0.2$.
There, we find that the critical temperature does not depend largely on $K/A$.
There are several ways to estimate the temperature dependence of $K(T)$.\cite{KTsakuma} 
For example,
the temperature dependence of the anisotropy field $H_{\rm A}(T)=K(T)/M$ is defined to be the magnetic field at which the magnetization curve in the easy axis $m_z(H)$ and an extrapolated magnetization in the hard direction $m_x(H)$ meet.\cite{matsumoto}
In the present paper, we define $K(T)$ from the zero field transverse susceptibility, which is explained in Appendix A, where the temperature dependence of the order parameter and the anisotropy $K(T)$ is given for various values of $K(0)$. 
%\begin{figure}[t]
%  \centering
%   {\includegraphics[width=0.4\textwidth]{TC202020H0-eps-converted-to.pdf}}\                 
%  \caption{Temperature dependence of  $\langle m_z^2\rangle$ (closed circle), and $K(T)$ (closed triangle) for a system of $N=20^3$ spins and $D=0.2$.
%The quantity $K_1\times \langle m_z^2\rangle^{3/2}$ is plotted by open circles.} 
%  \label{TCK}
%\end{figure}
There, we find that the temperature dependence of the anisotropy is more significant than that of the spontaneous magnetization $m_s(T)$.
The Callen-Callen law\cite{Callen}, $K(T)\propto m_s(T)^3$, holds for a wide range of temperatures. 
Indeed, for the case $K(0)=0.2$, this relation holds approximately for all temperatures.
%The temperature dependence of $K(T)/2$, plotted by closed triangles, almost agrees with that of  $\langle m_z^2\rangle^{3/2}$, which is plotted by open circles in Fig.~\ref{TCK}.

To study the field thresholds for the system depicted in Fig.1, we apply a uniform external magnetic field $H_i(t)=H$ in the negative $z$ direction in the system with the following two initial conditions:\\
(1) In order to study the process of nucleation in region II, we start our simulations from the configuration in which all the spins point in the positive $z$ direction $(+++)$, \\
and\\
(2) In order to study the process of domain wall propagation, we start our simulations from the configuration in which the spins in region III  point in the negative $z$ direction, while the rest of the spins point toward the positive $z$ direction $(++-)$. \\
%The initial condition (1) allows us to study the threshold field for nucleation, while the initial condition (2) represents a situation in which the domain wall propagates to the left.

%In the present paper, we set the parameter for the noise amplitude
%%check!!
%$D_i=0.1\alpha k_{\rm B}T/\gamma$ with $\alpha/\gamma=1$. 
%%
%%miya0806 Note(1)
%In present paper, we set the parameter for the noise amplitude and the damping constant, respectively in the following forms: 
%$D_i= D^0 k_{\rm B}T$, and $\alpha= D^0\gamma$, which satisfies the relation (13). 
%Because we have set $S_i=1$ and $\gamma=1$, and thus $\alpha=D^0$. Here adopted $D^0=0.1$ which gives the stable simulation.
%%
%%miya0811
In the present paper, we set the parameter for the noise amplitude 
$D_i=\alpha_i k_{\text{B}}T/\gamma$ with $\alpha_i=0.1$ and $\gamma = g\mu_\text{B}=1$.
We take $A_1$ as the unit of energy and measure the temperature in this unit.
To determine the threshold fields, we perform simulations for $5\times 10^5$ updates with time steps of value $\Delta t=0.01$.
In the present notation the period of precession is of order O(1), and $\Delta t=0.01$ is small enough to simulate the situation. 
%In order to keep the spin length constant we use normalization of spin length every 2000 updates.

If we simulate for a longer time the results may change, but we regard the observation time $t=5\times 10^3$ to be enough to grasp the dependence of the threshold on its parameters. 
Usually, an observation time of 1s corresponds to a simulation time of order $t\sim 10^{12}$, which is much longer than $t=5\times 10^3$.
However, the change of relaxation time around the threshold is very sharp as is usual in critical phenomena. 
In Appendix B, we show examples of the field dependence of the relaxation time observed in long simulations. %$t=5\times 10^4$.
Thus, we expect that  the estimation of the thresholds of the field do not depend largely on the observation time.

We classify the final configurations by specifying the signs of the magnetization in the three regions
$ (m_{\rm I},m_{\rm II},m_{\rm III})$.
For example, $(+++)$ denotes the configuration where no nucleation occurs, $(+-+)$ denotes the case where  nucleation occurs but the reversed magnetization does not propagate, and $(---)$ denotes the case where nucleation occurs and the reversed magnetization propagates. 
There also exist the case of $(+--)$ when we start from $(++-)$ to study the domain wall propagation phenomenon.

We define $h_{\rm NCII}(T)$ as the boundary between the fields for which the final configurations are  $(+++)$ and  $(+-+)$,
$h_{\rm NC}(T)$ as the boundary between the fields for which the final configurations are  $(+-+)$ and  $(---)$ or between 
 $(+++)$ and  $(---)$, and $h_{\rm DWP}(T)$ as the boundary between the fields for which the final configurations are  $(+--)$ and  $(---)$.
 
\begin{figure}[t]
  \centering
$$
\begin{array}{cc}
%  {\includegraphics[width=0.4\textwidth]{T00F07E007H02v2.png}}&
%  {\includegraphics[width=0.4\textwidth]{T00F07E007H03v2.png}}\\
  {\includegraphics[width=0.4\textwidth]{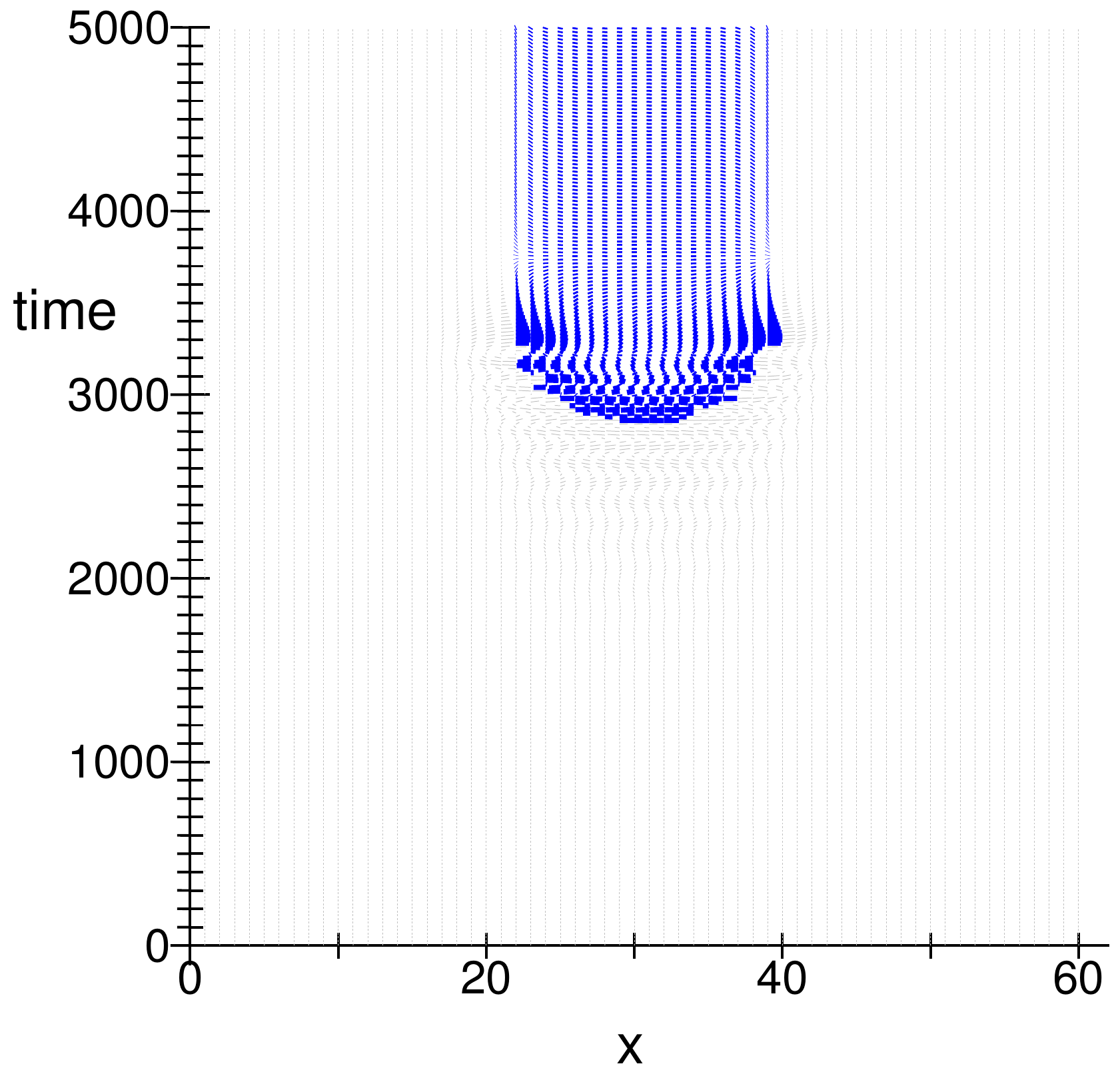}}&
  {\includegraphics[width=0.4\textwidth]{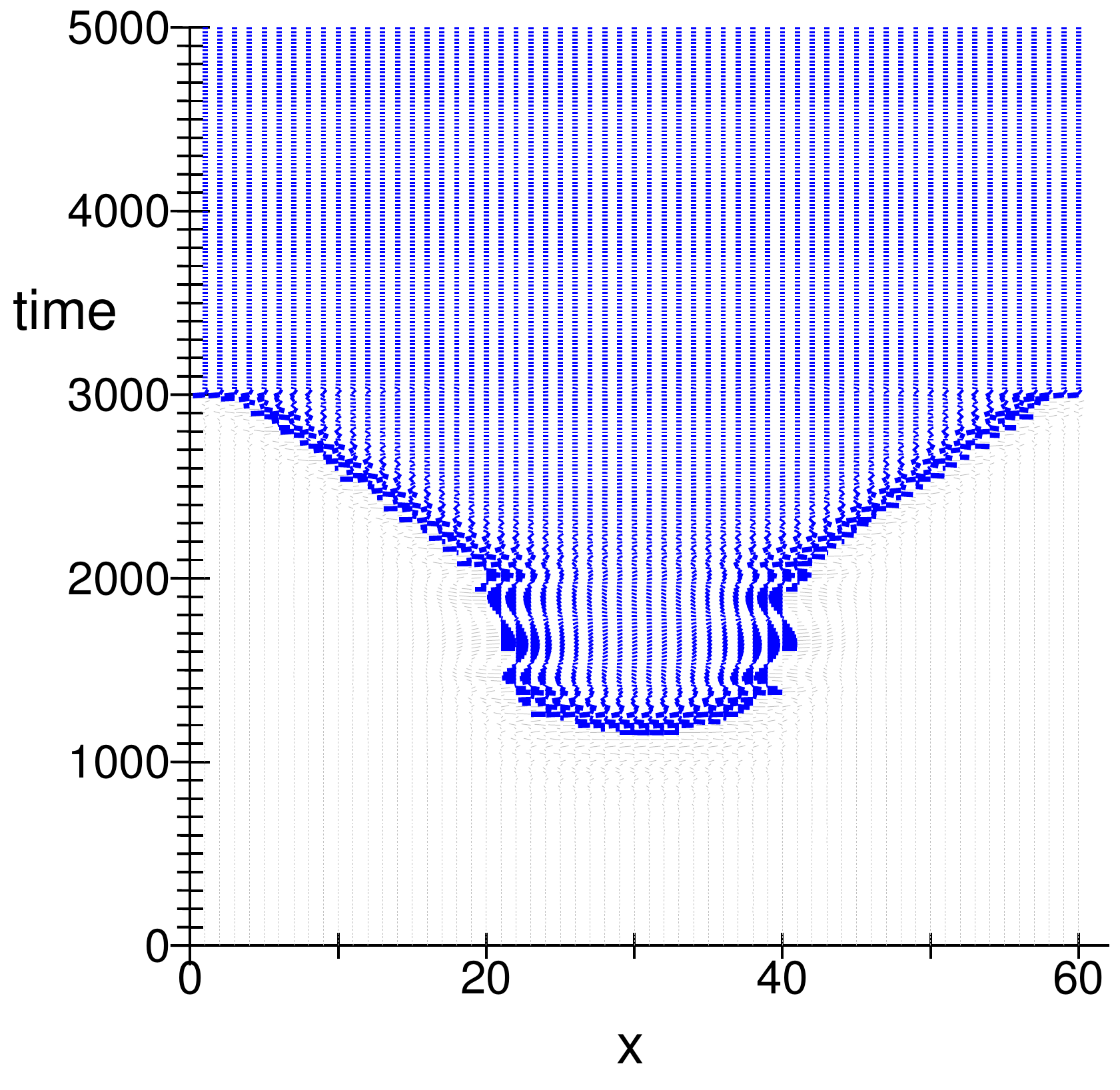}}\\
  {({\rm a})} &  {({\rm b})}
  \end{array}
$$ 
\caption{Time evolution of magnetization for $F=0.7$, $E=0.07$ and (a) $h=0.2$, and 
  (b) $h=0.3$ at $T=0$.
  Each row denotes a configuration of spins at the site $(x,4,4),x=1,\cdots 60$ 
at a time $t$.
  The vertical axis denotes the time.  
%The direction of spin at the site $(x,4,4)$ at the time is shown by an arrow. 
  Spins of positive and negative $S_z$ are plotted by thin gray bar and bold blue bar, respectively. This notation is used in other plots of configurations in this paper.}    
  \label{confF07E007h}
  \end{figure}

We give examples of $(+-+)$, and  $(---)$ in Fig.~\ref{confF07E007h}(a) and (b), respectively. 
There, we depict the time evolution of the spin configuration.
To represent the configuration we draw spins at a line of the system $(x, y=4,z=4), x=1,\ldots 60$ at a set time in rows, and the vertical axis represents the time evolution.  

%\section{Simulation results}

\section{Temperature dependence of nucleation}\label{sectionnucleation}

Starting from the initial condition (1) with $(+++)$, we investigate the threshold fields for the nucleation in region II. 
We studied the time evolution of systems with $F=A_2/A_1<1$ and $E=FK_2/K_1 \le 1$, for $t=5\times 10^3$ and various values of $h=H/2K_1$ to find the threshold fields.

\subsection{$T=0$}

First, we study the phase diagram at $T=0$.
Figures~\ref{confF07E007h}(a) and (b) present instances of processes of nucleation and propagation from the nucleated reversed magnetization, respectively. 
For small values of $h$, nucleation does not occur (not shown), and as $h$ increases, nucleation begins to occur.
In Fig.~\ref{confF07E007h}(a),
for $h=0.2$ we find nucleation at around $t=2700$. There, the reversed magnetization remains inside the defect region.
Thus, $h=0.2$ is between $h_{\rm NCII}(0)$ and $h_{\rm NC}(0)$.
 For a larger field, the reversed magnetization propagates into the bulk hard magnetic region.
 We depict an example in which $h=0.3$ in Fig.~\ref{confF07E007h}(b).
 
\begin{figure}[!t]
$$
\begin{array}{ccc}
  {\includegraphics[width=0.3\textwidth]{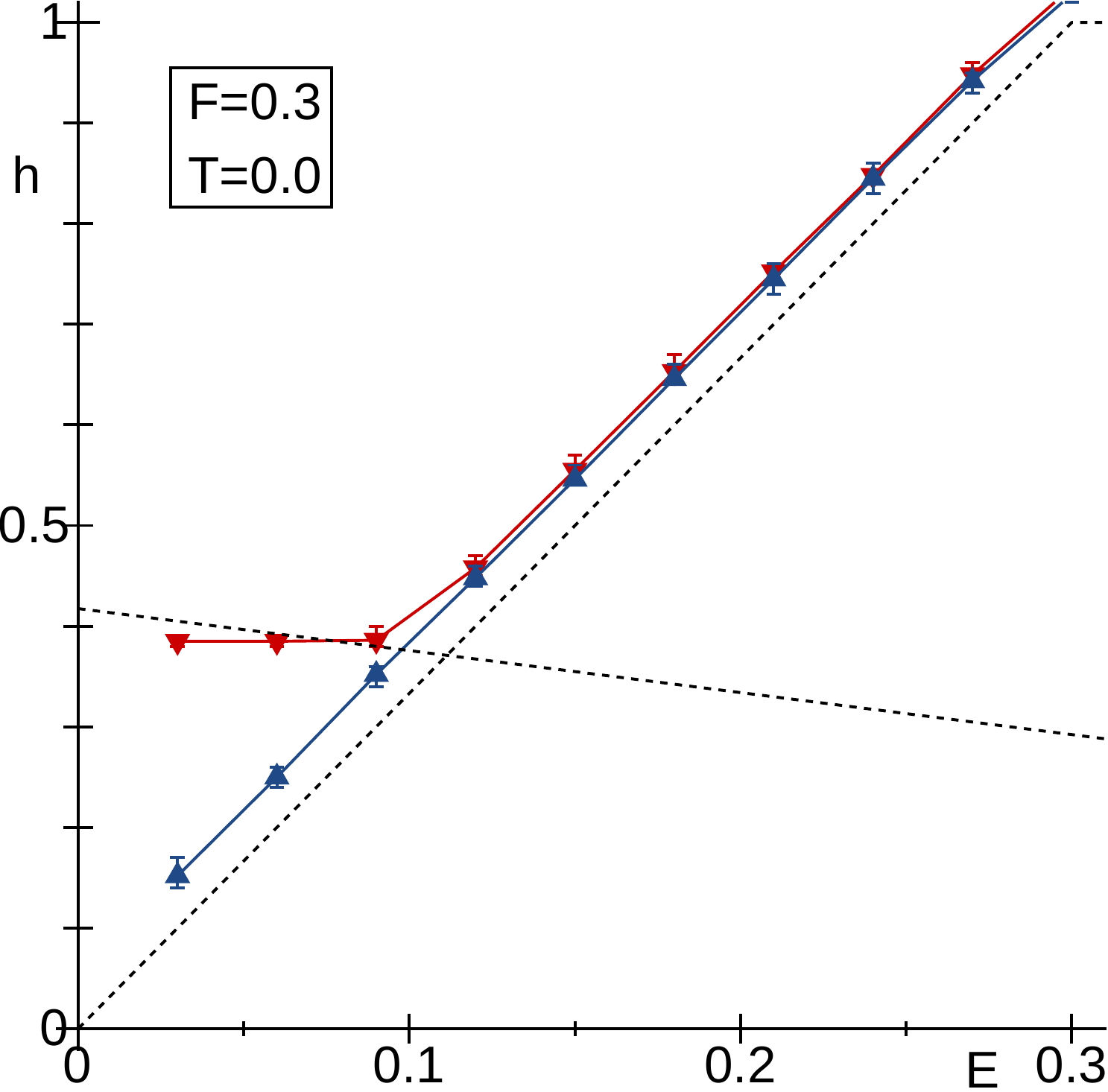}}&               
  {\includegraphics[width=0.3\textwidth]{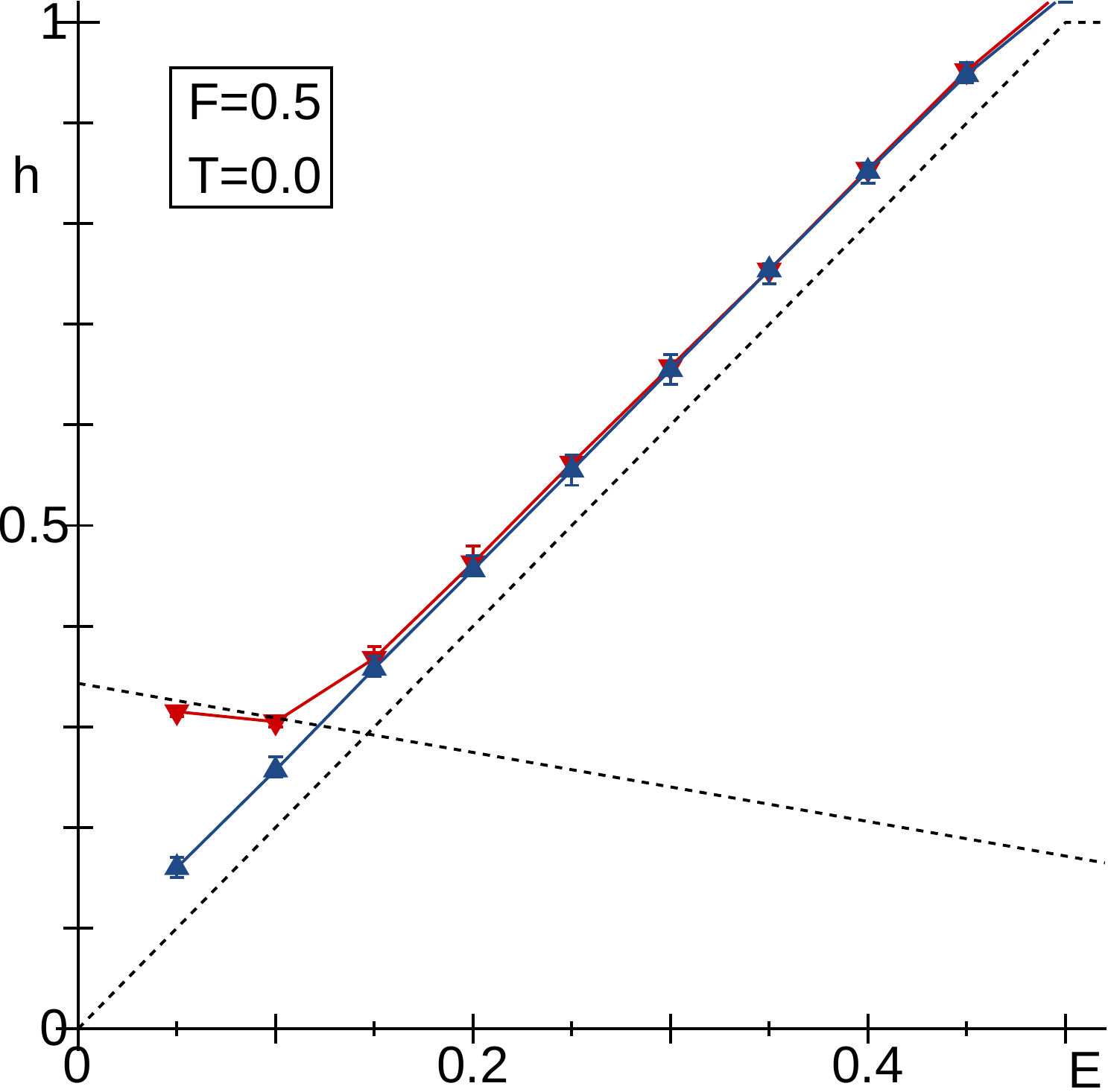}}& 
  {\includegraphics[width=0.3\textwidth]{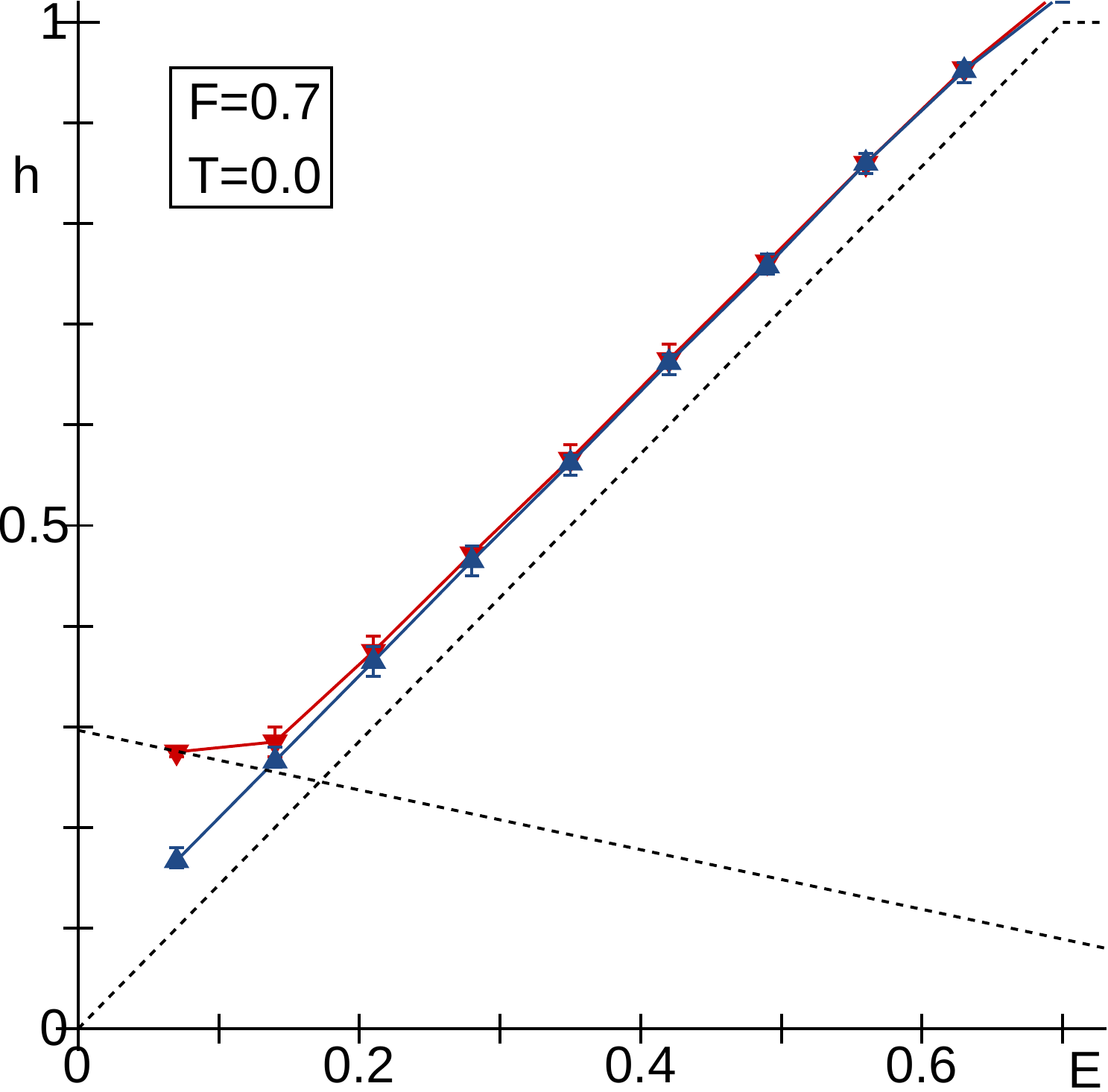}}\\
  {({\rm a})} &  {({\rm b})} &  {({\rm c})}
  \end{array}
 $$ 
\caption{
Phase diagram ($T=0$) of the final configuration starting from the initial condition ($+++$) for 
(a) $F=0.3$, (b) 0.5 and (c) 0.7.
The border between $(+++)$ and $(+-+)$ is given by blue upward triangles. The upper limit of the error bar denotes the field above which $(+-+)$ appears
 and the lower limit denotes the field below which $(+++)$ appears.
 Similarly, the border between $(+-+)$ and $(---)$ is given by red downward triangles, and $(---)$ appears above the upper limit of the error bars, while $(+-+)$ appears below the lower limit.
 The blue and red lines are guides for the eye.
  The dotted lines denote the analytical estimation for the threshold nucleation field, Eq.~(\ref{hn0}), and the threshold domain-wall propagation field, Eq.~(\ref{hp}), for the continuum system.\cite{sakuma}
} 
\label{NucleationT0}
\end{figure}

In Fig.~\ref{NucleationT0}, the phase diagrams for $T=0$ for $F=0.3, 0.5$, and 0.7 are shown.
%The dashed lines denotes the analytical estimation for nucleation (Eq.~\ref{hn0}), and 
The dotted lines show the threshold fields for the nucleation, $h_{\rm NCII}(0)$, and 
domain wall propagation, $h_{\rm NC}(0)$, given by Eq.~(\ref{hn0}) and Eq.~(\ref{hp}), respectively.\cite{sakuma}
The borders between the cases of final configurations $(+++)$, and  $(+-+)$ are plotted by blue upward triangles, 
and the borders between  $(+-+)$ and $(---)$ are plotted by red downward triangles.
Here at $T=0$, the error bars denote the step size of the external field, $\Delta h=0.01$.
%minimum field of $(+-+)$ and the maximum field of $(+++)$ which we investigated.
The small deviation of the threshold of nucleation to the theoretical estimation is due to the fact that the width of region II is fixed to be $W=20$. 
Namely, there is correlation from the hard magnet in the defect region due to its finite size, and this correlation stands in the way of the nucleation phenomenon. Consequently, the threshold in the simulation is larger than the analytical estimation, but the overall features are well reproduced.

It should be noted that at zero temperature, if we start from the completely aligned initial configuration, the initial state remains unchanged because it is an unstable stationary state. To avoid this situation, we introduced a small fluctuation to the angle of the magnetization $(\theta,\phi)$ with a Gaussian distribution of 
standard deviation $\sqrt{\langle (\theta-\theta_0)^2\rangle}=0.01$ with $\theta_0=0.03$[radian]. 
We confirmed that our results have little dependence on the choice of $\theta_0$.

\begin{figure}[!b]
$$             
  {\includegraphics[width=0.4\textwidth]{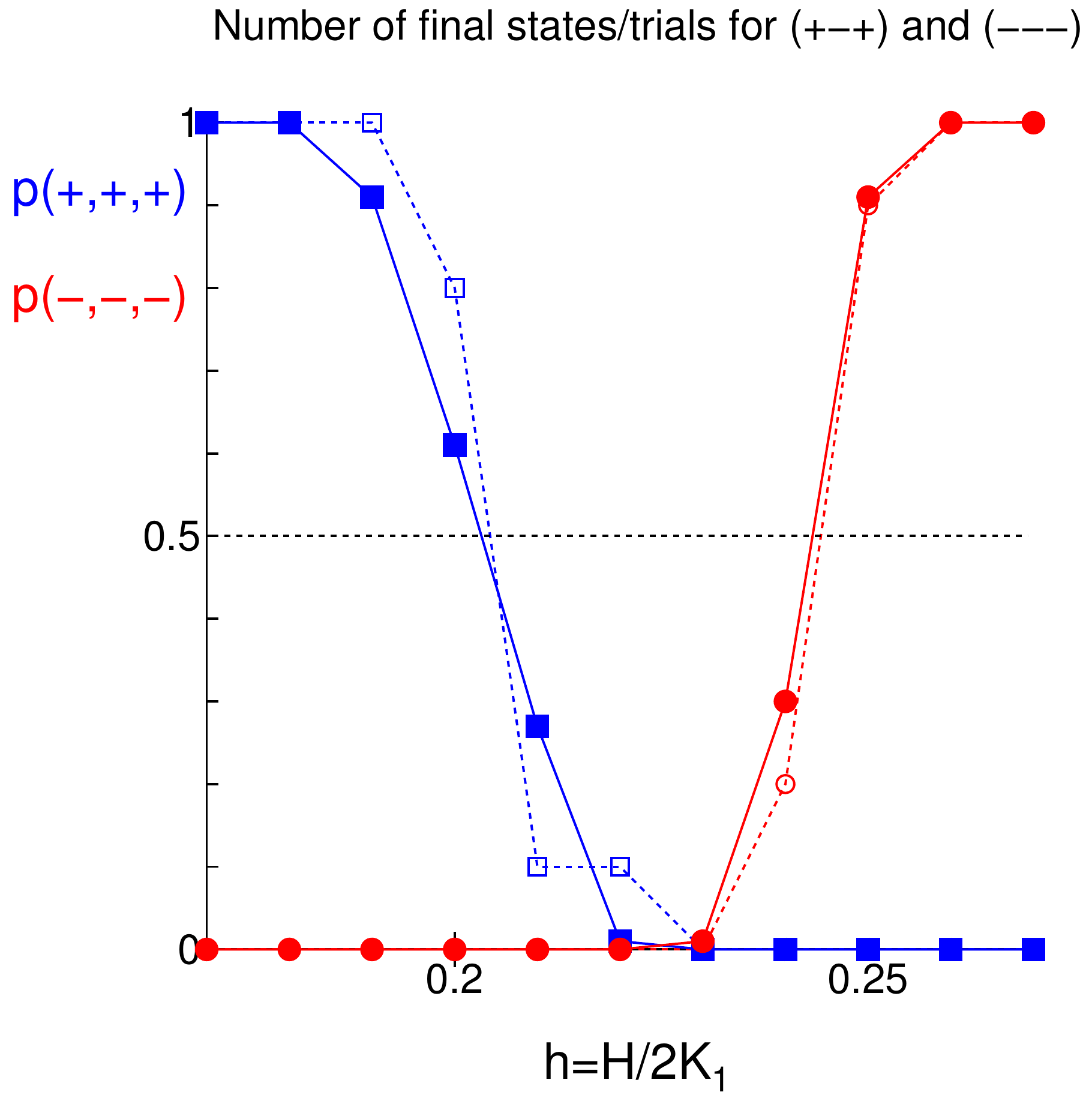}}
 $$
  \caption{
The frequency $p(+++,h)$ of cases in which the final state ($t=5000$) has the configuration $(+++)$ is plotted by squares. The closed squares denote the frequency for each $h$ obtained from 100 samples, while the open squares denote that obtained from 10 samples.
The frequency $p(---,h)$ of cases in which the final state ($t=5000$) has the configuration $(---)$ is plotted by circles. The solid circles denote the frequency for each $h$ obtained from 100 samples, while the open circles denote that obtained from 10 samples.
The frequency $p(+-+,h)$ is given by $1-p(+++,h)-p(---,h)$.
Here we find that the threshold field is well estimated by the sampling of 10 samples, and 
we define the threshold to be the field at which the dotted line crosses 0.5, and define the error bars to be the interval between the fields for $p(m_1,m_2,m_3)=1$ and $p(m_1,m_2,m_3)=0$.
}
\label{Fig5}
\end{figure}

\subsection{$T>0$}

Now, we study the temperature dependence of the phase diagram.
%We fixed the duration of the simulation to be $t=5\times 10^3$, because the change in the threshold field does not change significantly at longer observation times (see Appendix~\ref{AppC}). 
%Region II has a weaker interaction constant, $A_1\geq A_2$, and thus nucleation occurs in this region as shown in Fig.~\ref{confF07E007h}.
Nucleation occurs stochastically in region II, and the corresponding waiting time obeys a Poisson distribution. 
In order to determine the threshold field, we made a histogram of the number of events. 
Namely, we performed simulations of 10 samples for each parameter, and counted the number of cases in which the system showed nucleation within the observation time ($t=5\times 10^3$). 

In Fig.~\ref{Fig5}, we show an example of the rate $p(---,h)$ of samples in which the final configuration was $(---)$, denoted by red open circles, 
and the rate $p({+++},h)$ of the number of samples in which the final configuration was $(+++)$, denoted by blue open squares, for $F=0.7$ and $E=0.14$ at $T=0.1$ when taking 10 samples.
We also performed 100 samples, and show the rates by closed circles and closed squares.
We find that a larger sample number does not change the estimation of the threshold point significantly, and thus we took the histograms with 10 samples in other cases.

We assign an error bar which extends from the point where there are ten (all) occurrences to the point where there are zero occurrences of the configuration in question.  
%It is evident that the threshold point obtained from taking 100 samples well agree with the error bars we propose. 

We identify the threshold as the point where the interpolated line of the histogram crosses $p=0.5$.
For example, the error bar of the threshold field $h_{\rm NCII}(0.1)$ between $p({+++},h)$ and $p({+-+},h)$ is from 0.19 to 0.23, and the threshold point obtained from taking 100 samples lies at $h=0.203$.
Similarly, the error bar of $h_{\rm NC}(0.1)$ between $p({+-+},h)$ and $p({---},h)$ is from 0.23 to 0.26, with the 100-sample threshold estimation lying at $h=0.242$.

\begin{figure}[!t]
%  \vspace*{-20mm}
  \centering             
  {\includegraphics[width=0.3\textwidth]{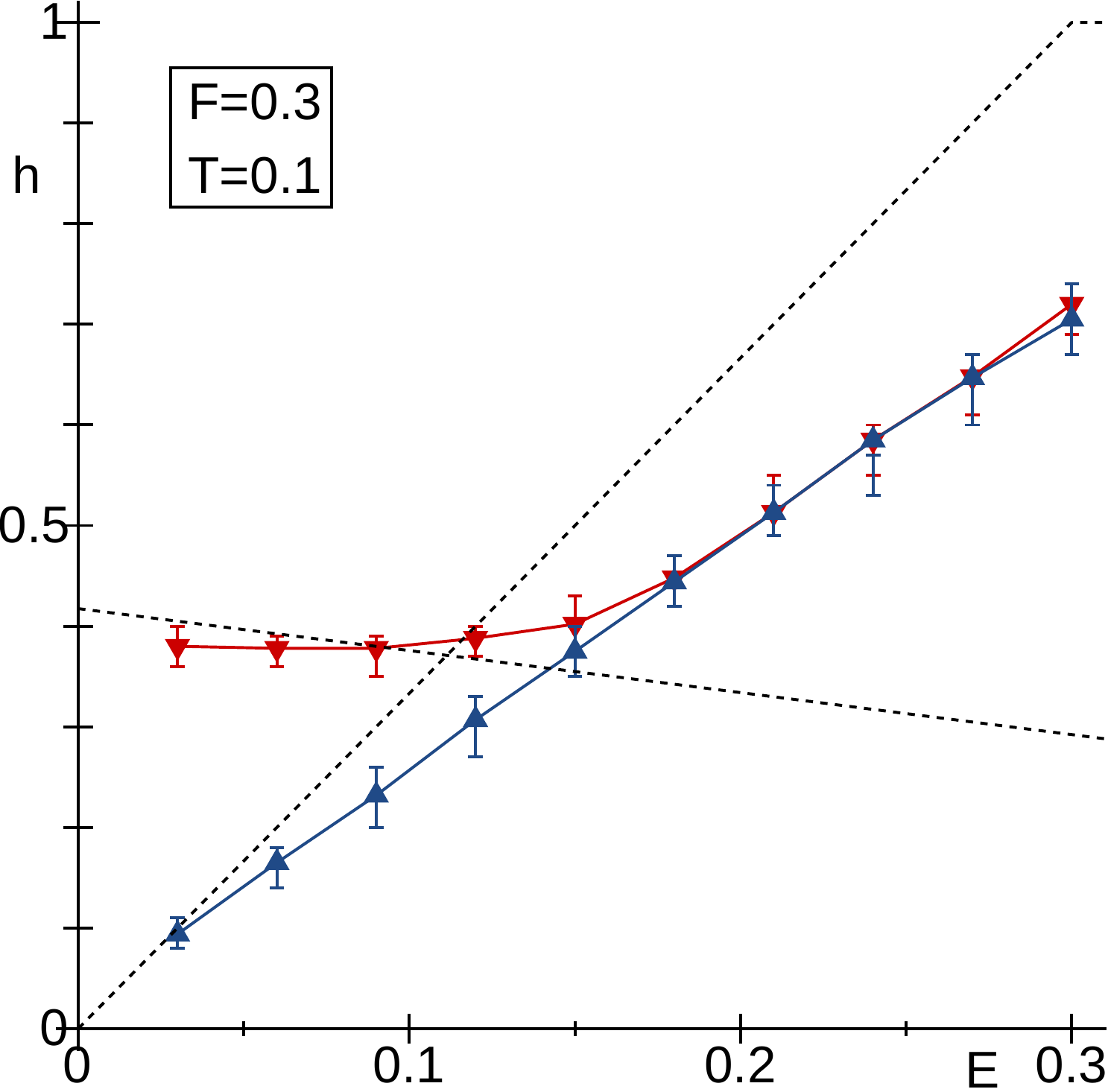}}\ 
  {\includegraphics[width=0.3\textwidth]{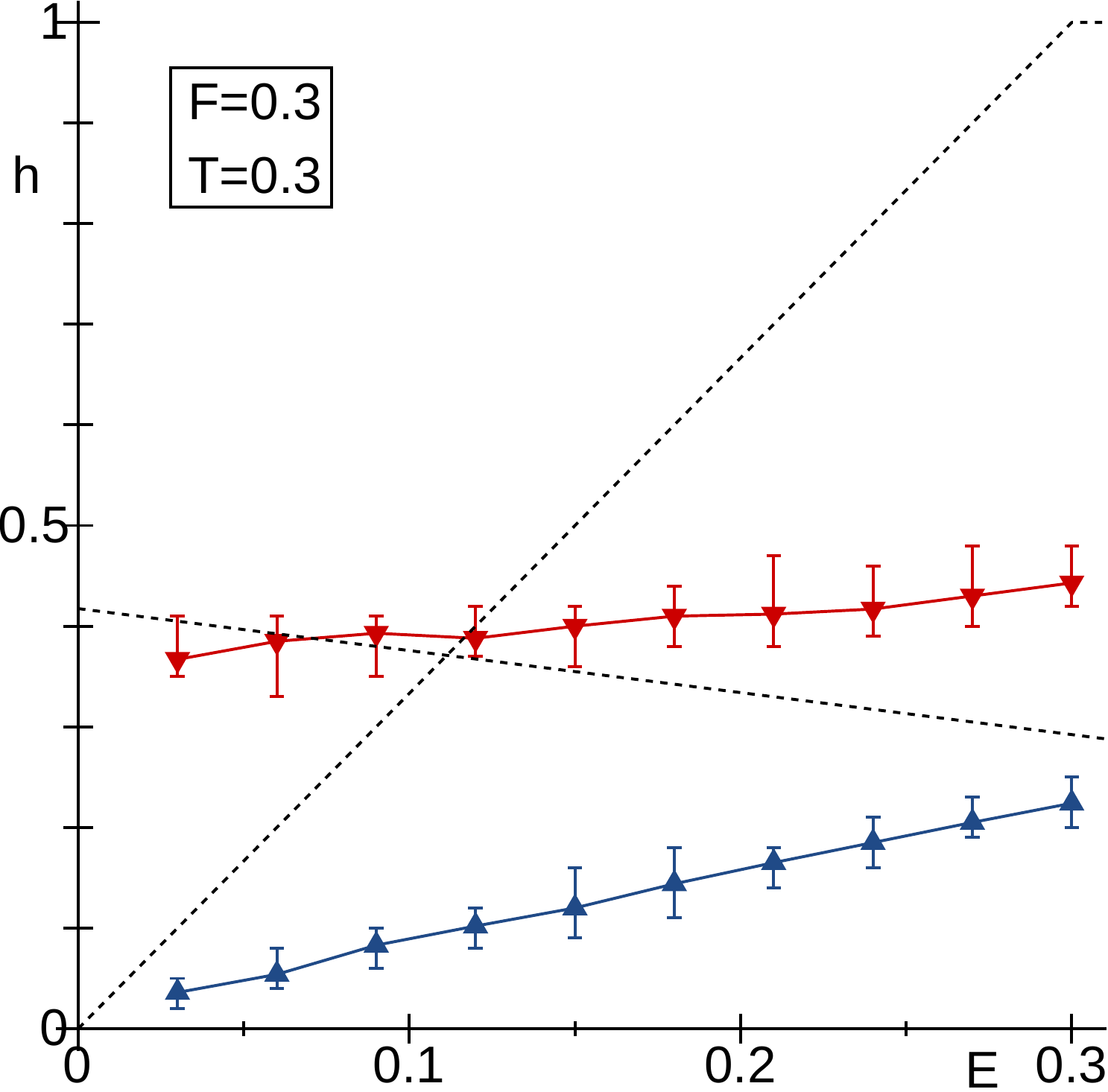}}\ 
  {\includegraphics[width=0.3\textwidth]{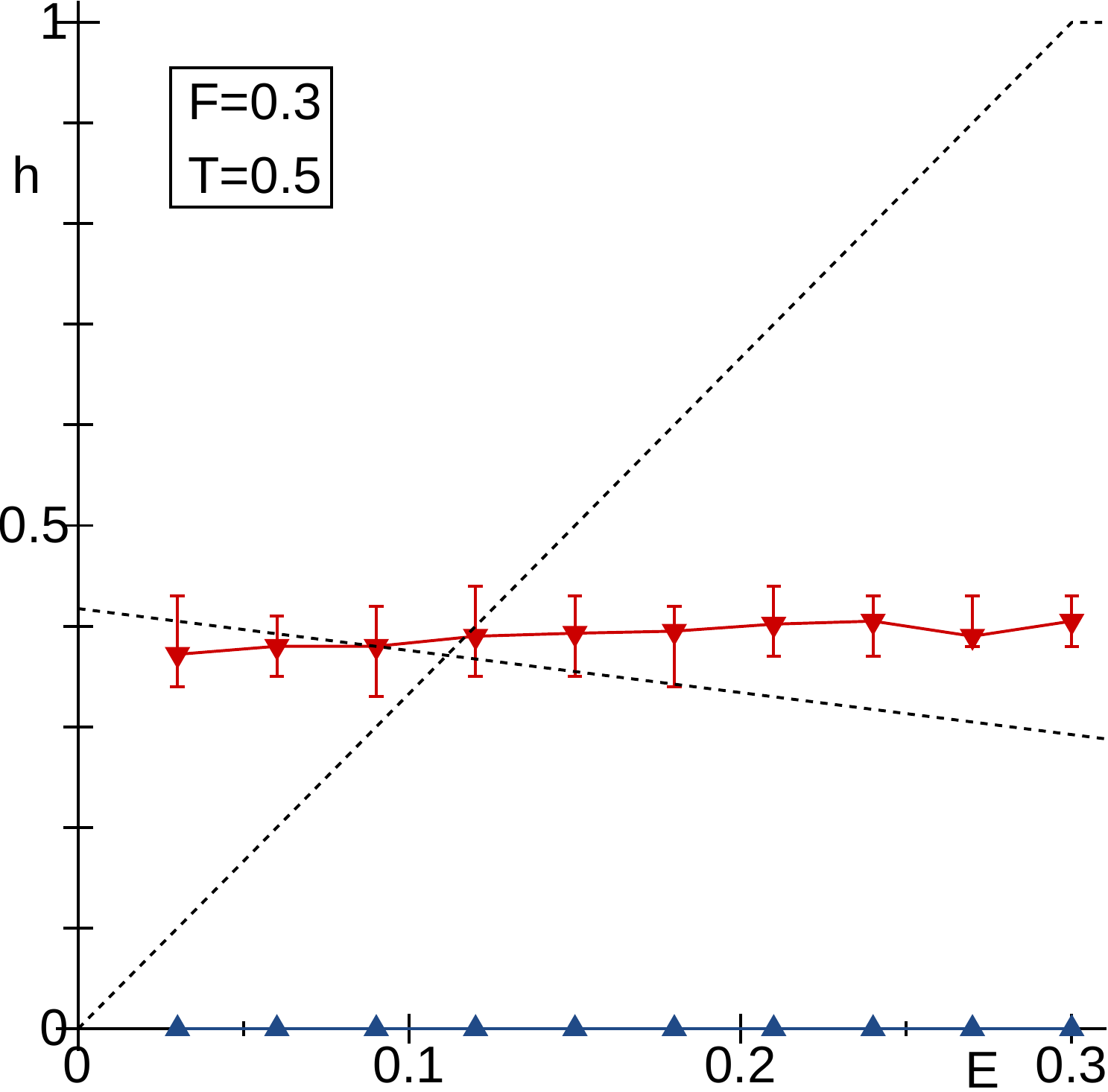}}\\
  
  {\includegraphics[width=0.3\textwidth]{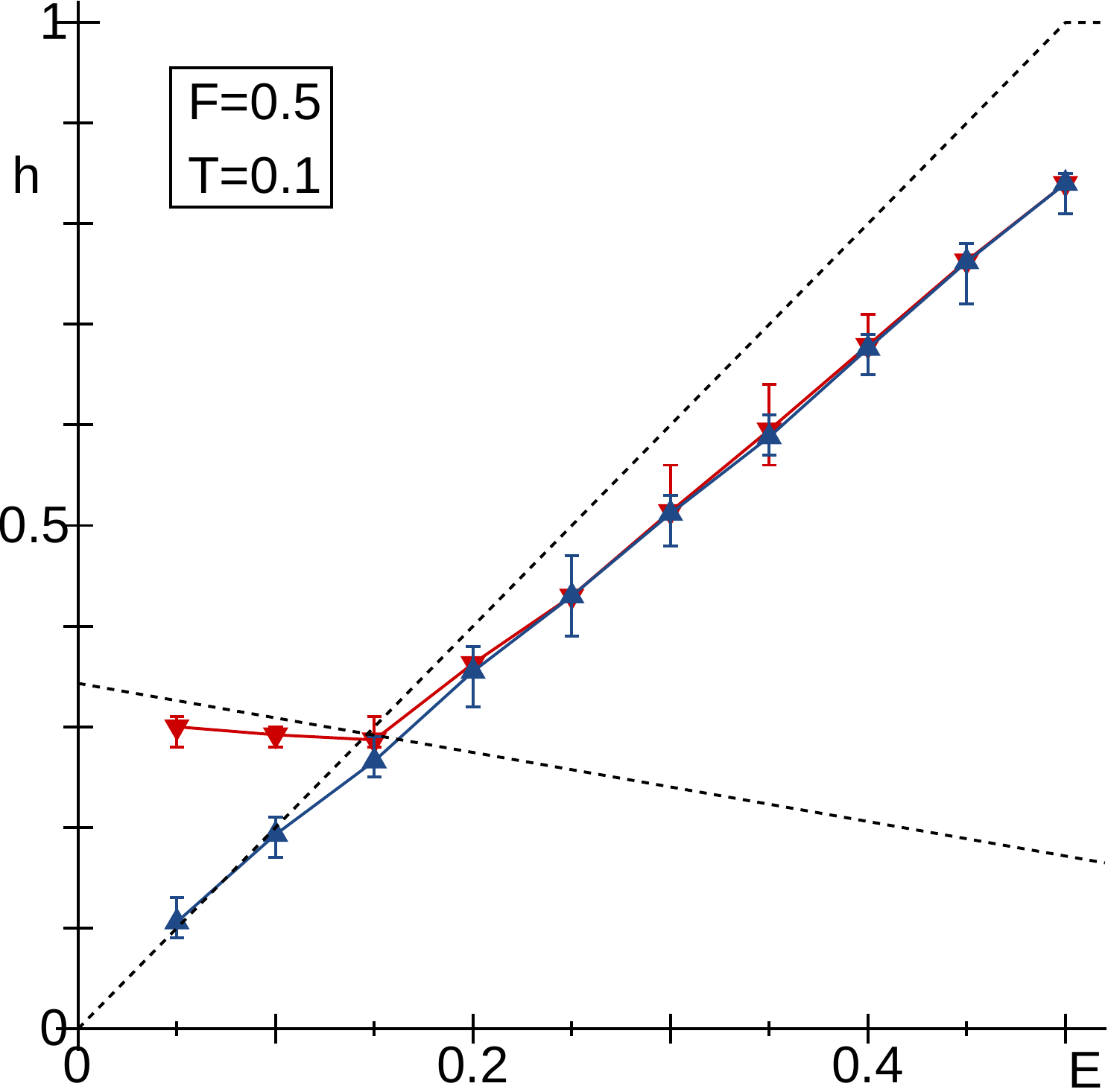}}\ 
  {\includegraphics[width=0.3\textwidth]{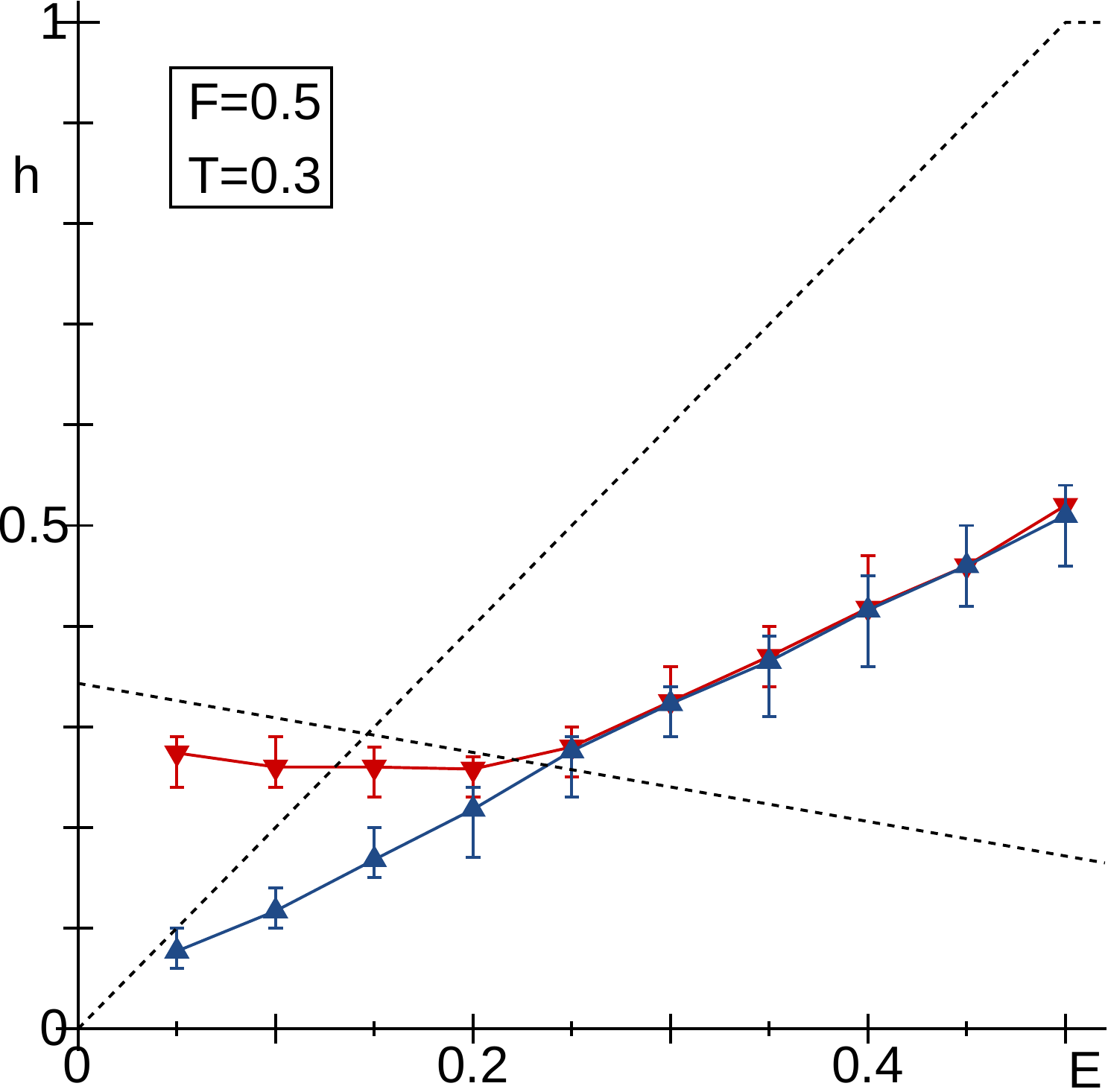}}\ 
  {\includegraphics[width=0.3\textwidth]{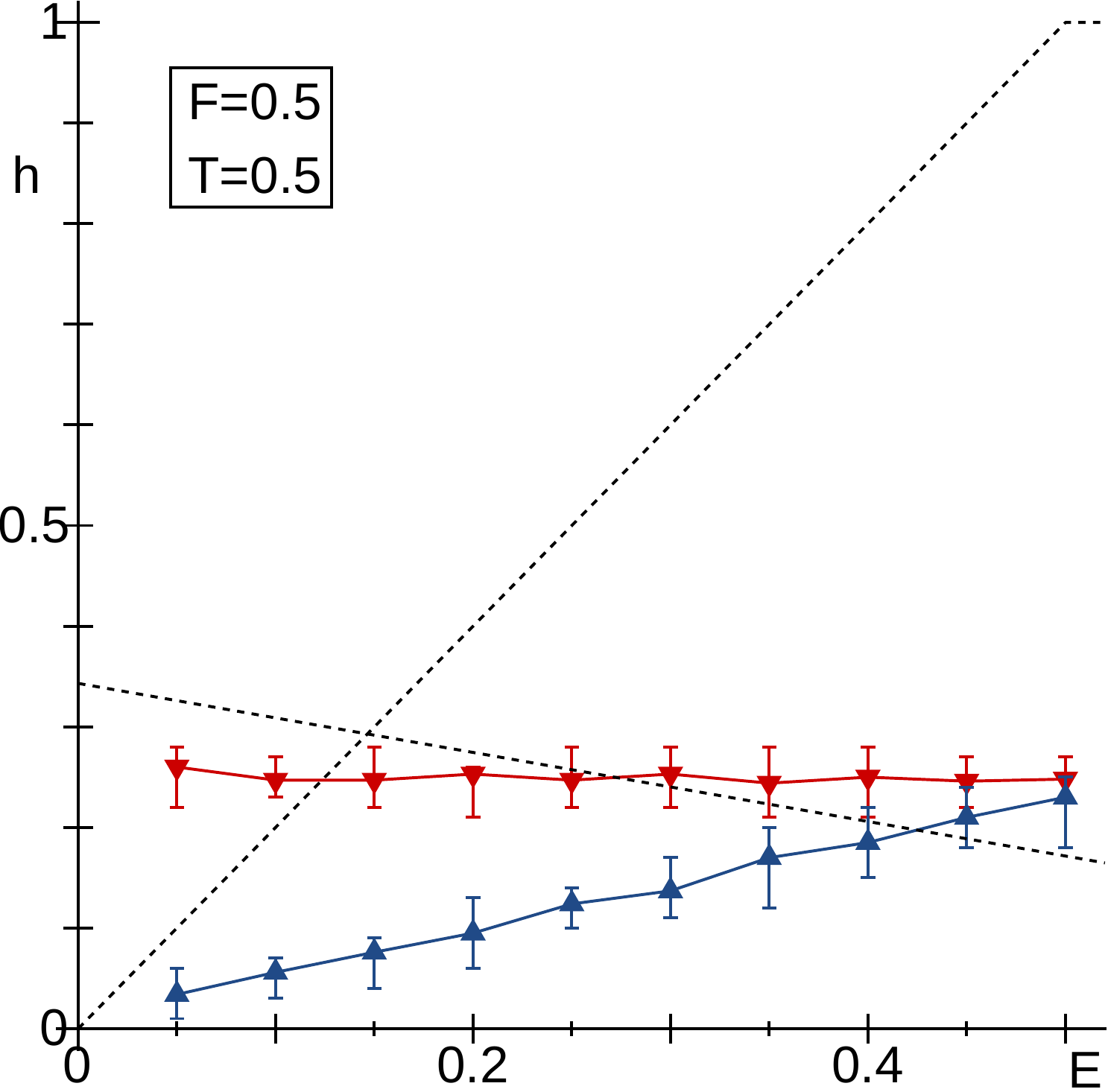}}\\

  {\includegraphics[width=0.3\textwidth]{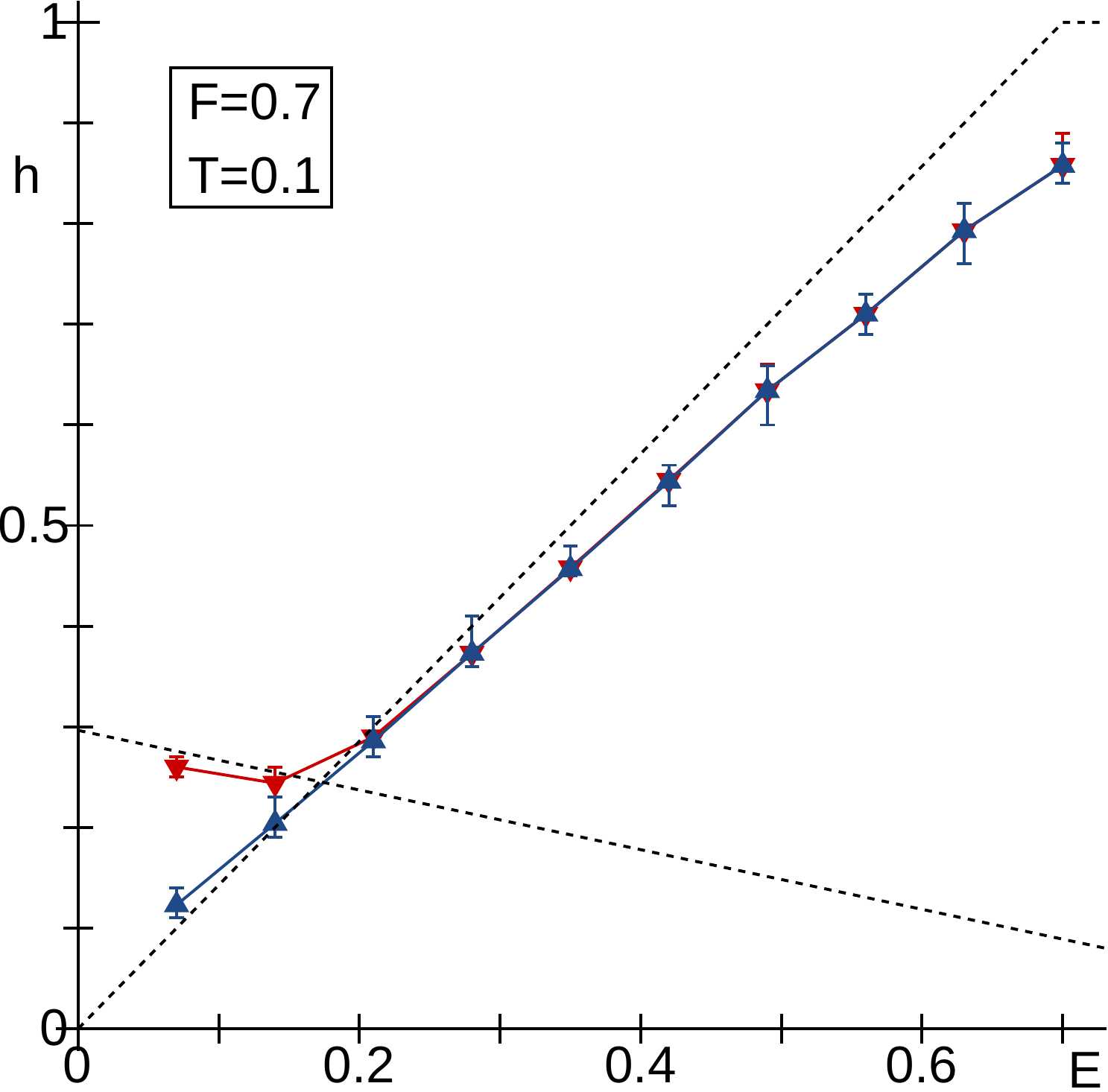}}\ 
  {\includegraphics[width=0.3\textwidth]{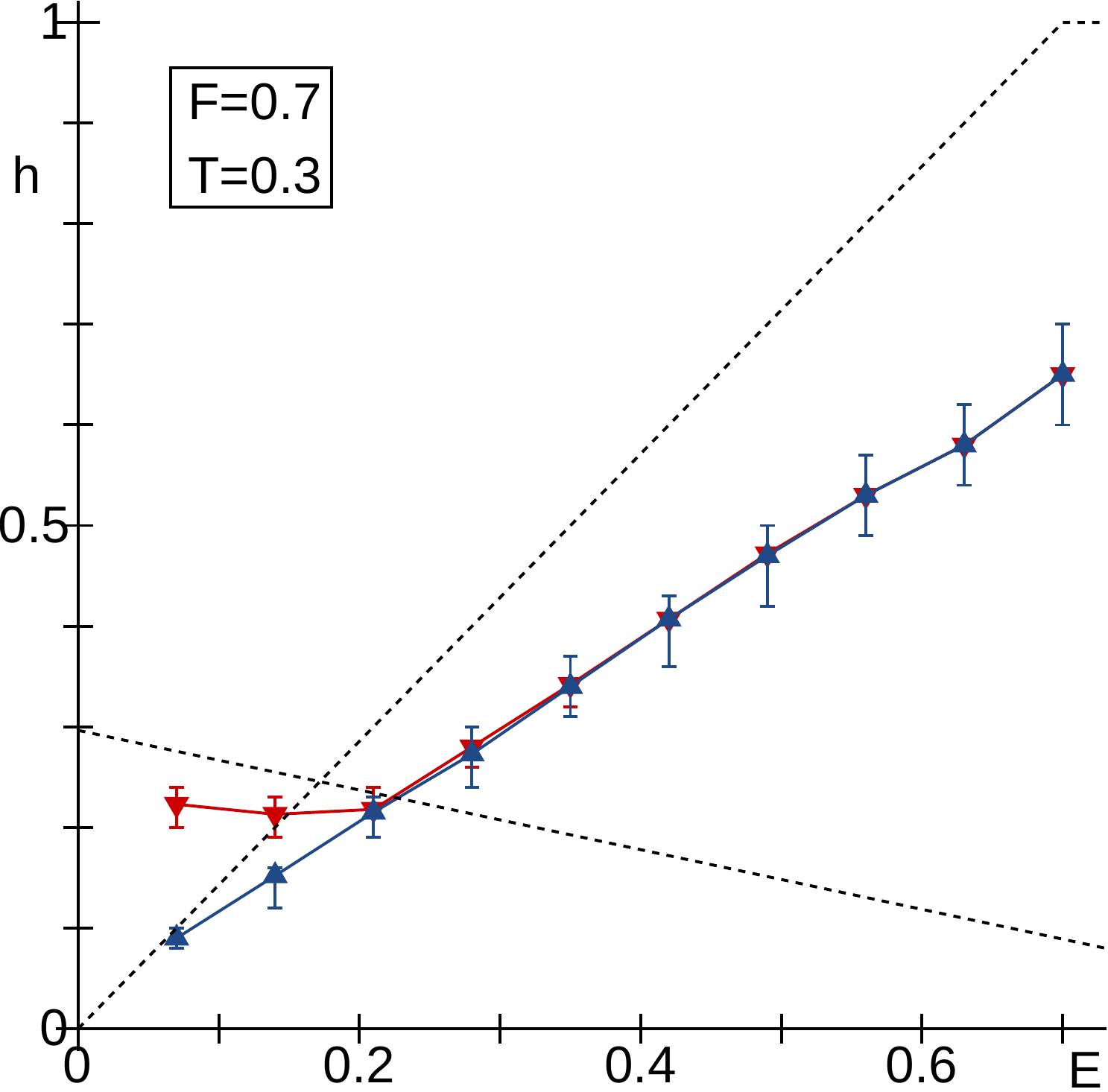}}\ 
  {\includegraphics[width=0.3\textwidth]{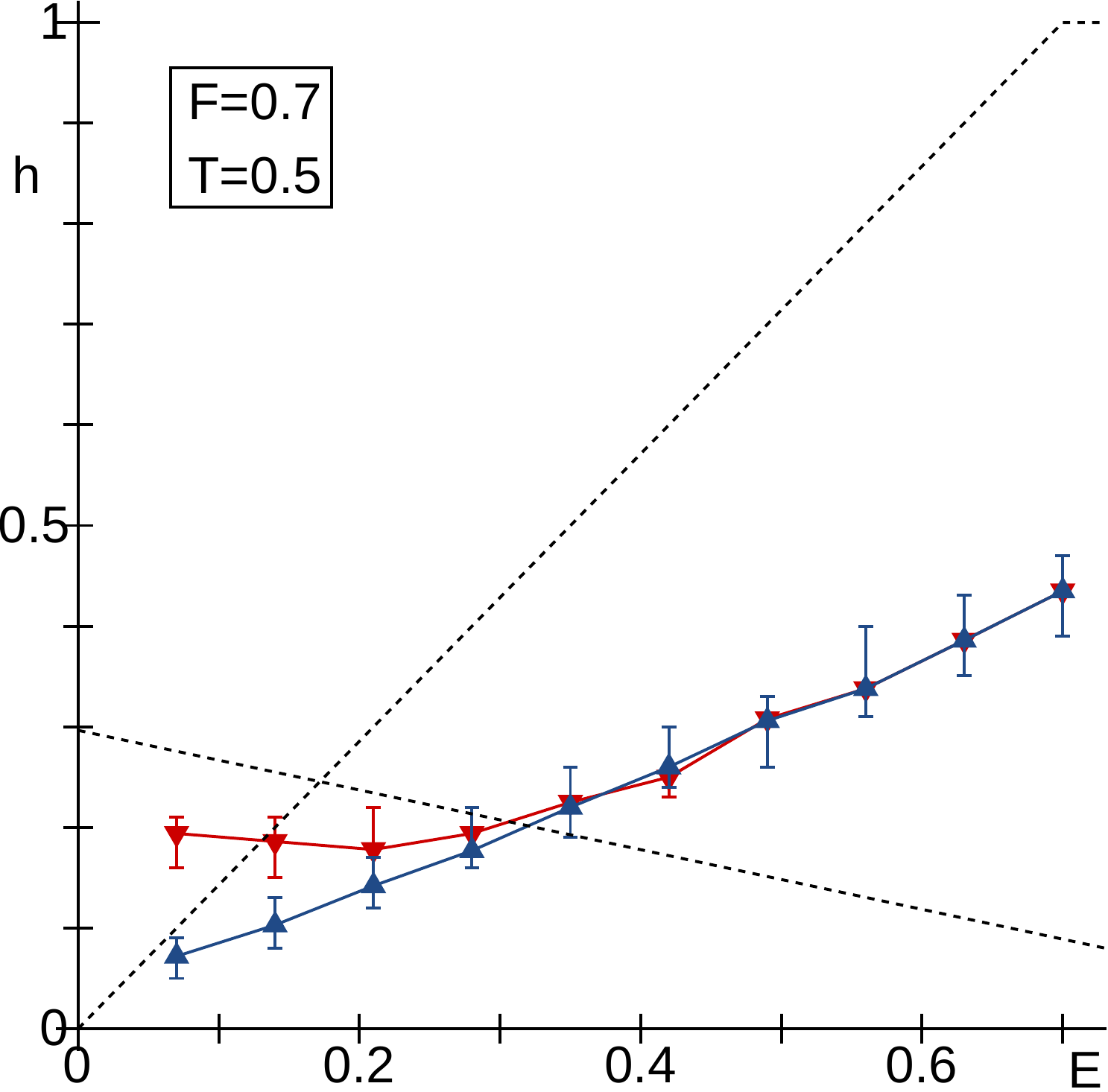}}\\
%  {\includegraphics[width=0.3\textwidth]{FigNCF03T01-eps-converted-to.pdf}}\ 
%  {\includegraphics[width=0.3\textwidth]{FigNCF03T03-eps-converted-to.pdf}}\ 
%  {\includegraphics[width=0.3\textwidth]{FigNCF03T05-eps-converted-to.pdf}}\\
%  
%  {\includegraphics[width=0.3\textwidth]{FigNCF05T01-eps-converted-to.pdf}}\ 
%  {\includegraphics[width=0.3\textwidth]{FigNCF05T03-eps-converted-to.pdf}}\ 
%  {\includegraphics[width=0.3\textwidth]{FigNCF05T05-eps-converted-to.pdf}}\\
%
%  {\includegraphics[width=0.3\textwidth]{FigNCF07T01-eps-converted-to.pdf}}\ 
%  {\includegraphics[width=0.3\textwidth]{FigNCF07T03-eps-converted-to.pdf}}\ 
%  {\includegraphics[width=0.3\textwidth]{FigNCF07T05-eps-converted-to.pdf}}\\
%  \vspace*{20mm}
  \caption{Dependences of the threshold fields between $(+++)$ and $(+-+)$ (blue upward triangle) and between $(+-+)$ and $(---)$
  on $E$ for $T=0.1$, 0.3 and 0.5 with $F=0.3,$ 0.5 and 0.7. The straight lines are guides for the eye, and the dotted lines correspond to the analytical estimation at $T=0$. 
} \label{NucleationT}
\end{figure}

In Fig.~\ref{NucleationT}, we show the dependence of the threshold fields of nucleation at finite temperatures. 
We show diagrams for $T=0.1, 0.3$ and 0.5 at $F=0.3$, 0.5 and 0.7.
We find that the nucleation field decreases significantly as the temperature rises.
%for large values of $E$ (and $\cD_2$), and regions of $(+-+)$ increases.

When $E$ is small, i.e., $K_2$ is small, $h_{\rm NC}(T)$ and $h_{\rm NCII}(T)$ decrease with temperature. 
This dependence is naturally understood as a consequence of thermal fluctuations.
On the other hand, for large $E$ (where $K_2$ is large) $h_{\rm NC}(T)$ and $h_{\rm NCII}(T)$ separate.
In the case where $F=0.3$, the reduction of the threshold $h_{\rm NCII}(T)$ is significant.
As we show in Appendix~\ref{AppA}, the effective anisotropy decreases rapidly at finite temperatures, and
for $F=0.3$, i.e., $A_2=0.3$, the effective anisotropy falls substantially at $T=0.3$. At $T=0.5$, region II is in the paramagnetic 
state, where the concept of nucleation does not apply.  
However, regions I and III with $A_1=1$ are still robust against the external field, which keeps $h_{\rm NC}(T)$ at high values.

%%%%%%%%%%%%%%%%%%%%%%%%%%%%%%%%%%%%%%%%%%%%%%%%%%%%%%%%%%%%%%%%%%%%%%%%
\subsubsection{Temperature dependence of the threshold field $h_{\rm NCII}(T)$}

\begin{figure}[!t]
  \centering
  $$
\begin{array}{cc}
  {\includegraphics[width=0.3\textwidth]{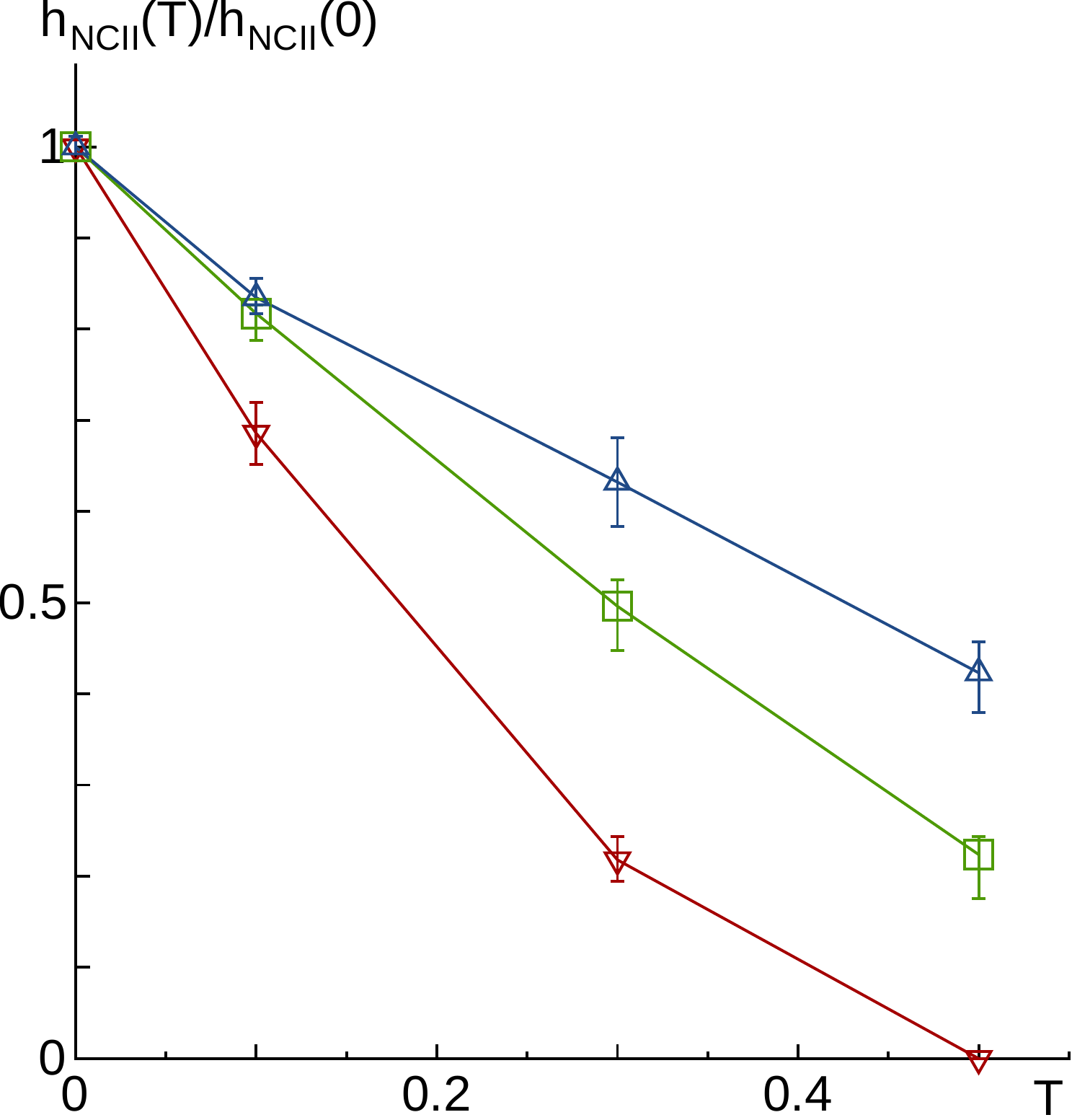}}&               
  {\includegraphics[width=0.3\textwidth]{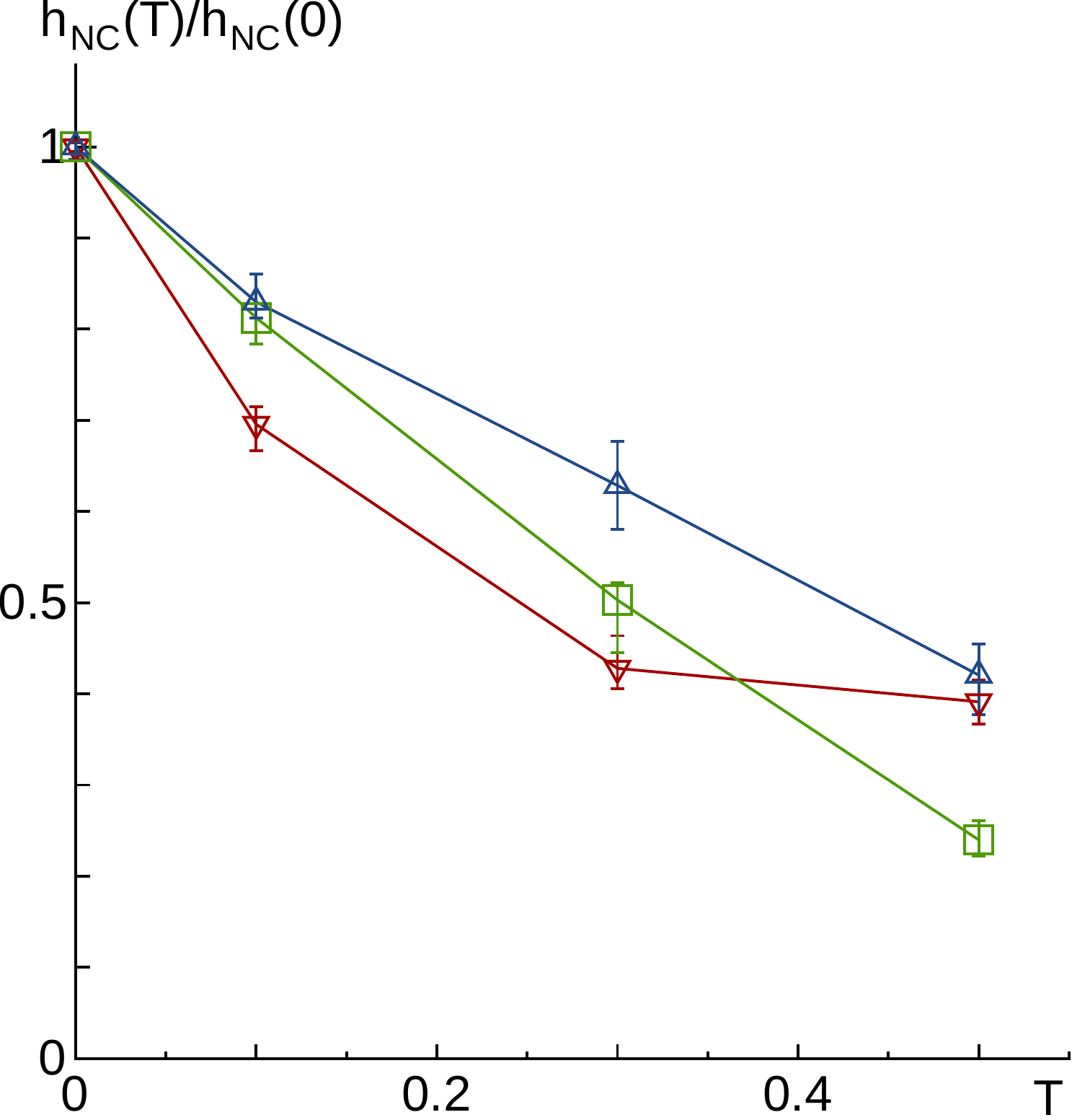}}\\
  {({\rm a})} &  {({\rm b})}
  \end{array}
 $$       
  \caption{(a)
Dependence on temperature of the threshold field between patterns $(+-+)$ and $(+++)$,
 (b) dependence on temperature of the threshold field for nucleation, i.e., between patterns $(---)$ and $(+-+)$, 
 for $E=F=0.3$ (red downward triangle), $0.5$ (green square) and $0.7$ (blue upward triangle).
 The lines between data points are guides for the eye.} 
 \label{NucleationatF}
\end{figure}

In Fig.~\ref{NucleationatF}(a),
we plot the temperature dependence of the threshold field $h_{\rm NCII}(T)/h_{\rm NCII}(0)$  at $E=F$ (i.e., $K_2=K_1$) for $F=0.3, 0.5$ and 0.7. 
The threshold field monotonically decreases with rising temperature.
We also find that $h_{\rm NCII}(T)/h_{\rm NCII}(0)$ decreases monotonically when $F$ decreases. 
That is, if $A_2$ decreases, nucleation in the region becomes easier and 
$h_{\rm NCII}(T)/h_{\rm NCII}(0)$ decreases with $F$, which is natural.

Because $h_{\rm NCII}(T)/h_{\rm NCII}(0)$ is the temperture dependence of the threshold of 
the nucleation in region II, 
we may relate $h_{\rm NCII}(T)/h_{\rm NCII}(0)$ to the temperature dependence of 
the bulk anisotropy energy $K_2(T)$ estimated in Appendix A (Fig.~\ref{MZTKTall}(b)).
That is, from the definition  (\ref{hscaled}), the field $h$ is proportional  to $2K_2$, 
and we may expect the following temperature dependence:
$h_{\rm NCII}(T)/h_{\rm NCII}(0)=K_2(T)/K_2(0)$.

In Fig.~\ref{MZTKTall}(b), we have data of $K(T)$ for a system with $A=1$ for various values of $K(0)$ ($K(0)=0.0, 0.2,\ldots, 1.0$). 
The value of exchange energy in region II is given by $A_2 (=F)$.
We can estimate the temperature dependence $K_2(T)$ of the system with $A_2$ 
by making use of the following scaling for the parameters in the Hamiltonian.
Because in Fig.~\ref{MZTKTall}(b), the parameters are scaled by $A$, i.e., $T/A$ and $K/A$,
we replace the parameters as
\beq
K_{\rm scaled}=K_2\times \left({A\over A_2}\right), \quad 
T_{\rm scaled}=T\times \left({A\over A_2}\right).
\eeq
The temperature dependence $K_2(T)$ is estimated as
\beq
K_2(T)=K_{A=1}\left(T\left({A\over A_2}\right), K_2\left({A\over A_2}\right) \right)\times \left({A_2\over A}\right),
\eeq
where $K_{A=1}(T,K)$ is the dependence given in Fig.~\ref{MZTKTall}(b).

If we adopt this transformation, the ratio $h_{\rm NCII}(T)/h_{\rm NCII}(0)$ is given by
\beq
{h_{\rm NCII}(T)\over h_{\rm NCII}(0)}=
K_{A=1}
\left( T\left({A\over A_2}\right), K_2\left({A\over A_2}\right) \right)
\times \left({A_2\over A}\right)\times{1\over K_2(0)}
\eeq
The estimated value of $K_2(T)$ for the system of $F=0.5$, $T=0.3$, $E=F$ and $K_2=0.2$ 
is obtained by putting $A_2=0.5, K_2(0)=0.2$:
\beq
{h_{\rm NCII}(0.5)\over h_{\rm NCII}(0)}=
K_{A=1}
\left( 
0.3 \left( {1\over 0.5} \right), 
0.2 \left( {1\over 0.5} \right) 
\right)
\times \left({0.5\over 1}\right){1\over 0.2}
=K_{A=1}(0.6, 0.4)\times 2.25.
\eeq
In Fig.~\ref{MZTKTall}(b) we find that $K_{A=1}(0.6, 0.4)\simeq 0.25$, and thus
${h_{\rm NCII}(0.5)/h_{\rm NCII}(0)}\simeq 0.6$.
On the other hand, in Fig.~\ref{NucleationatF}(a), we find ${h_{\rm NCII}(0.5)/ h_{\rm NCII}(0)}\simeq 0.5$.
Thus, we find that the estimation of $h_{\rm NCII}(T)$ (Fig.~\ref{NucleationatF}(a)) is lower than the estimation from the temperature dependence of the anisotropy energy $K_2(T)$ obtained from the data in Fig.~\ref{MZTKTall}(b).
For other points, we find the same tendency, and we conclude that the thermal effects are stronger than the estimation from the temperature dependence of the anisotropy energy.

%%%%%%%%%%%%%%%%%%%%%%%%%%%%%%%%%%%%%%%%%%%%%%%%%%%%%%%%%%%%%%%%%%%%%
\subsubsection{Temperature dependence of the threshold field $h_{\rm NC}(T)$}

In Figs.~\ref{NucleationatF}(b),
The threshold field $h_{\rm NC}(T)$ at $E=F$, i.e., $K_2=K_1$, is shown.
%************
In contrast to $h_{\rm NCII}(T)/h_{\rm NCII}(0)$, 
the threshold field $h_{\rm NC}(T)/h_{\rm NC}(0)$ shows a non-monotonic dependence on $F$ at $T=0.5$.

\begin{figure}[t]
  \centering
  $$
\begin{array}{cc}
  {\includegraphics[width=0.3\textwidth]{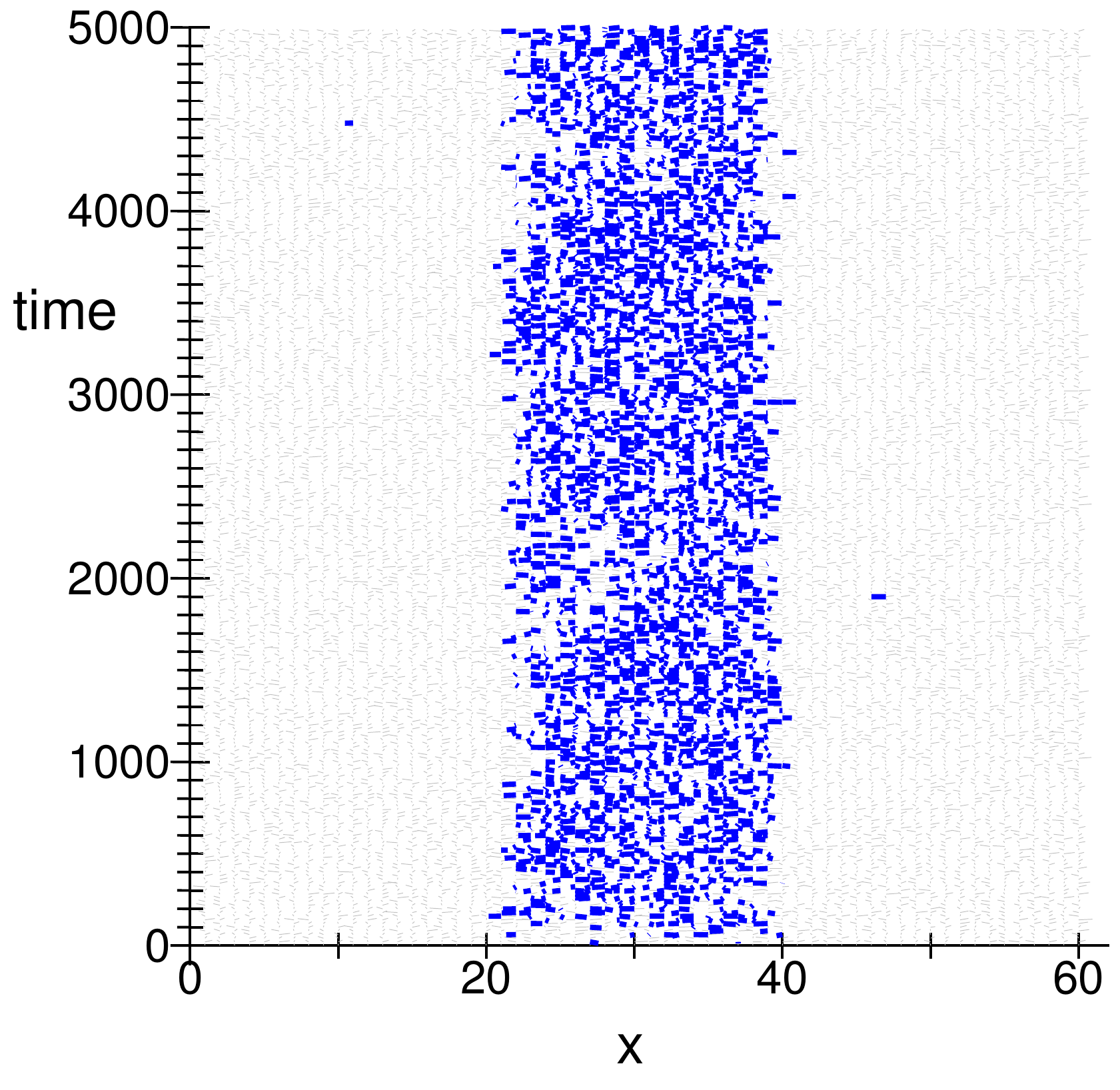}}&               
  {\includegraphics[width=0.3\textwidth]{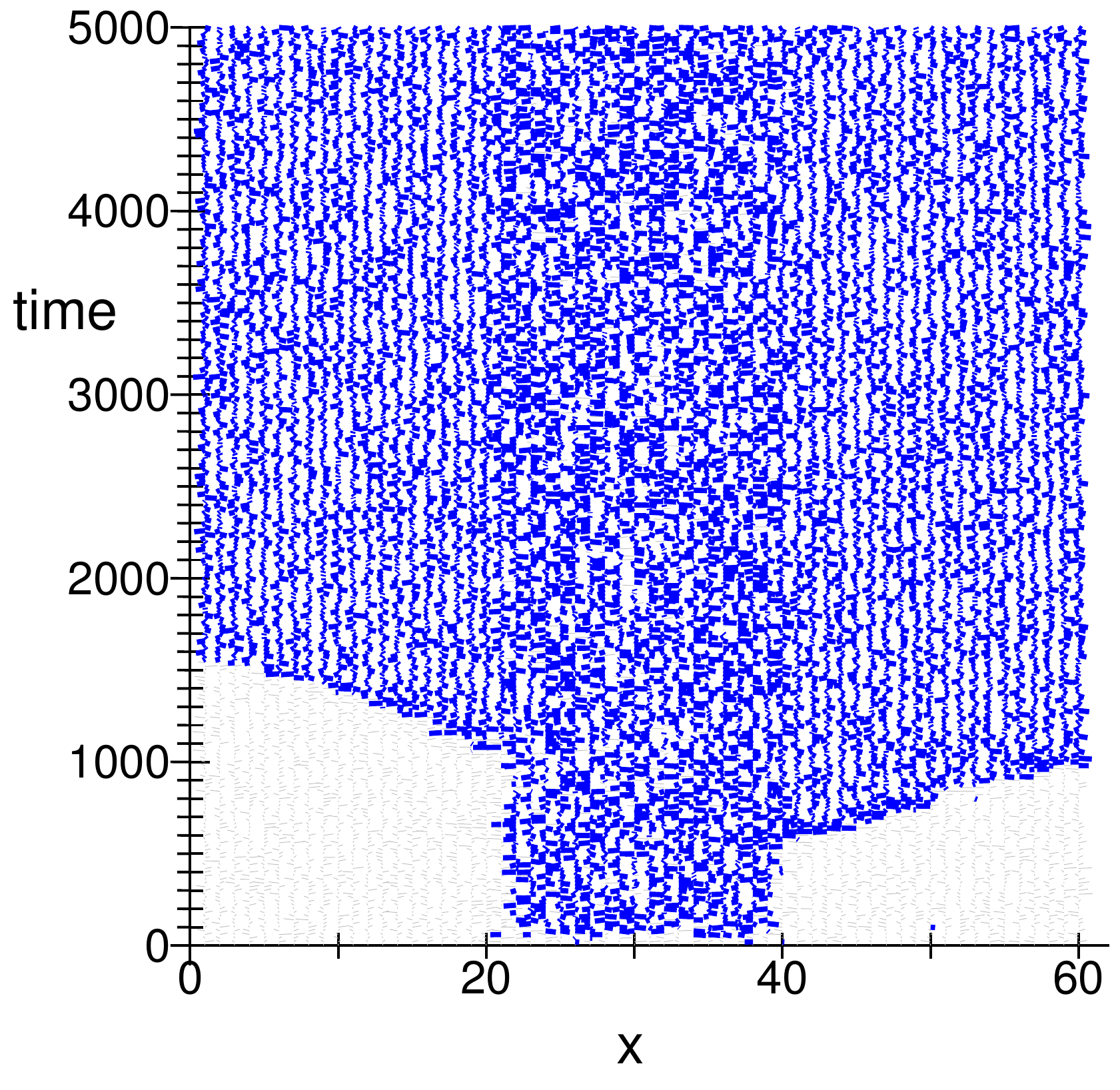}}\\
  {({\rm a})} &  {({\rm b})} 
  \end{array}
 $$
  \caption{Configurations at (a) $h=H/K_1=0.1$, and (b)  0.5 for $F=E=0.3$ and $T=0.5$.} 
  \label{conf030305}
\end{figure}

To understand the magnetic reversal processes of the case where  $F=E=0.3$ at $T=0.5$, 
we depict time evolutions of the line configurations at various values of $h$
in Fig.~\ref{conf030305}.
At this temperature, we find that the magnetization of region II with $A_2=0.3$ is always reversed, which should not be regarded as a nucleation process, but it should be regarded as a paramagnetic state with a field induced negative magnetization. 
Thus, the reversal of the hard magnets (regions I and III) occurs as a surface nucleation under a field consisting of the external field and the molecular field from region II:
\beq
%check
h_{\rm surface}=-H+A_1\langle S_{i,z}\rangle_{\rm  region II}.
\label{hsurface}
\eeq

Now let us consider the dependence of $h_{\rm NC}(T)$ on $F$.
Because $F=E$, we have $K_1=K_2$ and $A_2=F$. 
Therefore, the dependence on $F$ is the same as that on $A_2$.
%%%This should be rewritten
Increasing $A_2$ causes the increase of the threshold $h_{\rm NCII}(T)$ in region II
as we see in Fig.~\ref{NucleationatF}(b). 
As long as there is no nucleation in region II, the reversal of regions I and III does not take place. 
Thus, the increase of the threshold $h_{\rm NCII}(T)$ with $A_2$ causes an increase of $h_{\rm NC}(T)$.
%This effect causes the decrease of  $h_{\rm NC}(T)$ when $F$, i.e., $A_2$ decreases from 0.7 to 0.5.
For $F=0.5$ and 0.7, just after the nucleation takes place the whole system reverses.
Thus $h_{\rm NCII}(T)\simeq h_{\rm NC}(T)$ and they decrease with $A_2$.

On the other hand, at $F=0.3$, the magnetization in region II, $|\langle S_{i,z}\rangle_{\rm  region II}|$, is small. 
Then, the magnitude of the second term 
$A_1\langle S_{i,z}\rangle_{\rm  region II}(<0)$ is small, and thus 
the spring effect decreases with the temperature, which causes an increase of the threshold.
%Thus, because of the relation (\ref{hsurface}), the negative field from region II weakens, and then $|h_{\rm surface}|$ decreases. 
Consequently, to reverse region I, a kind of surface nucleation at the surface of the hard magnets (regions I and III) must take place.
On the other hand, the thermal fluctuation reduces the robustness of regions I and III. 
These mechanisms compete with each other, and  $h_{\rm NC}(T)$ shows a weak dependence on temperature.
Because of the above mentioned mechanisms,  $h_{\rm NC}(T)$ has a
non-monotonic dependence on $F$. 
%%%%%%%

\section{Temperature dependence of domain wall propagation}\label{sectiondomainwallpropagation}

Next, we use the initial condition (2) where a domain wall exists in the system and we study whether the domain wall can propagate to region I. 
In the present situation, the magnetization of region III is already reversed, and thus the threshold of the domain wall propagation $h_{\rm DWP}(T)$ should be smaller than 
$h_{\rm NC}(T)$.
Indeed, once nucleation occurs in region II,
$h_{\rm DWP}(T)$ should be the same as $h_{\rm NC}(T)$.
However, in Fig.~\ref{NucleationT}, $h_{\rm NC}(T)$ was raised by $h_{\rm NCII}(T)$,
where the dependence of the domain wall depinning was masked by $h_{\rm NCII}(T)$.
The parameter dependence of the domain wall propagation itself is important, and thus in this section 
we study $h_{\rm DWP}(T)$ as a function of the parameters $F$, $E$ and $T$. 

\subsection{$T=0$}

\begin{figure}[h]
$$
\begin{array}{ccc}
  {\includegraphics[width=0.3\textwidth]{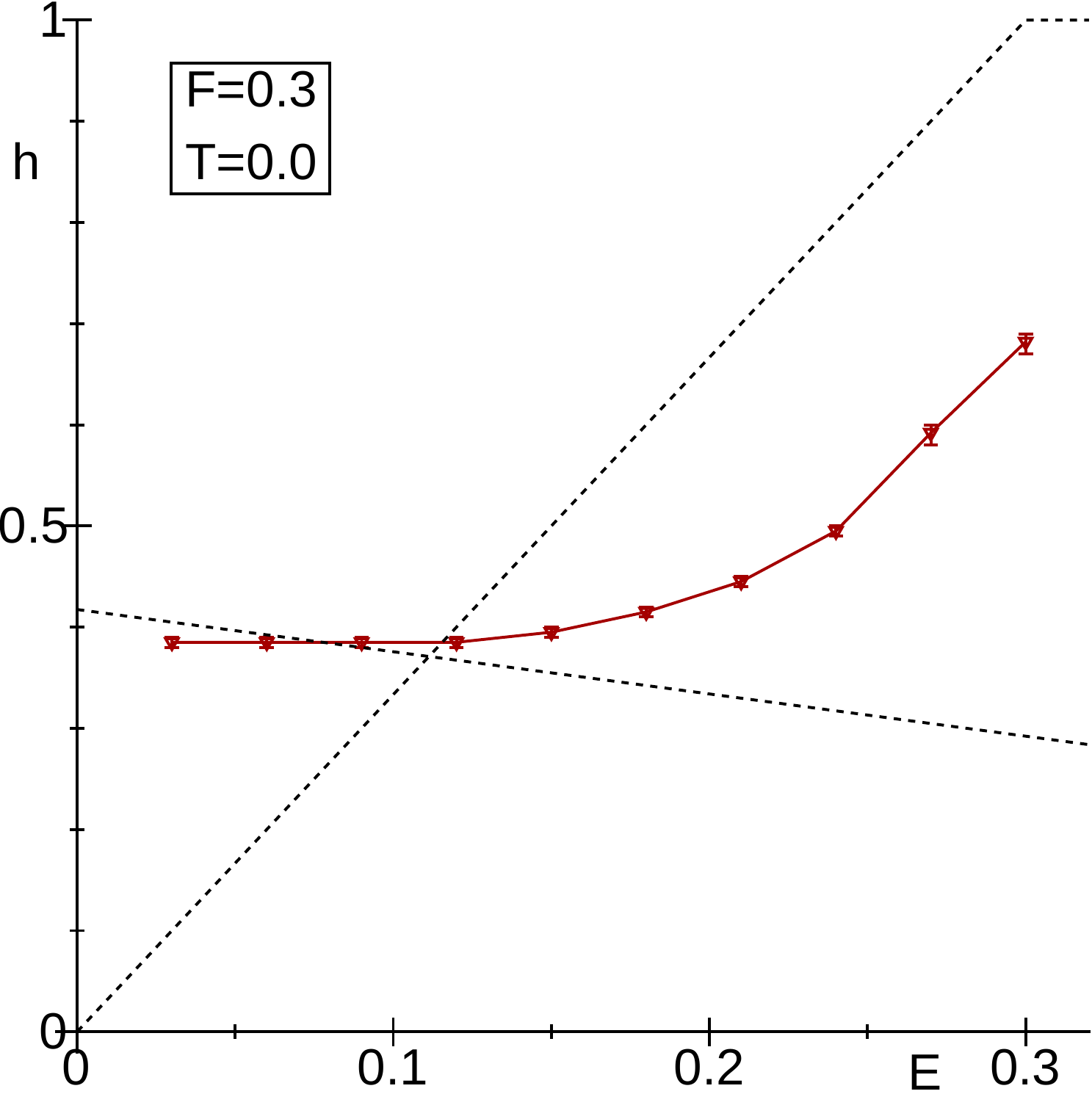}}&               
  {\includegraphics[width=0.3\textwidth]{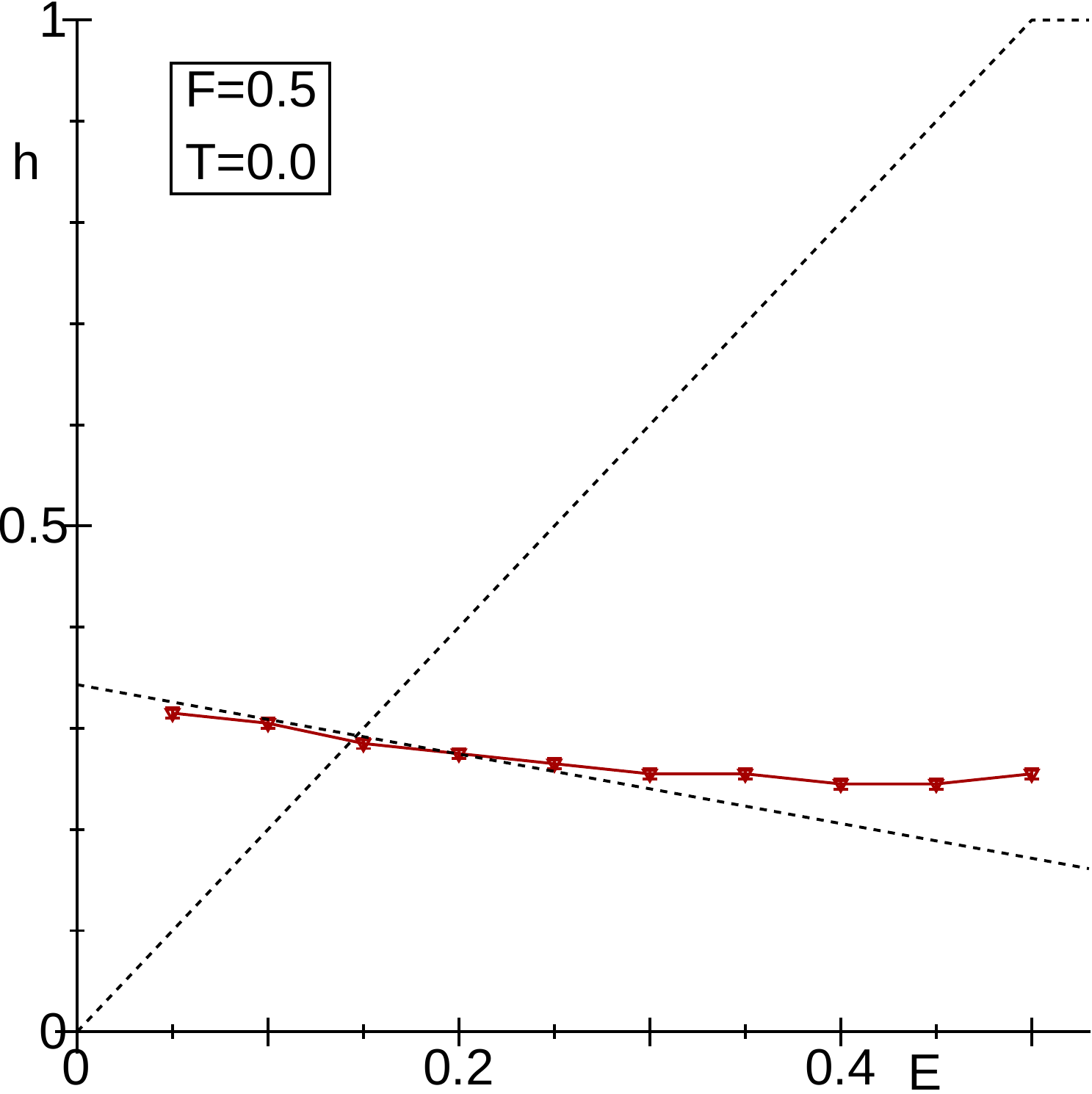}}& 
  {\includegraphics[width=0.3\textwidth]{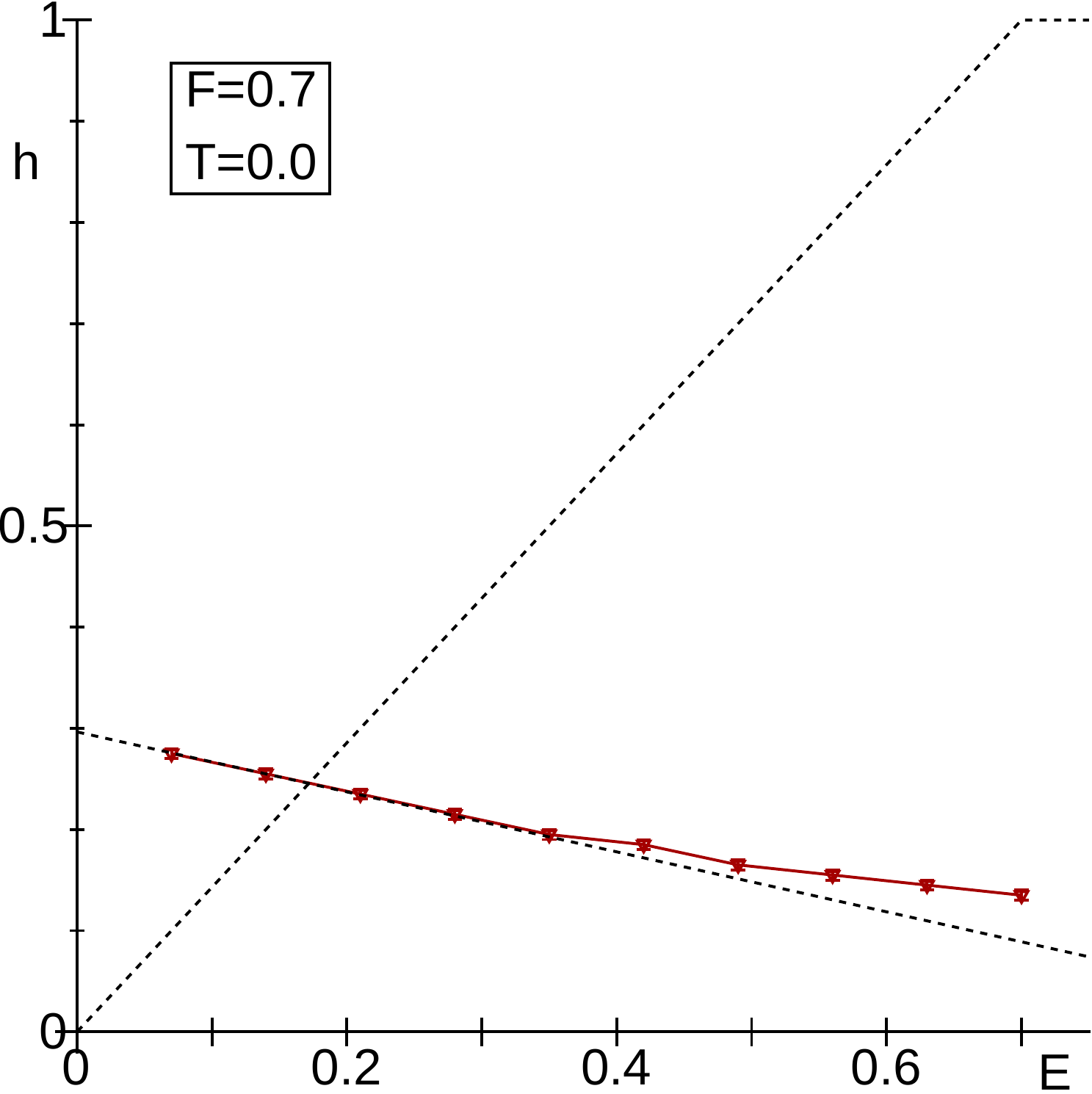}}\\
  {({\rm a})} &  {({\rm b})} &  {({\rm c})}
  \end{array}
 $$ 
\caption{Dependence of $h_{\rm DWP}(T)$ at $T=0$ for $F=0.3, 0.5$ and 0.7, compared with those obtained analytically.
 The error bars denote the threshold between the final state $(---)$ and $(+--)$, and their length is given by the step size on $h$, i.e., $\Delta h=0.01$. The solid lines are guides for the eye, and the dotted lines denote the analytical estimation.
 } \label{DomainWallPropResultsT0}
\end{figure}

First, we compare the results at $T=0$ with those obtained analytically.\cite{sakuma}
The dependence of $h_{\rm DWP}(0)$ on $E$ is shown in Fig.~\ref{DomainWallPropResultsT0}. 
For relatively large values of $F$, e.g., $F=0.5$ and 0.7, we find good agreements with the analytical result which shows that $h_{\rm DWP}(0)$ decreases with $E$.
However, for the small value $F=0.3$, it is found that  $h_{\rm DWP}(0)$ increases with $E$. 

If $F$ is small, that is, if the interaction in region II ($A_2$) is small, the correlation length of the magnetization is short.
If $E$ becomes large, that is, if $K_2/A_2$ becomes large, the domain wall width becomes short. 
Thus, we understand that the effect of the reversed magnetization on region III is well shielded in region II, and thus a larger external field is necessary to reverse region I by a surface nucleation process.
%check
This effect did not appear in the analytical estimation, where a continuous change of spins is assumed and the configuration is of the Bloch type. 
In the case for large values of $K/A$, the usual Bloch wall does not appear, and the so-called narrow domain wall\cite{narrowDW} appears with a discontinuous change, which is explained in Appendix~\ref{AppB}.

\begin{figure}[!t]
  \centering     
  {\includegraphics[width=0.3\textwidth]{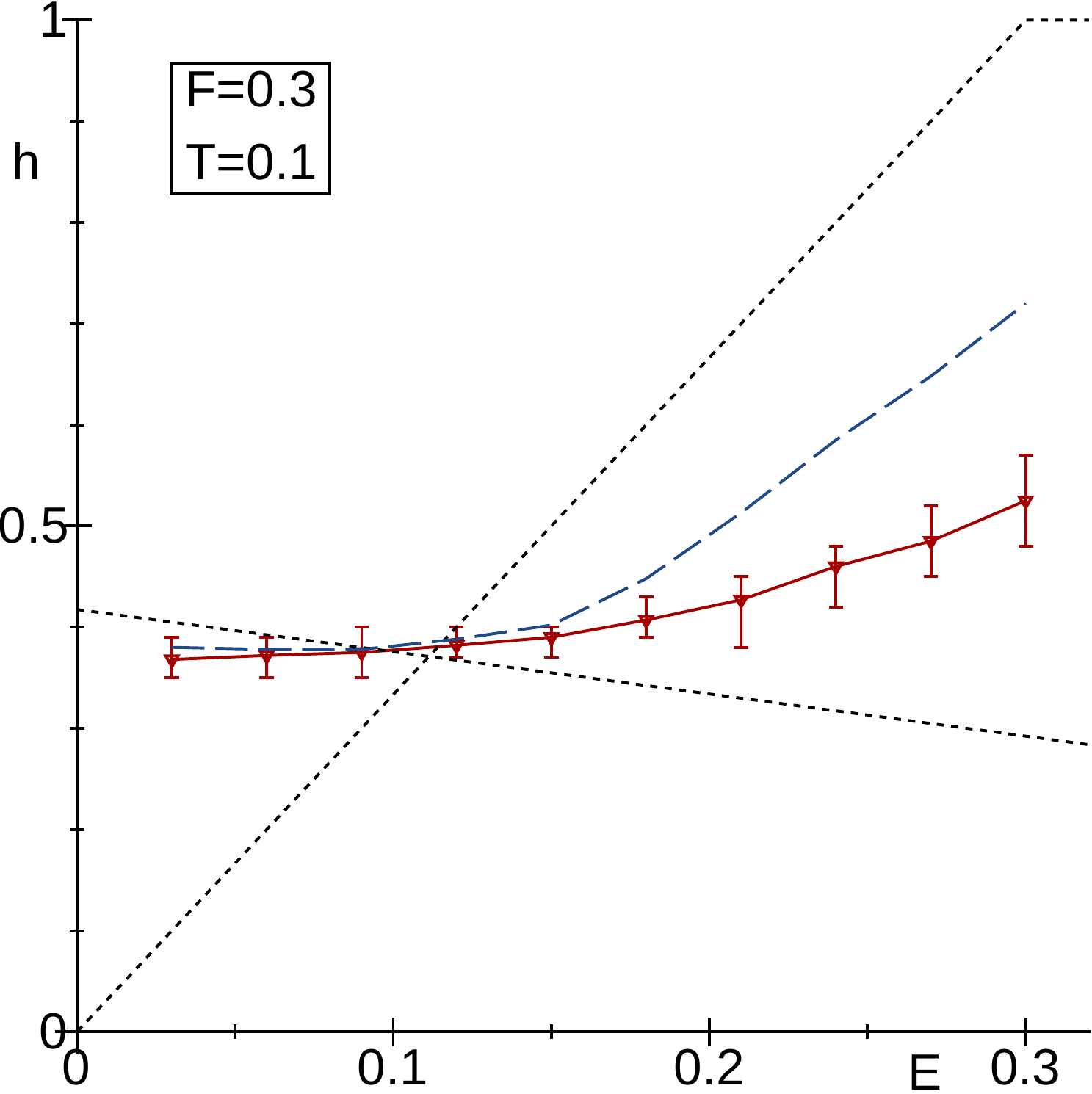}}\ 
  {\includegraphics[width=0.3\textwidth]{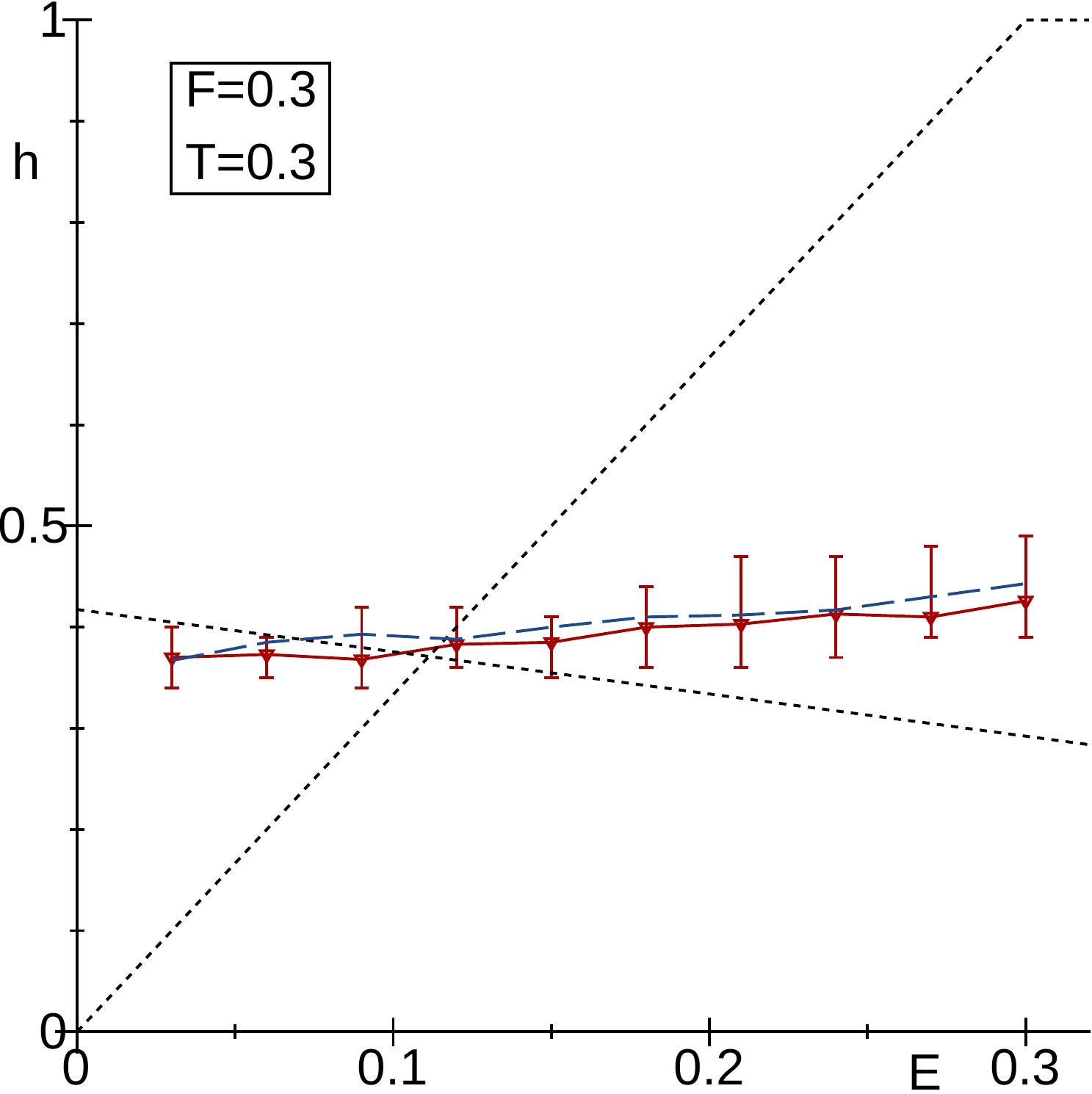}}\ 
  {\includegraphics[width=0.3\textwidth]{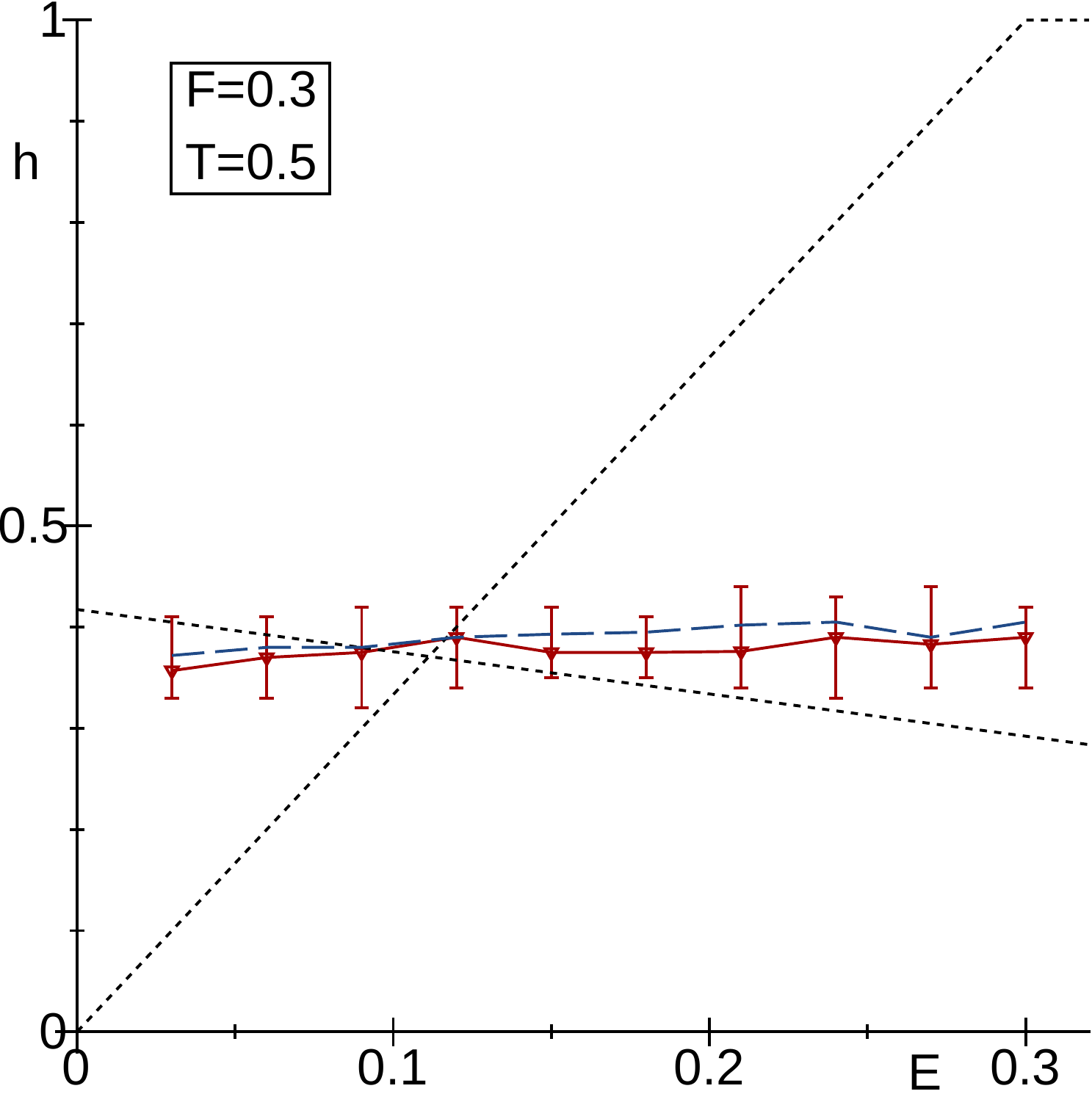}}\\
  
  {\includegraphics[width=0.3\textwidth]{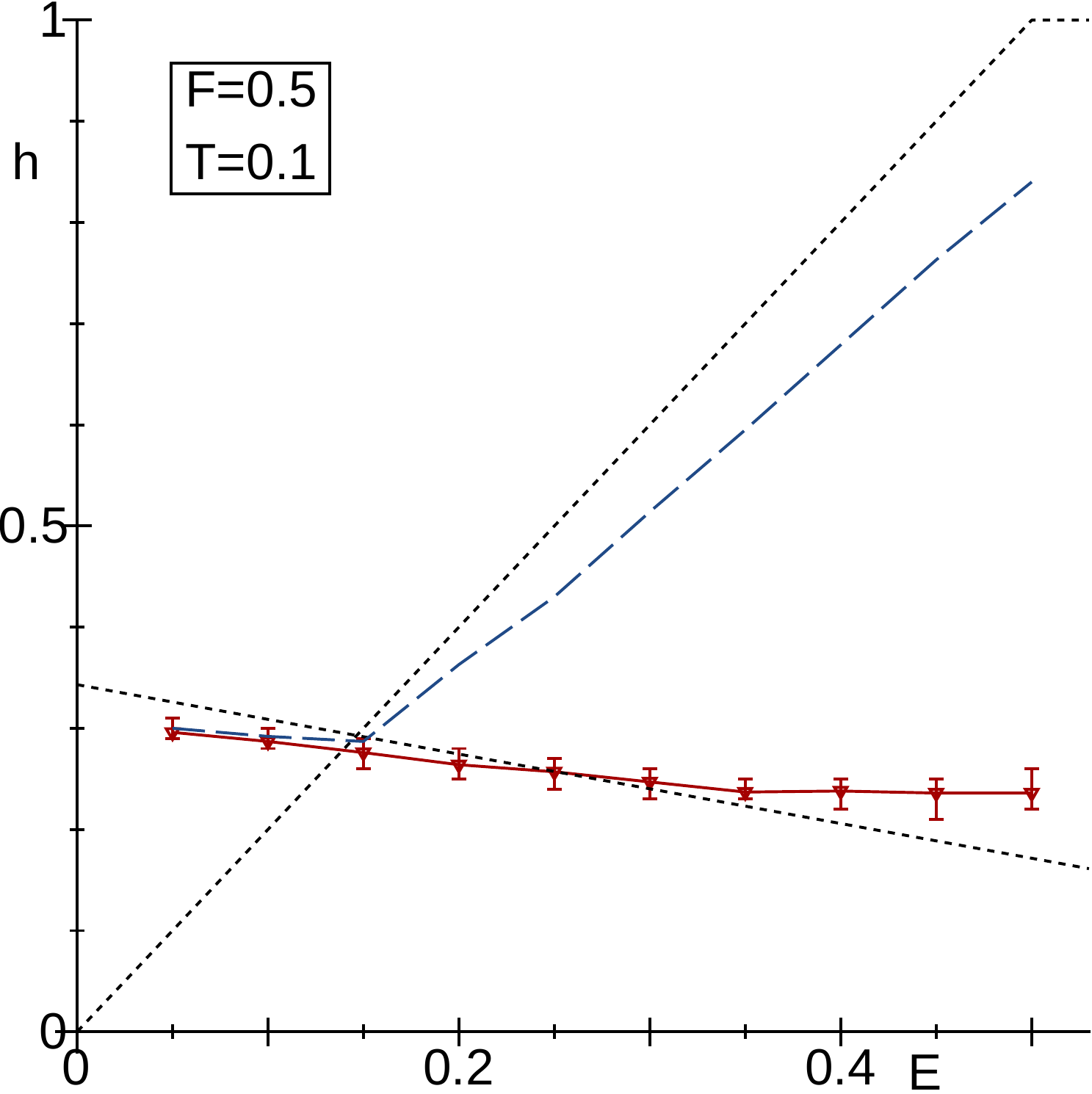}}\ 
  {\includegraphics[width=0.3\textwidth]{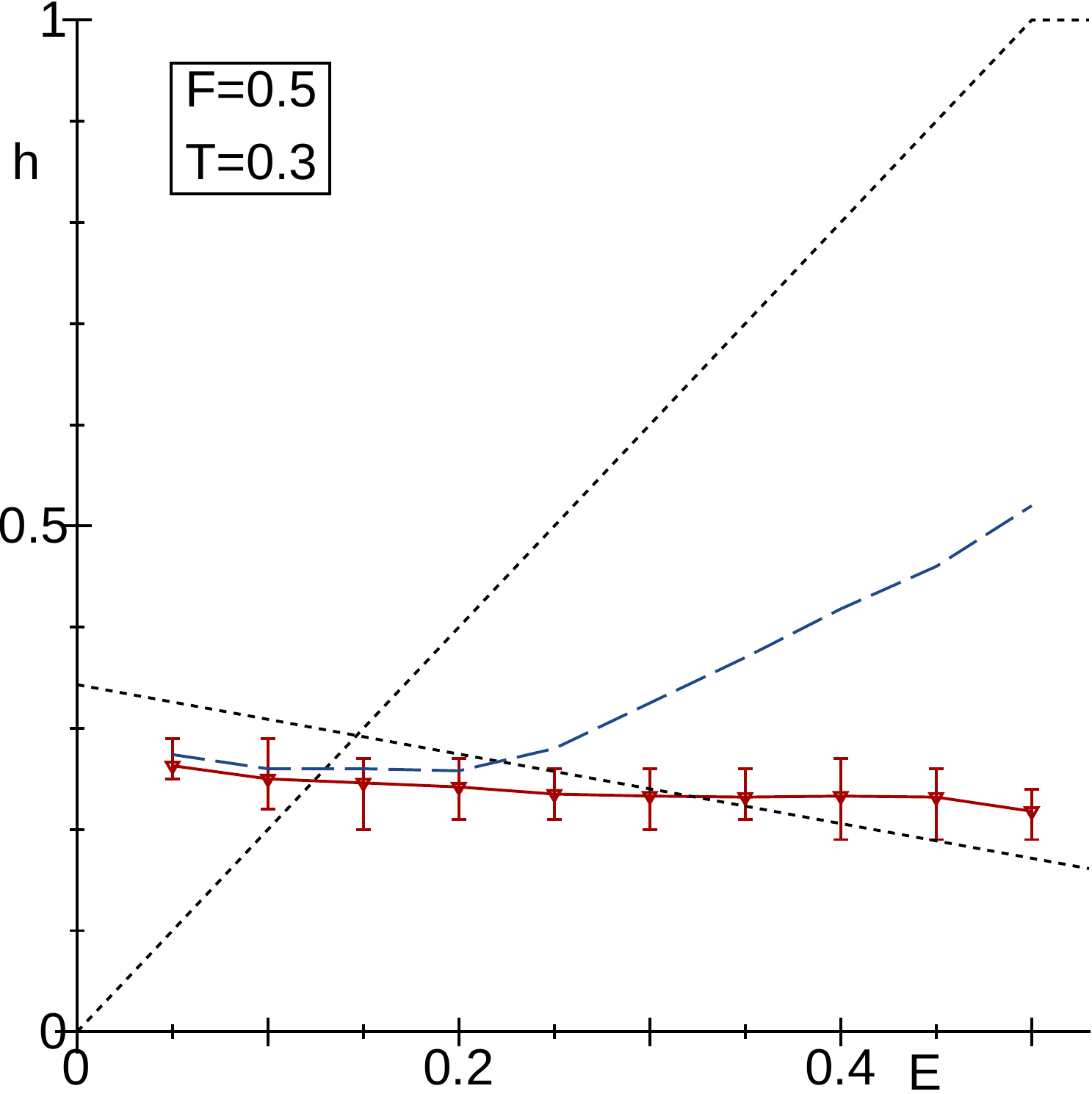}}\ 
  {\includegraphics[width=0.3\textwidth]{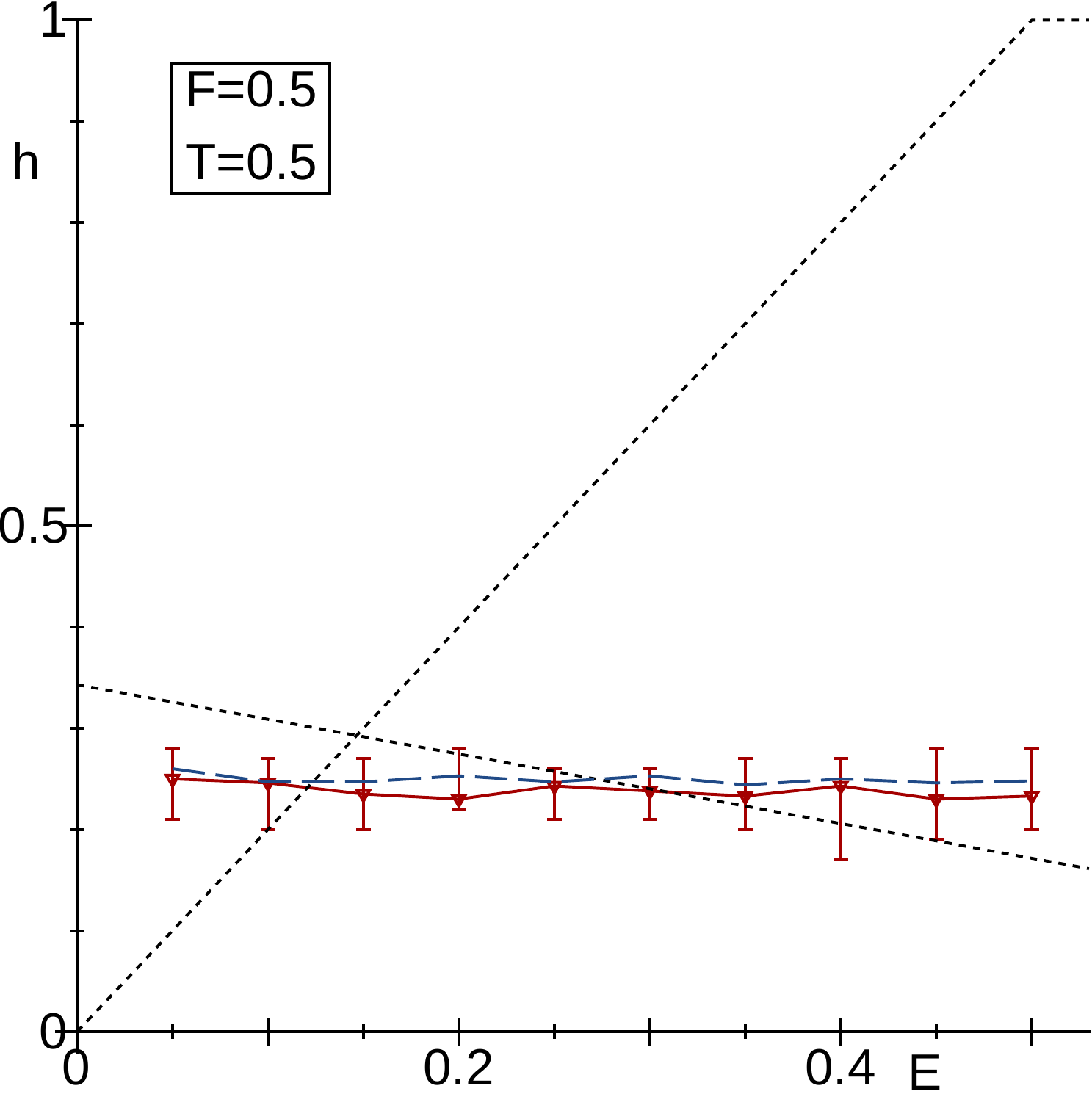}}\\

  {\includegraphics[width=0.3\textwidth]{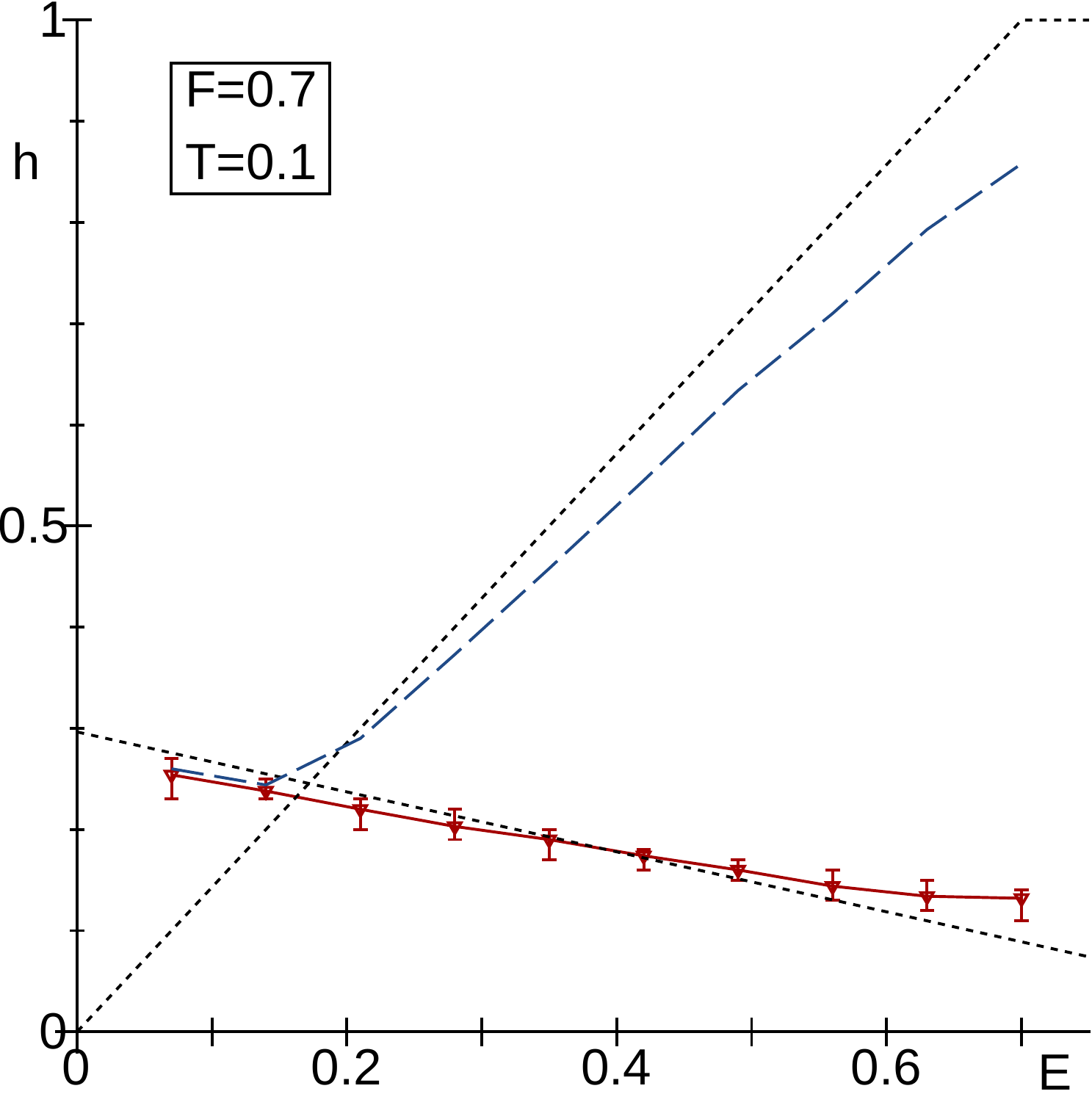}}\ 
  {\includegraphics[width=0.3\textwidth]{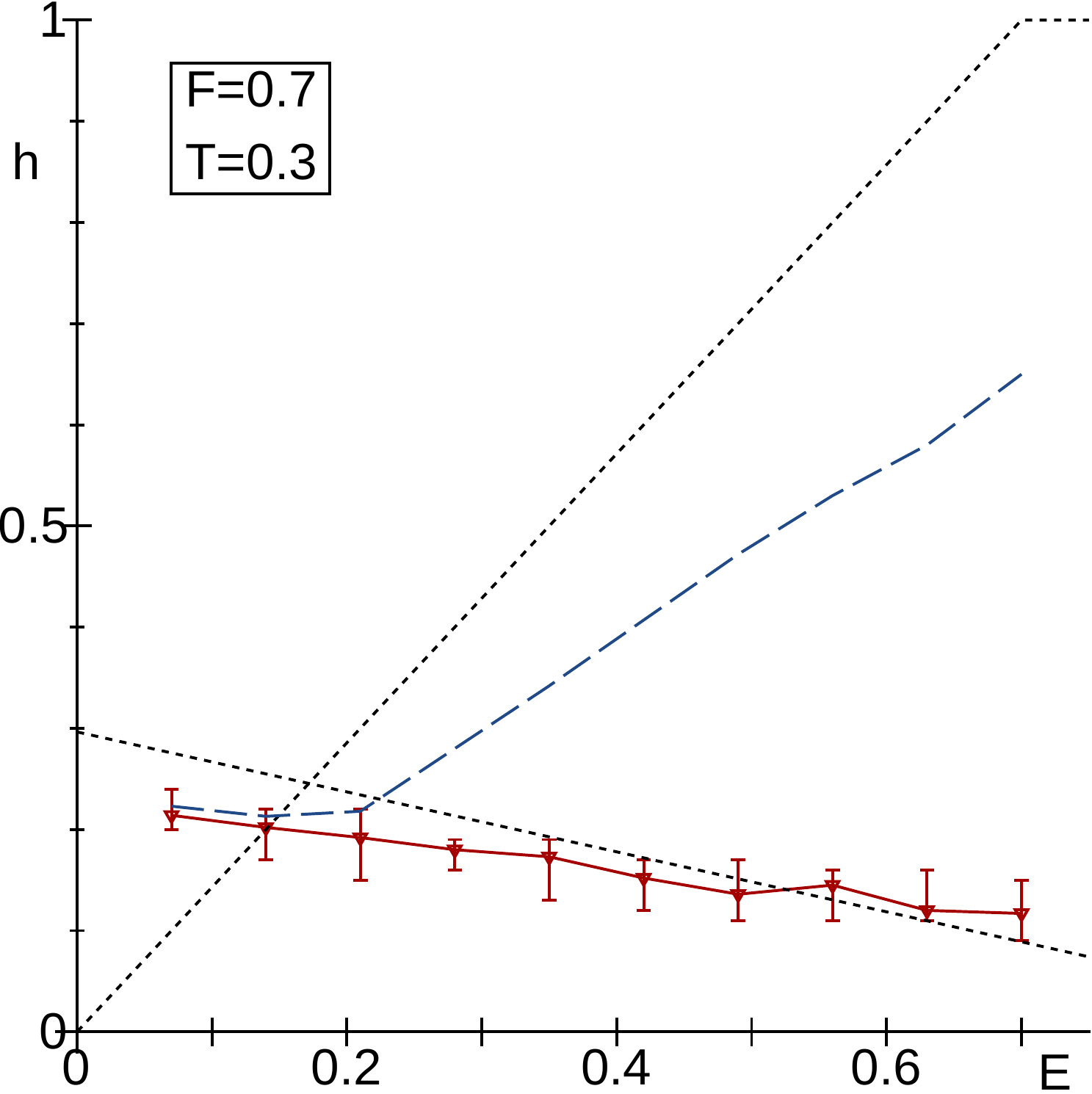}}\ 
  {\includegraphics[width=0.3\textwidth]{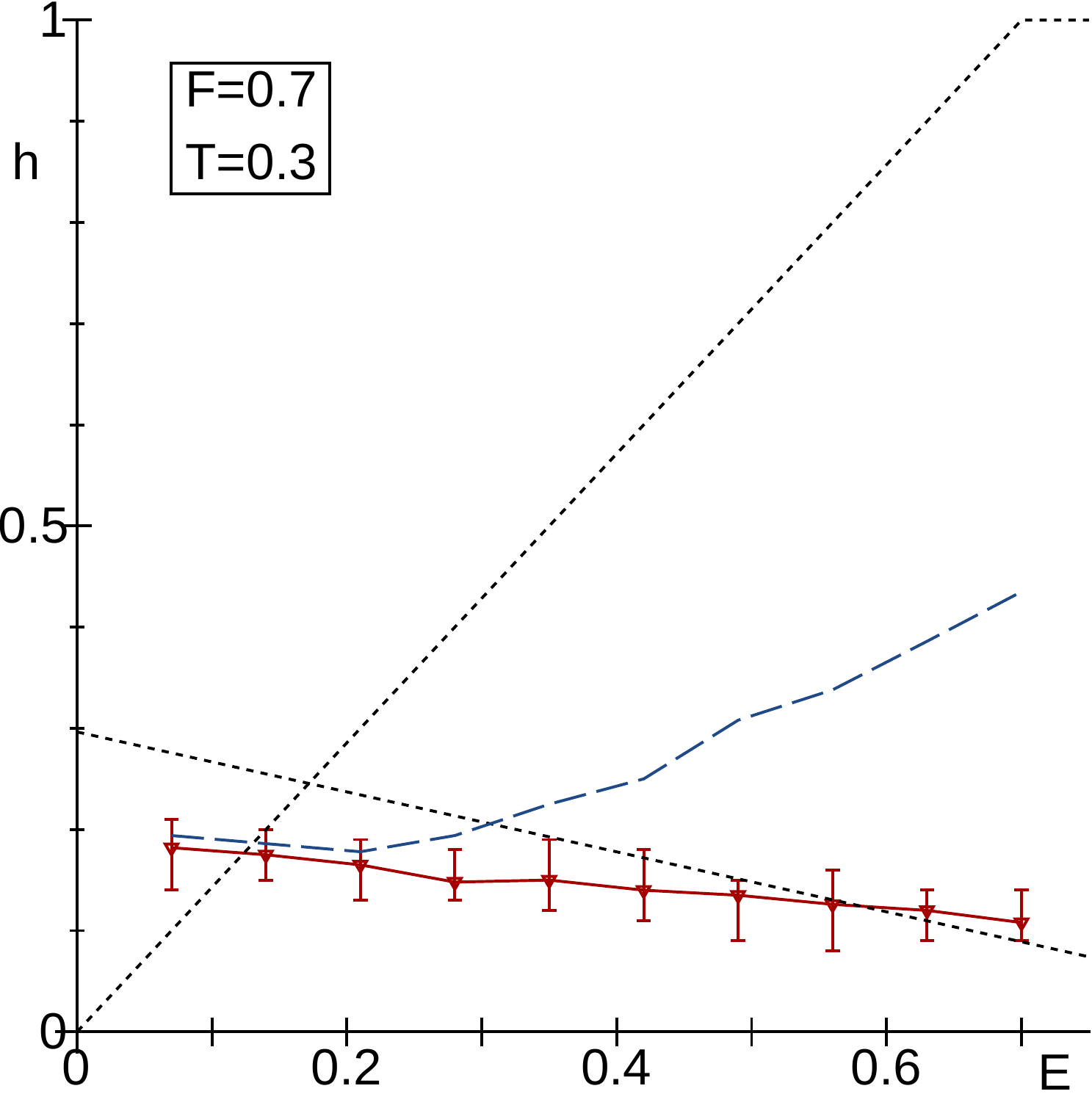}}\\        
% {\includegraphics[width=0.3\textwidth]{FigDWNCF03T01-eps-converted-to.pdf}}\                 
% {\includegraphics[width=0.3\textwidth]{FigDWNCF03T03-eps-converted-to.pdf}}\
% {\includegraphics[width=0.3\textwidth]{FigDWNCF03T05-eps-converted-to.pdf}}\\
%  
% {\includegraphics[width=0.3\textwidth]{FigDWNCF05T01-eps-converted-to.pdf}}\
% {\includegraphics[width=0.3\textwidth]{FigDWNCF05T03-eps-converted-to.pdf}}\
% {\includegraphics[width=0.3\textwidth]{FigDWNCF05T05-eps-converted-to.pdf}}\\
%
% {\includegraphics[width=0.3\textwidth]{FigDWNCF07T01-eps-converted-to.pdf}}\
% {\includegraphics[width=0.3\textwidth]{FigDWNCF07T03-eps-converted-to.pdf}}\
% {\includegraphics[width=0.3\textwidth]{FigDWNCF07T05-eps-converted-to.pdf}}\\

\caption{Domain wall propagation field for $F=0.3$, 0.5 and 0.7, and  $T=0.1$, 0.3 and 0.5. The lines joining the data points are guides for the eye. The dotted lines represent the fields given by Eqs.~\eqref{hn0} and \eqref{hp}, and the dashed lines represent the eyeguides for the total nucleation field depicted in Fig.~\ref{NucleationT}.} 
\label{DomainWallT}
\end{figure}

\subsection{$T>0$}
\begin{figure}[h]
  \centering
  $$
\begin{array}{ccc}
  {\includegraphics[width=0.3\textwidth]{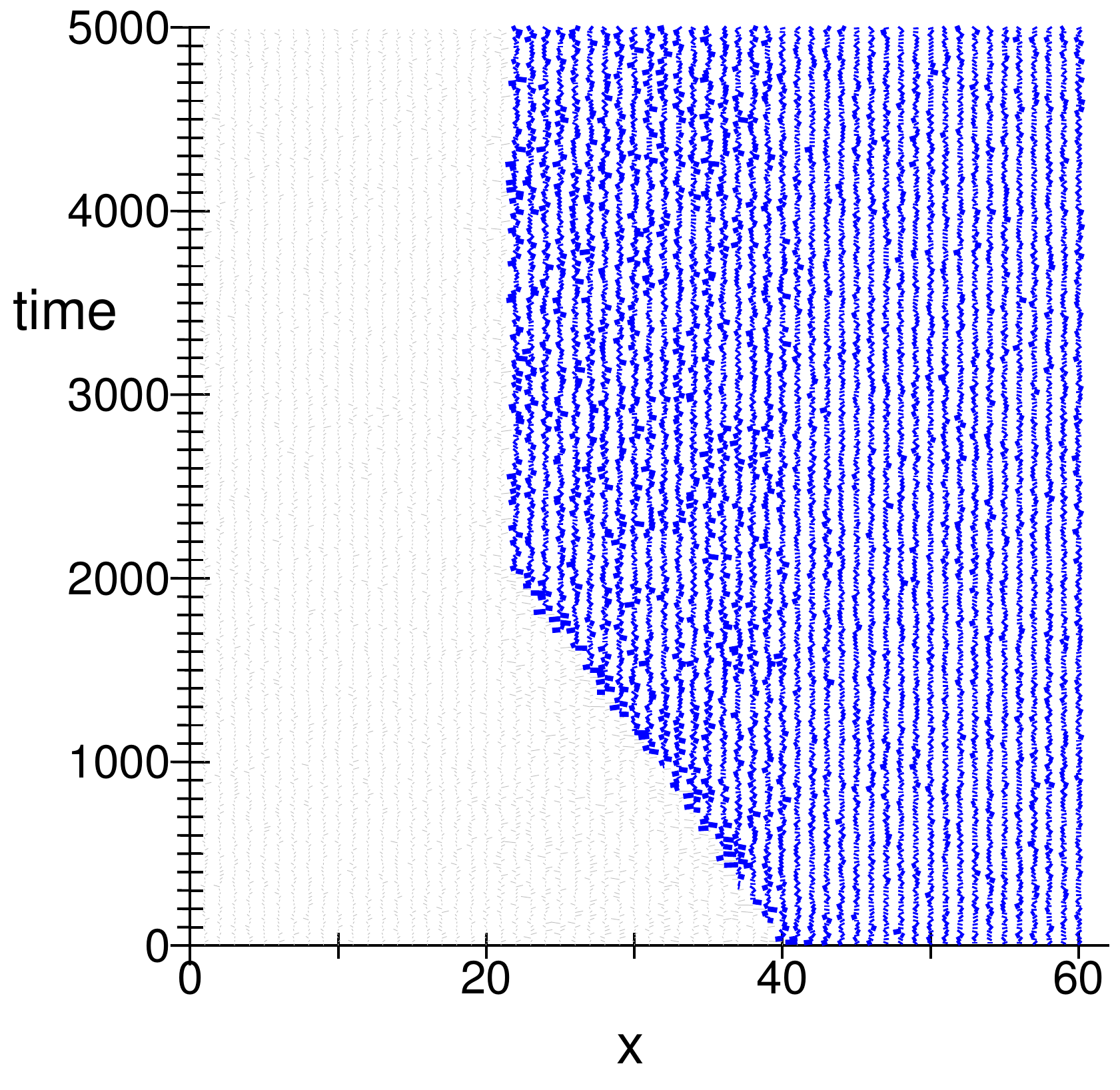}}&
  {\includegraphics[width=0.3\textwidth]{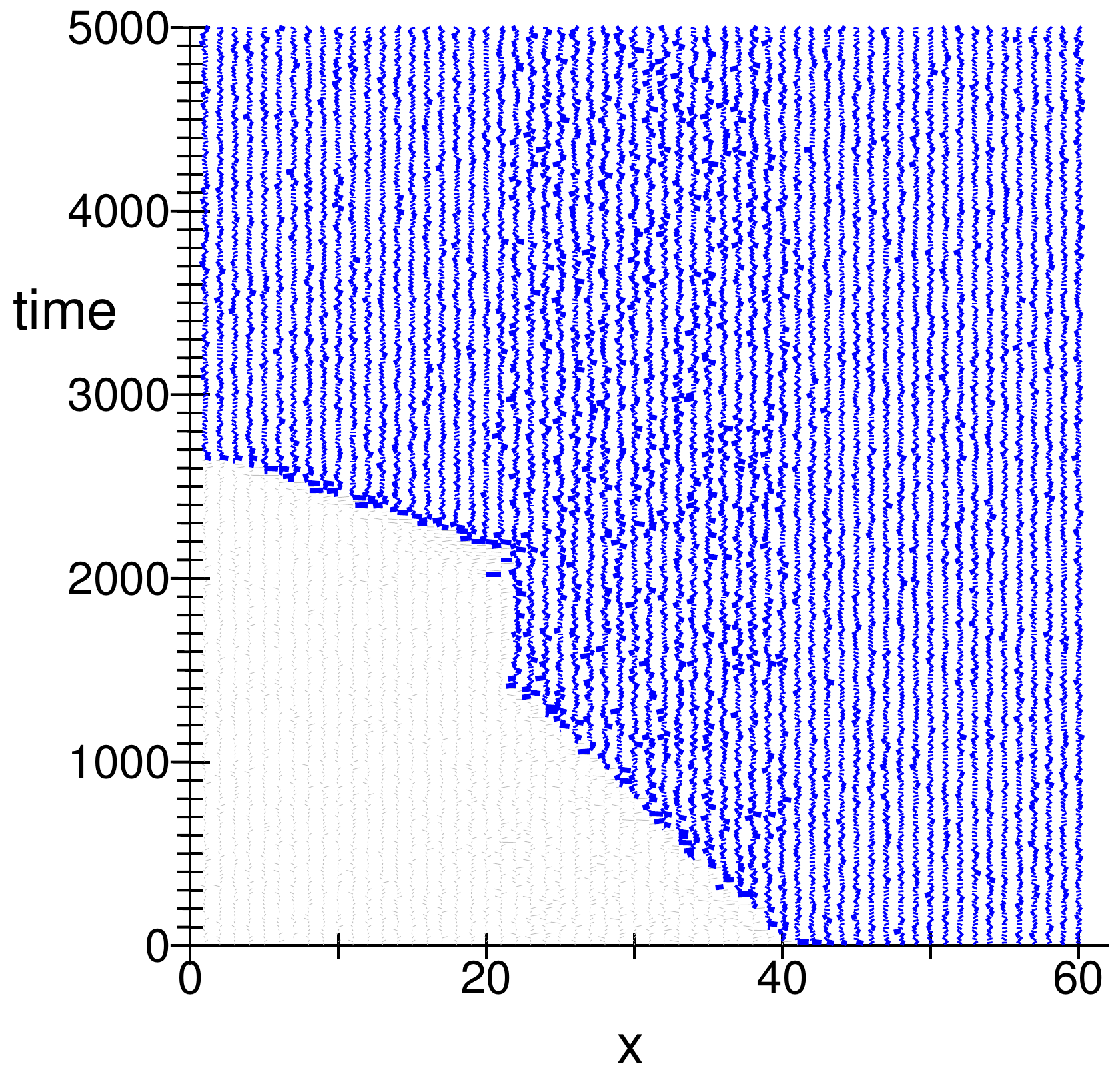}}\\
  {({\rm a})} &  {({\rm b})} &  
  \end{array}
 $$ 
\caption{Time evolution of the magnetization reversal process for $F=0.3$, $E=0.27$ and (a) $h=0.4$, and 
  (b) $h=0.6$ at $T=0.1$. 
  Each row denotes a configuration of spins at the site $(x,4,4),x=1,60$ at a time $t$.
  The vertical axis denotes time.  }    
  \label{confF03Eh}
  \end{figure}

Next, we study the temperature dependence of $h_{\rm DWP}(T)$, which is shown in Fig.~\ref{DomainWallT}.  
As the temperature rises, $h_{\rm DWP}(T)$ is reduced. 
But, the dependence is much weaker than in the case of $h_{\rm NC}(T)$ 
except for the case $F=0.3$ where $h_{\rm DWP}(T)$ shows a similar dependence to that of $h_{\rm NC}(T)$.
For $F=0.3$, $h_{\rm DWP}(T)$ increases with $E$ at low temperatures. 
The time evolutions of domain wall propagation for $E=0.27$ at $h=$0.40 and 0.60 are shown 
in Fig.~\ref{confF03Eh}.
For $h=0.4$, the domain wall is pinned at the border between regions I and II.
When the field is increased up to $h=0.6$ the domain wall penetrates into region I.
%At a strong field $h=0.55$ a kind of nucleation occurs at the border and the down domain proceeds into region I. 
As we discussed above, we may again understand this phenomenon as a kind of surface nucleation of hard magnets.

\begin{figure}[h]
  \centering
  $$
\begin{array}{cc}
  {\includegraphics[width=0.3\textwidth]{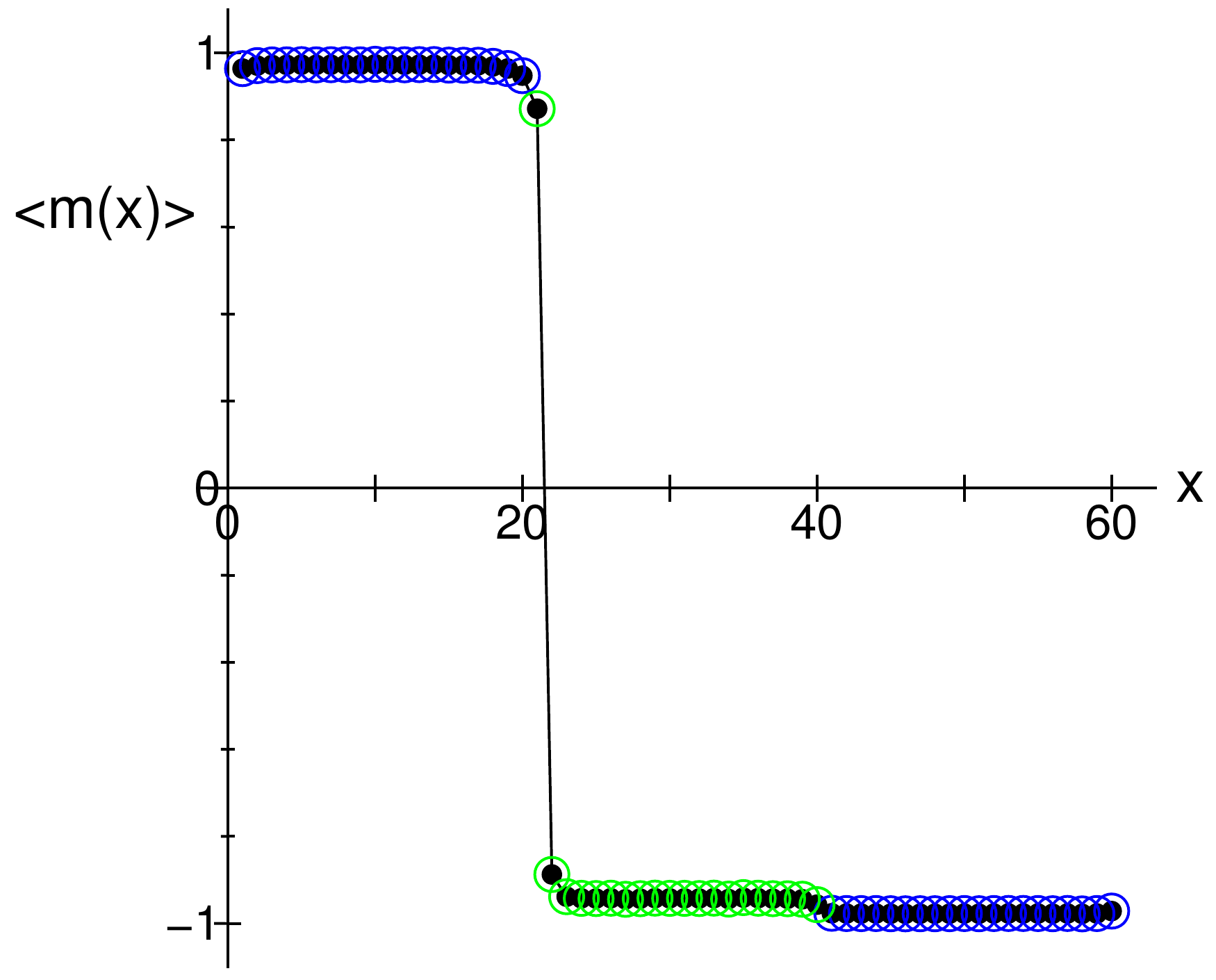}}&
  {\includegraphics[width=0.3\textwidth]{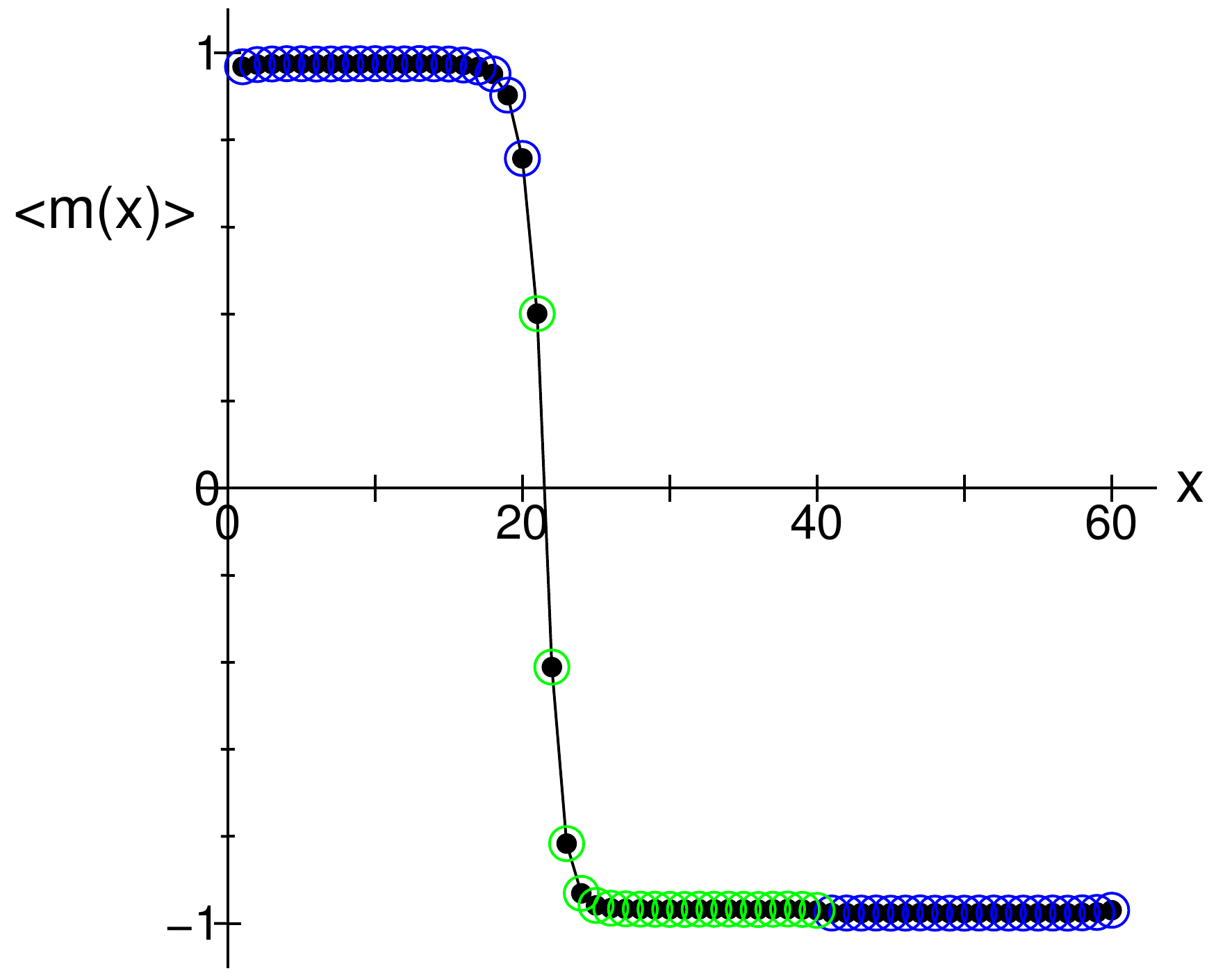}}\\
  {({\rm a})} &  {({\rm b})}
  \end{array}
 $$ 
\caption{Magnetization profiles $\langle m(x)\rangle$ for (a) $F=0.3$, $E=0.27$ and $h=0.45$ at $T=0.1$, and (b) $F=0.7$, $E=0.63$ and $h=0.10$ at $T=0.1$. The lines are guides for the eye.}    
  \label{magprof}
  \end{figure}

In the case of a narrow domain wall, the effect of the reversal of regions II and III is masked. 
In order to see this situation, we plot the magnetization profiles ($\langle m(x)\rangle$) in Fig.~\ref{magprof}. 
These are obtained by averaging in the steady state of the $(+--)$ type after the domain wall is pinned at the border of regions I and II.

For $F=0.3$ and $E=0.27$ (Fig.~\ref{magprof}(a)), we find a sharp change of the spin direction.
In this case, as we discussed above,
$A_2=0.3$ and $K_2=K_1 \times 0.9=0.2\times 0.9=0.18$, and the width of the Bloch domain wall in the defect region, 
$\xi=\sqrt{0.3/(2\times0.18)}\simeq 0.745$, is less than unity, so the continuous approximation is not adequate. 
 For comparison, the magnetic profile at $T=0.1$ for $F=0.7$ and $E=0.63$ with $h=0.1$ shows a smooth profile. 
 In this case, the increase of $h_{\rm DWP}(T)$ does not take place as we see in Fig.~\ref{DomainWallT}.  
%The correlation from region I penetrates in the defect region with a couple of layers. 
  
%This observation indicates that the increase of $h_p$ for small $F$ is due to the narrow domain width(Appendix B).

%Above the critical ratio of $D/J=2/3$ for the narrow domain wall, the spins are completely aligned to the easy axis(see Appendix B\cite{narrowDW}). In the above case $K_2/A_2=0.6$ is close to the critical value and we find a sharp change in the magnetization profile. On the other hand, for $F=0.7$ and $E=0.63$, the ratio is
%$K_2/A_2=0.18/0.7\simeq 0.26$ and we see a smooth change. There, the continuous approximation of the type of Bloch domain wall is appropriate.

\begin{figure}[h]
$$
     {\includegraphics[width=0.3\textwidth]{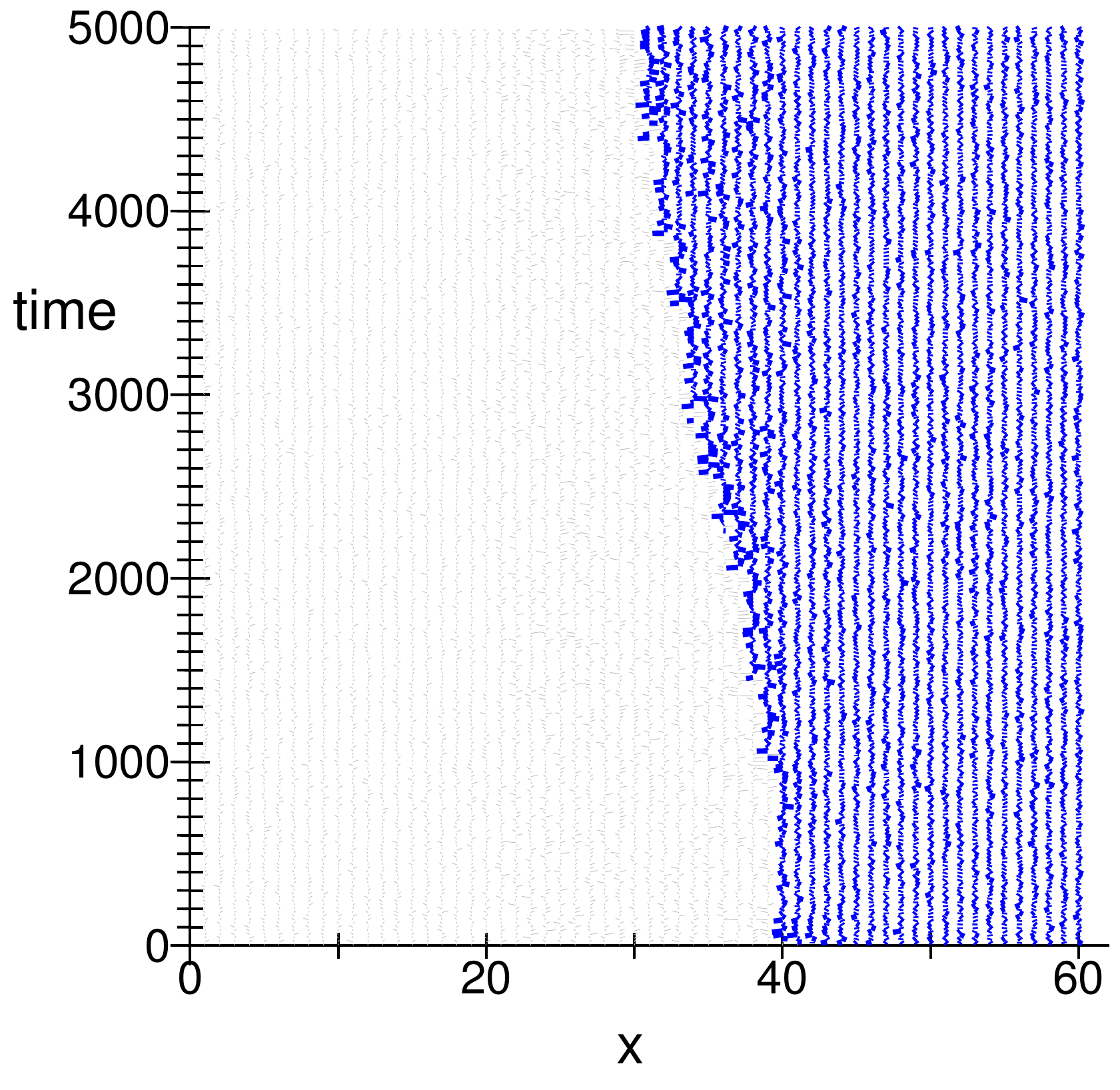}}
$$   
\caption{Time evolution of the magnetization reversal process for $F=0.3$, $E=0.27$ and $h=0.05$ at $T=0.1$.
  Every row denotes a configuration of spins at the site $(x,4,4),x=1,60$ at a time $t$.
  The vertical axis denotes the time.  }    
  \label{confF03E005h}
  \end{figure}

 In Fig.~\ref{confF03E005h}, we show the domain wall motion at $T=0.1$ for $h=0.05$, $F=0.3$ and $E=0.27$. There, we observe a narrow domain wall, and it is temporally trapped at the right border and also at some intermediate points in region II. 
However, it finally moves to the left border and remains trapped there until the end of the observation time (not shown).
  
%\section{Comparison between nucleation and domain wall propagation as functions of temperature}\label{sectioncomparison}
%
%From the previous discussions, it is of interest to study the variation in the nucleation and pinning field curves as a function of $T$. The results are shown in Fig.~\ref{DomainWallT}. 
%In every case, as the temperature rises the pinning and nucleation lines approach each other until they join. 
%This means that at high temperatures, the nucleation in region II takes place easily and the difference 
%between the initial configurations $(+++)$ and $(++-)$ becomes irrelevant for the reversal of region I.
%%the external field magnitude required to force the spins in the system to point in the $-z$ direction is the same regardless of the initial conditions (1) or (2).
%%Note that as the temperature rises, the nucleation and pinning field curves move closer together until they become one curve.

 Indeed, as we discussed in previous sections, if $T>1.4A_2$ then the defect region is paramagnetic. 
 In this case the magnetic reversal of the hard magnets can be regarded as that of isolated magnets approximately.
% At finite field the magnetization reversal of the left region can be regarded as a surface nucleation 
% with an effective field.
% \beq
% H_{\rm eff}=H+J_2m_z^{(II)}(T,H).
% \label{Heff}
% \eeq
 In all cases, the domain wall propagation can be regarded as a surface nucleation with the relation \eqref{hsurface}.
% The parameter dependence of the threshold field is attributed to $m_z^{(II)}(T,H)$ as a function of the magnetization of the right hard magnet. In cases that the spin correlation in the defect region is strong,
%$m_z^{(II)}(T,H)$ depends on the magnetization in the right region. However, in cases that the correlation is weak, the reversal of the right magnet does not depend on the magnetization at the right side, where the thresholds of nucleation and depinning becomes the same.
%At high temperatures we find this tendency. 

\section{Summary and Discussion}\label{sectionsummary}

The temperature dependences of the threshold fields for nucleation ($h_{\rm NCII}(T)$ and $h_{\rm NC}(T)$) and domain-wall pinning ($h_{\rm DWP}(T)$) were studied in the system depicted in Fig.~\ref{ModelImage}, where region II has a weaker exchange interaction $F=A_2/A_1< 1$ and a weaker anisotropy $K_2/K_1=E/F <1$ (or $K_2< K_1$) and it is sandwiched between the hard magnets.

We found that the threshold fields of nucleation phenomena  ($h_{\rm NCII}(T)$ and $h_{\rm NC}(T)$) strongly depend on temperature.
The threshold $h_{\rm NCII}(T)$ decreases monotonically with the temperature and also with $A_2$ (Fig.6(a)).
This is due to the reduction of magnetic order in the soft magnet region II. 
We estimated the reduction of anisotropy energy of region II ($K_2(T)$) by making use of the bulk information given in Appendix A, and we found that the reduction of the threshold is faster than that estimated from $K_2(T)$.
As for the threshold of deconfinement of the nucleated magnetization $h_{\rm NC}(T)$, we found a similar dependence for relatively large $F$ (i.e., $F=0.5$ and 0.7).
 But, for $F=0.3$ the spring effect from region II is suppressed and a saturation behavior was found  (Fig.6(b)). 
There, the process can be regarded as a surface nucleation phenomenon at the hard magnet region. 

The domain-wall pinning 
%%miya0806
shown in Fig. 9
has a similar mechanism to that for the deconfinement of the nucleated negative magnetization $h>h_{\rm NC}(T)$.
That is, the threshold $h_{\rm DWP}(T)$ is the same as $h_{\rm NC}(T)$ if the nucleation already takes place.
But, the process of $h_{\rm NC}(T)$ must occur after nucleation occurs in region II, i.e., $h_{\rm NC}(T)> h_{\rm NCII}(T)$,
and the intrinsic dependence of the depinning process is not observed when $h_{\rm NCII}(T)$ is large at large $E$.
Thus, we studied the threshold of depinning $h_{\rm DWP}(T)$ starting from the initial configuration $(++-)$.
%%miya0806
The comparison of $h_{\rm NCII}(T)$ and $h_{\rm DWP}(T)$ was given in Fig.~9.
As observed in the analytical estimation at $T=0$, the general tendency is that a
small $F$ (interaction $J_2$)  makes the threshold of the domain wall propagation large, while a large $F$ causes the threshold of the nucleation in the soft magnet to become large. 
As new phenomena due to the temperature and the discreteness of the atomic structure,
we found the following properties.
The depinning threshold generally increases with decreasing of $F$ (i.e., with $A_2$) as given in the analytical estimation at $T=0$. 
For small values of $K_2$ the threshold becomes smaller than that estimated for the continuous model at $T=0$ (the dotted line), which is a natural temperature effect.
However, for small $F$, the threshold increases with $E$ (i.e., $K_2$) which we have 
attributed to the narrow domain wall effect.
Moreover, for small $F$, the threshold is robust against changes in temperature.
We have concluded that this robustness is due to the mechanism of surface nucleation phenomena at the surface of the hard magnet because the spring effect is reduced largely in those cases.

Thus for the domain wall pinning at high temperatures, the surface nucleation of the hard magnets would be important, and it is expected that the suppression of nucleation at the surface would help the increase of the coercive force at high temperature.
%%
%We found that $h_{\rm DWP}(T)$ is robust against changes in temperature in the region of large $E$, which is considered to be due to the mechanism of surface nucleation phenomena at the hard magnet region.
%It should be noted that even at $T=0$ the effect due to the narrow domain wall reflecting the discreteness of the lattice causes deviation from the analytic results on the continuum model even at $T=0$.
%%
In the present paper, we studied only the cases of soft magnet in the grain boundary, that is, in the parameter region $A_1>A_2, K_1\geq K_2$. 
But, in real situations other cases, such as $A_1\leq A_2$ and/or $K_1 < K_2$, also exist.\cite{Hirota}
In such cases, surface coating may assist the coercive force. We leave this case for future study.

\section*{Acknowledgments}

The authors would like to thank Professor Roy Chantrell for stimulating discussions and also Dr. Satoshi Hirosawa and Dr. Kazuhiro Hono for valuable information on magnets.
The present work was supported by
Grants-in-Aid for Scientific Research C (No. 25400391 and No. 26400324) from MEXT of Japan, and the Elements Strategy Initiative Center for Magnetic Materials under the
outsourcing project of MEXT. 
The numerical calculations were supported by the supercomputer center of the ISSP of the University of Tokyo.
S. Mohakud would like to acknowledge DST, Govt. of India for INSPIRE faculty award and research funding.

\appendix
\section{Temperature dependence of the effective anisotropy}\label{AppA}
The magnetic reversal of a single domain has been discussed in the relation of effective anisotropy.
At $T=0$, the coercive force is given by the Stoner-Wohlfarth mechanism, i.e.,
\beq
H_c=2K/M,
\eeq
where $K$ is the anisotropy energy and $M$ is the magnetization of the spin. At finite temperatures, one may characterize the properties of the system by introducing a temperature-dependent effective anisotropy $K(T)$.
The effective anisotropy has been studied extensively through various methods.\cite{Callen,sakuma,matsumoto} Here, we estimate $K(T)$ from the temperature dependence of the transverse magnetic susceptibility
\beq
\chi_{xx}=\left({\partial m_x\over\partial H_x}\right)_{H_{xx}=0}.
\eeq
 At $T=0$, all the spins are aligned and the angle of the magnetization is given by minimizing 
 the energy $E=K\sin^2\theta-H_xM\sin\theta$.  Thus, the transverse magnetization is given by
 \beq
 m_x=M\sin\theta_{\rm min}={H_x M^2 \over 2K} \quad \rightarrow\quad \chi_{xx}={M^2\over 2K}.\label{chixx0}
 \eeq
 At $T>0$, the susceptibility at $H_{xx}$ is given by the fluctuations of $M_x$,
 \beq
 \chi_{xx}={\langle M_x^2\rangle - \langle M_x\rangle^2\over TN}=
 {\langle M_x^2\rangle \over TN}, \quad M_x=\sum_{i=1}^NM\sin\theta_i.
 \eeq
Here, one may define an effective anisotropy $K(T)$ by analogy with \eqref{chixx0} using the values at finite temperatures, i.e., $K(T)$ and $m(T)^2$ as
 %$K(T)$ by eliminating $\chi_{xx}$ from the above relations, obtaining
 \beq
K(T)\equiv {m(T)^2\over 2\chi_{xx}}={TNm(T)^2\over 2\langle M_x^2\rangle},
 \eeq
where
\beq
m(T)\equiv \sqrt{\langle M_z^2\rangle+\langle M_y^2\rangle+\langle M_x^2\rangle}.
% \simeq  \sqrt{\langle M_z^2\rangle}.
 \eeq
 
\begin{figure}[h]
$$\begin{array}{ccc}
 \includegraphics[width=0.4\textwidth]{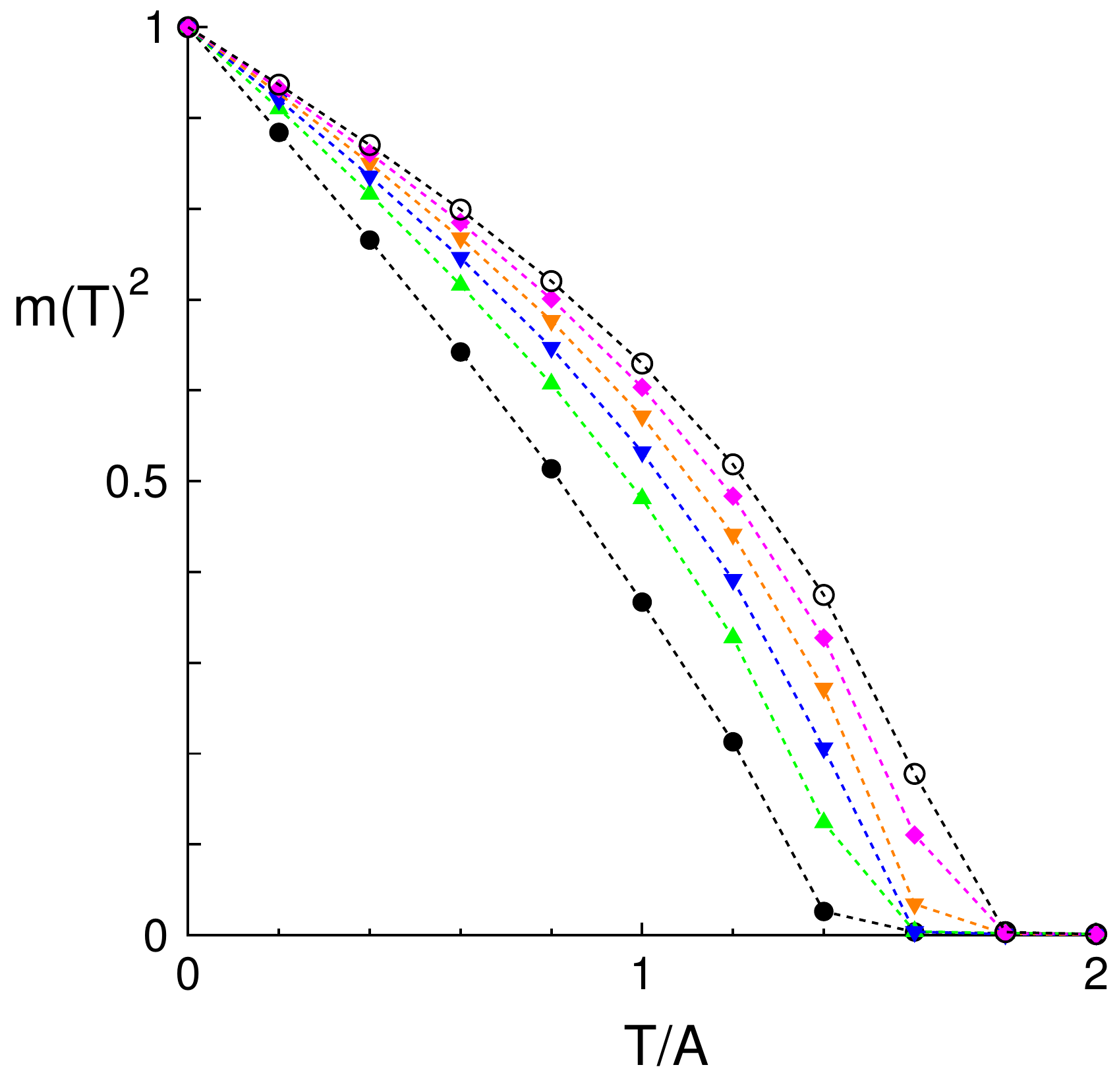}& \quad &  
 \includegraphics[width=0.4\textwidth]{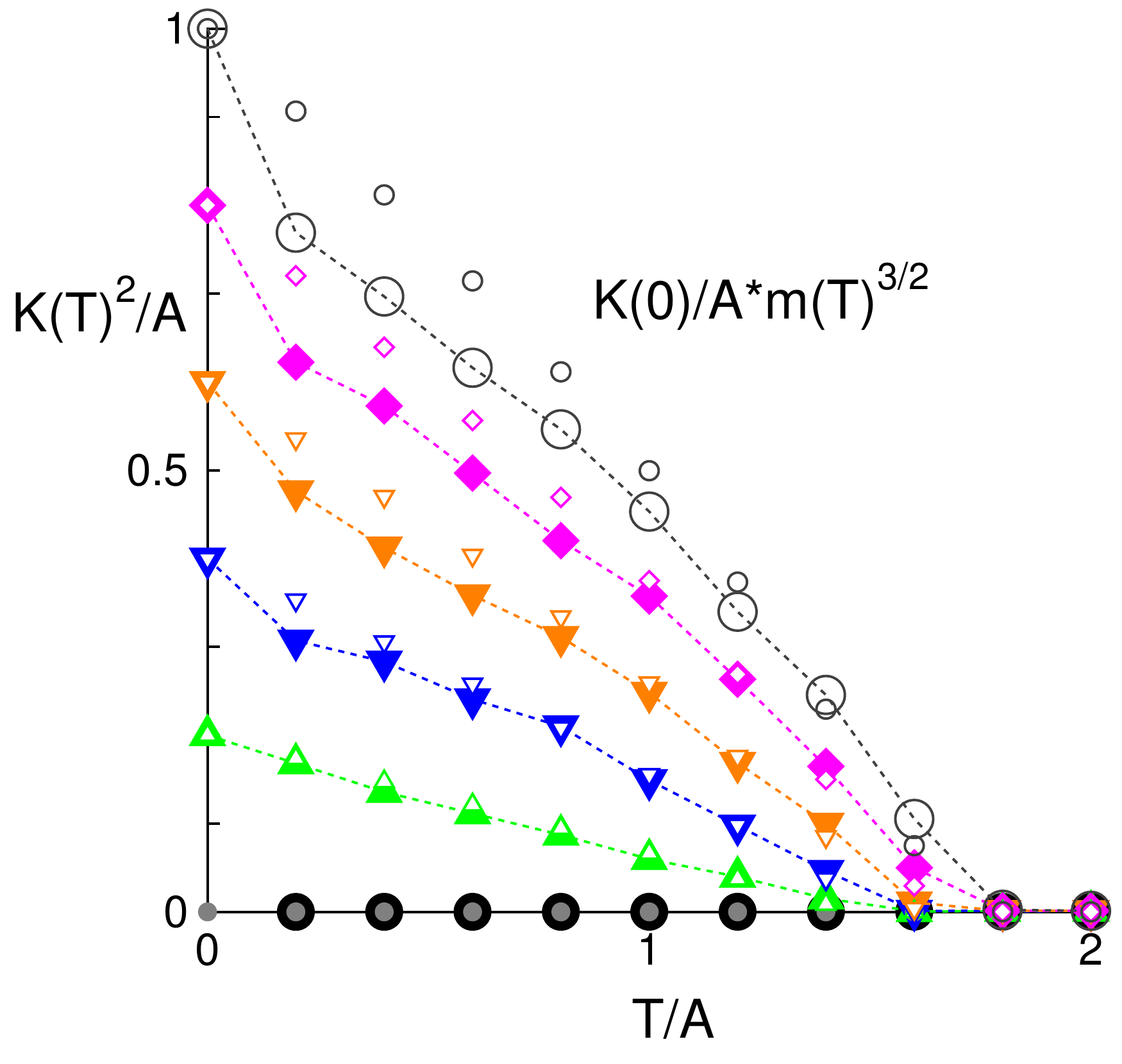}\\
 ({\rm a}) & & ({\rm b})\end{array}             
$$
%\vspace {5mm}
  \caption{
Temperature dependences of 
  (a) $\langle \left(\sum_{i=1}^{N} \mbold{S}_i \right)^2 \rangle /N^2\simeq m_{\rm s}(T)^2$
for various values of $K$ 
($K/A=0.0$, 0.2, 0.4, 0.6, 0.8 and 1.0), which are plotted by closed circle, upward triangle, square, 
 downward triangle, diamond, and open circle, respectively. 
  (b) Big symbols denote temperature dependences of $K(T)$ for various values of $K$. 
The small symbols denote $K(0)m(T)^{3/2}$ which agree with $K(T)$ at high temperatures.
The dotted curves are drawn as guides for the eye.
}
  \label{MZTKTall}
\end{figure}

 We show the temperature dependence of $\left(\sum_{i=1}^{N}\mbold{S}_i\right)^2/N^2$, which represents the square of the spontaneous magnetization ($\simeq m_{\rm s}(T)^2$)  approximately, and the above defined $K(T)$ for various values of $K/A$  in Fig.~\ref{MZTKTall}(a) and (b), respectively.
% \begin{figure}[h]
%$$\begin{array}{ccc}
% \includegraphics[width=0.4\textwidth]{TC202020H0DMZ-eps-converted-to.pdf}& \quad &  
% \includegraphics[width=0.4\textwidth]{TC202020H0DKT-eps-converted-to.pdf}\\
% ({\rm a}) & & ({\rm b})\end{array}             
%$$
%  \caption{Temperature dependence of 
%  (a) $\langle \left(\sum_{i=1}^{N} \mbold{S}_i \right)^2 \rangle /N^2\simeq m_{\rm s}(T)^2$, and
%  (b) $K(T)$. $D/J=0.0, 02,\cdots 1.0$.
% Data for $D/J=0.0$, 0.2, 0.4, 0.6, 0.8 and 1.0 are plotted by closed circle, upward triangle, square, 
% downward triangle, diamond, and open circle, respectively. 
%  }
%  \label{MK}
%\end{figure}

As we see in Fig.~\ref{MZTKTall}, the Callen-Callen law holds well for $K=0.2$. But, trivially it does not hold for $K=0$ and it also does not hold for large $K$.
 
 So far, we have considered the case $H=0$.  Now we consider the anisotropy for the case $|H|>0$.
 At $T=0$, the energy barrier between the metastable antiparallel state ($\theta=0$) 
 and the stable state ($\theta=\pi$) can be regarded as a quantity to measure the anisotropy.
 This quantity is obtained by studying the energy as a function of $\theta$:
 \beq
 E(\theta)=K\sin^2\theta-H\cos\theta.
 \eeq
 The energy of the metastable state for a negative field $H(<0)$ at $\theta=0$ 
is 
 \beq
 E(0)=-H=|H|
 \eeq
 and it has a  maximum at some angle $\theta_{\rm max}$, so the energy barrier is defined as
 \beq
\Delta E\equiv E(\theta_{\rm max})-|H|. 
 \eeq
 %The energy barrier $\Delta E(D, H)$ is given by Fig.\ref{DEH}.
 At $T>0$, we can estimate the free energy barrier in a mean-field approximation from the Hamiltonian \eqref{microscopichamiltonian}.
 Denoting the number of nearest neighbors by $z$ and choosing $m_y=0$ without loss of generality, the free energy is given by
  \beq 
  F(T,H,m_x,m_z)={zNA\over2}(m_x^2+m_z^2)-k_{\rm B}TN\ln Z(T,H,m_x,m_z) 
 \eeq
 with
 \begin{eqnarray}
 Z(T,H,m_x,m_z)&=&\int_0^{\pi}\sin\theta d\theta\int_0^{2\pi}d\phi\nonumber\\
&& \exp\left(
 \beta Az(m_z\cos\theta+m_x\sin\theta\cos\phi)+\beta K\cos^2\theta+\beta H\cos\theta
  \right).\quad
 \end{eqnarray}

\begin{figure}[h]
 $$\begin{array}{ccc}
 \includegraphics[width=0.4\textwidth]{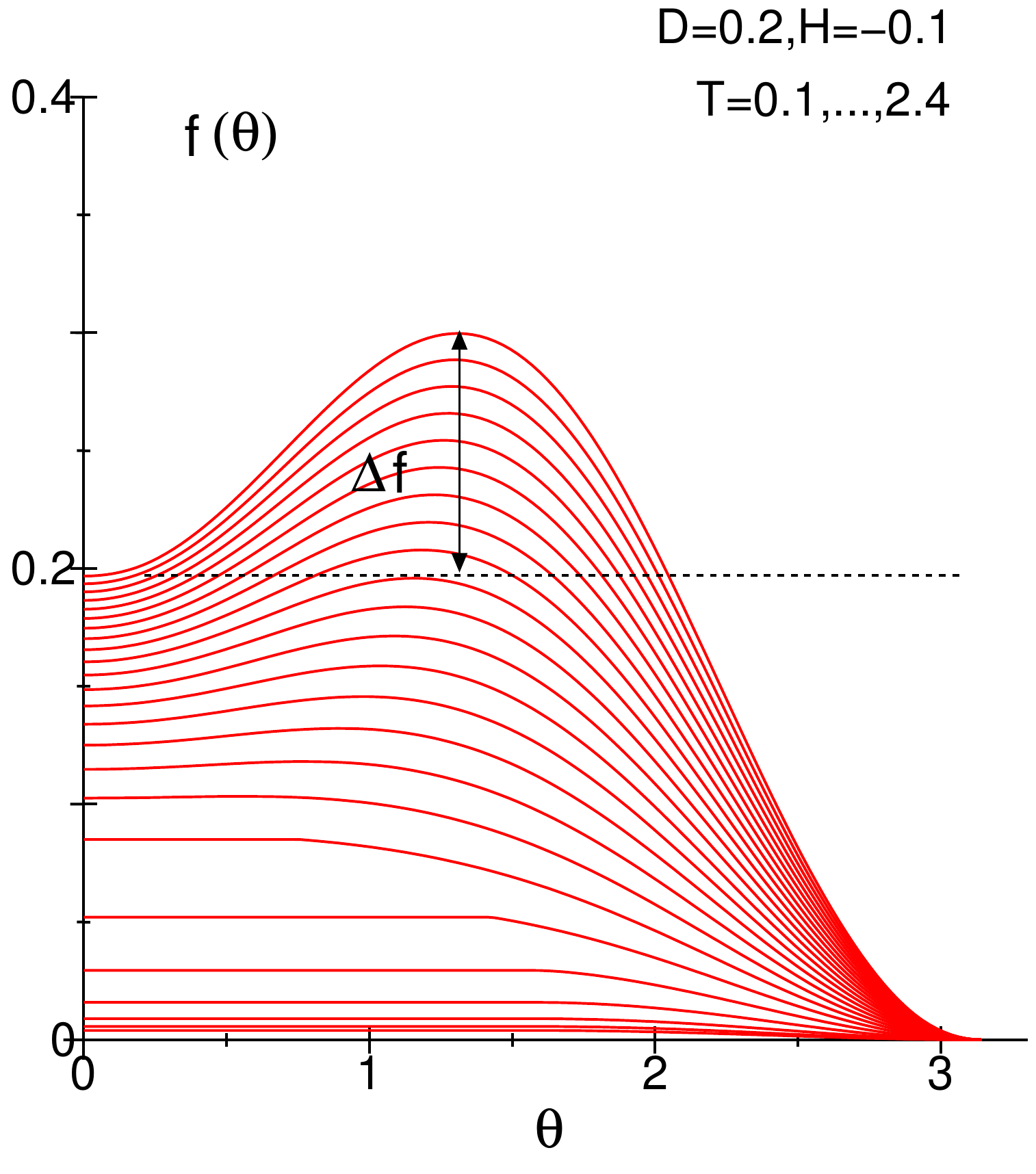} &
 \quad &
\includegraphics[width=0.4\textwidth]{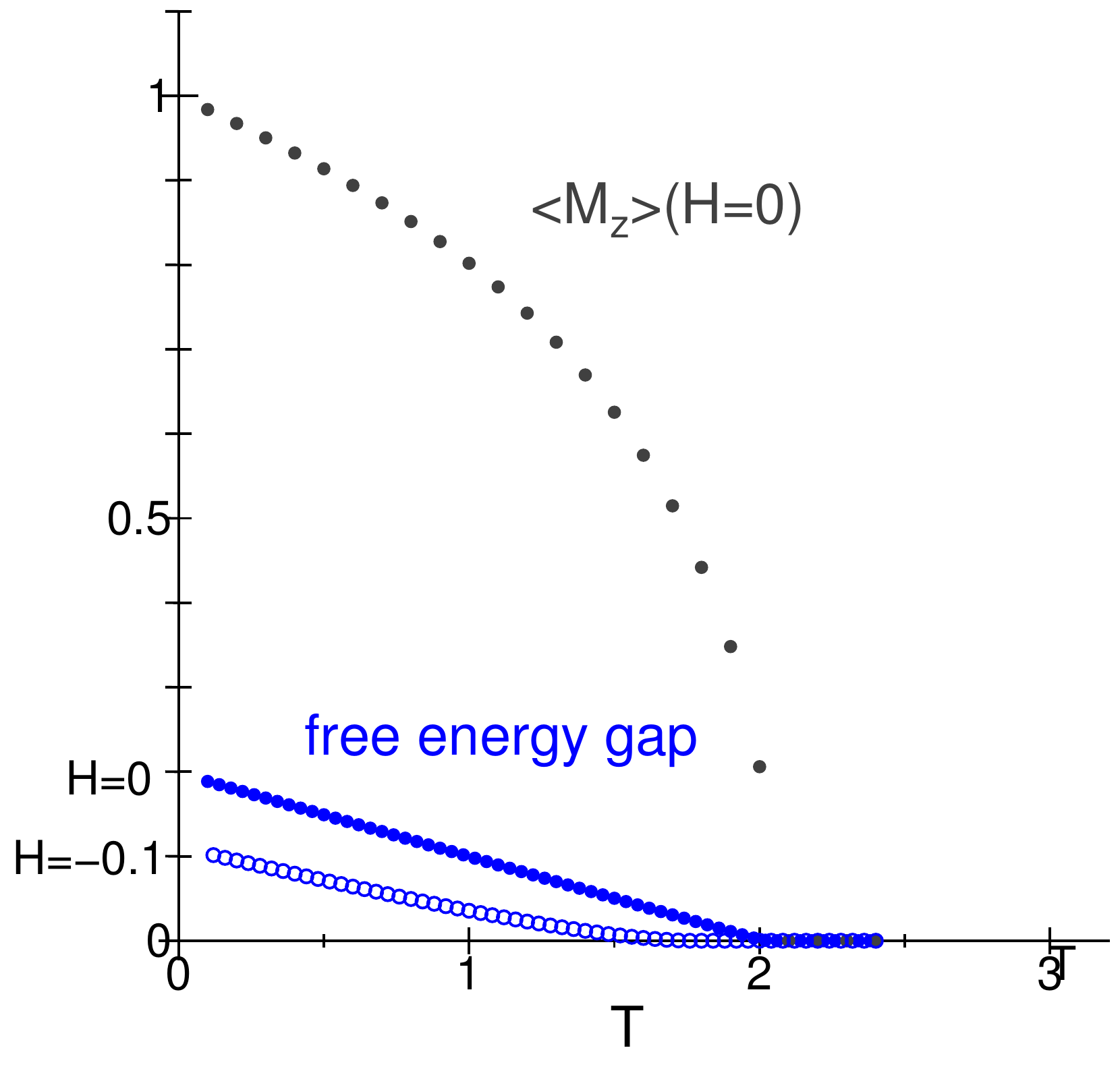}\\
({\rm a}) & & ({\rm b})\end{array}  
$$            
  \caption{
  (a) Angular dependence of the free energy gap $f=F/N$. $K=0.2$, $H=-0.1$, $T=0.1$, 0.2, $\ldots$ 2.4. (b) Temperature dependence of the free energy gap for $K=0.2$ at $H=0$ and $H=-0.1$.
  } 
  \label{DEH}
\end{figure}

 In Fig.~\ref{DEH}(a), we plot the angular  dependence of the free energy gap for $D=0.2$: 
 $\Delta f(\theta)=(F(T,H=-0.1,m_x,m_z)-F(T,H=-0.1,0,-1))/N$. 
 Here the angle $\theta$ is defined by
 $\theta=\tan^{-1}(m_z/m_x)$.
 This difference can be regarded as a kind of anisotropy.
We find that the gap disappears at around $T=1.6$. 

In Fig.~\ref{DEH}(b), we plot the temperature dependences of the spontaneous magnetization for $H=0$, and the free energy gap: $\Delta f=(F(T,H,m_x,m_z)-F(T,H,0,-1))/N$. 
 At finite magnetic field, the potential barrier due to the anisotropy is reduced significantly from that at $H=0$.

\section{Threshold field dependence on observation time}\label{AppC}

%\begin{figure}[h]
%$$\begin{array}{ccc}
%\includegraphics[width=0.4\textwidth]{RtimeF03T01D02k01-eps-converted-to.pdf}   & 
%\quad &
% \includegraphics[width=0.4\textwidth]{RtimeF03T01D02k02-eps-converted-to.pdf} \\
% ({\rm a}) & & ({\rm b}) \end{array}               
%$$
%%\vspace {5mm}
%  \caption{Dependence of the average time $t_{\rm NC}$ of domain wall depinning after nucleation on the reduced field $h=H/2K_1$.
%(a) $F=0.3,$ $T=0.1,$ $K_1=0.2,$  $K_2=0.02,$ $E=0.03$. 
%(b) $F=0.3,$ $T=0.1,$ $K_1=0.2,$  $K_2=0.04,$ $E=0.06$. After t=5000, 
%$t_{\rm NC}$ increases very rapidly, and thus we estimate $t_{\rm NC}$ from data obtained by simulations with $t_{\rm max}=5000$.
%  }
%  \label{RTF03E006k01}
%\end{figure}

%\begin{figure}[h]
%$$
%\includegraphics[width=0.4\textwidth]{RtimeF03T01D02k01-eps-converted-to.pdf}  
%$$ 
% \caption{Dependence of the average time $t_{\rm NC}$ of domain wall depinning after nucleation on the reduced field $h=H/2K_1$.
%$F=0.3,$ $T=0.1,$ $K_1=0.2,$  $K_2=0.04,$ $E=0.06$. 
%  }
%\label{RTF03E006k02}
%\end{figure}

\begin{figure}[h]
$$
%\begin{array}{ccc}
%\includegraphics[width=0.4\textwidth]{RtimeF03T01D02k01-eps-converted-to.pdf}   & 
%\quad &
 \includegraphics[width=0.4\textwidth]{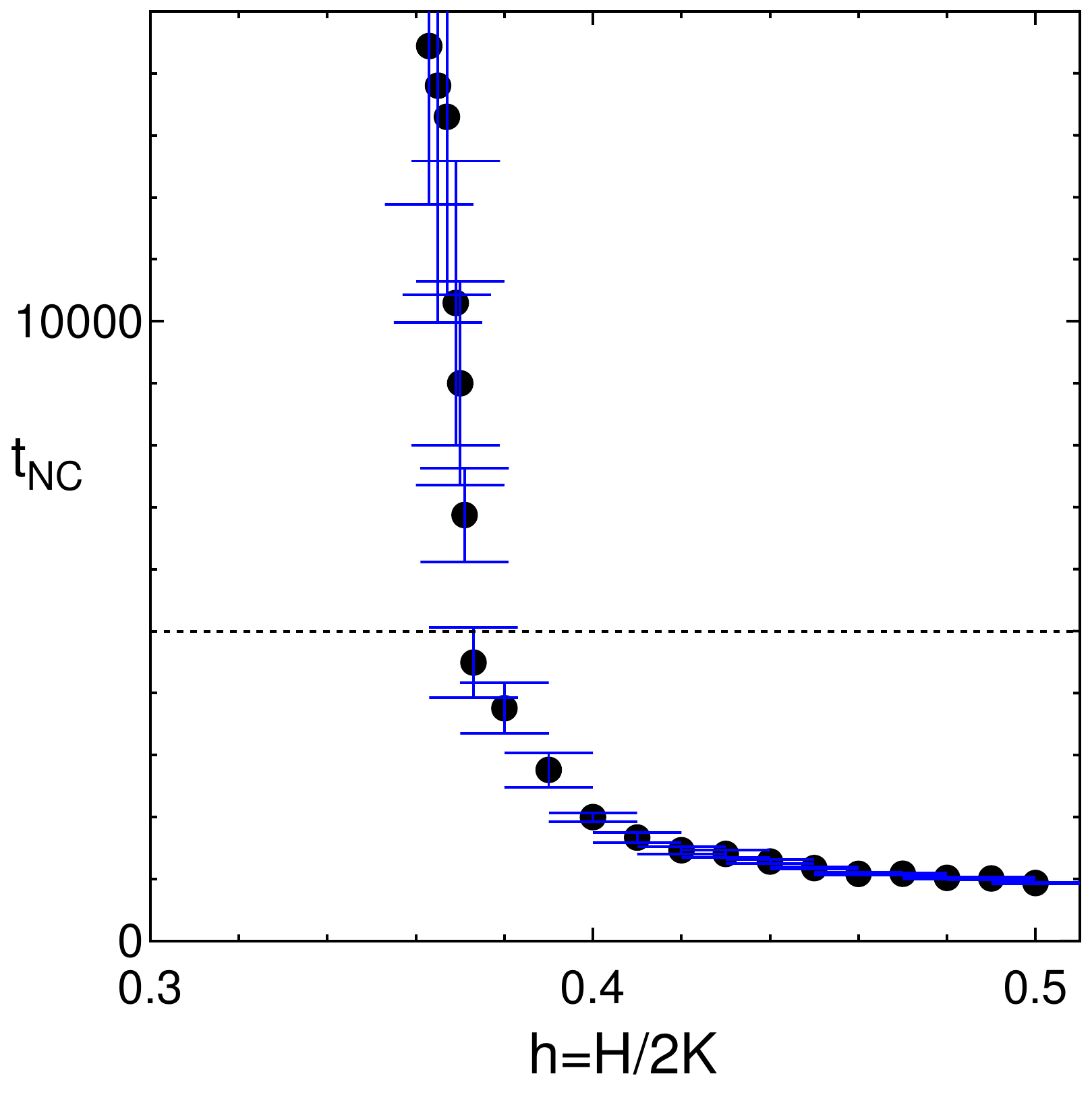} 
% ({\rm a}) & & ({\rm b}) \end{array}               
$$
%\vspace {5mm}
  \caption{Dependence of the average time $t_{\rm NC}$ of domain wall depinning after nucleation on the reduced field $h=H/2K_1$.
%(a) $F=0.3,$ $T=0.1,$ $K_1=0.2,$  $K_2=0.02,$ $E=0.03$. 
%(b)
 $F=0.3,$ $T=0.1,$ $K_1=0.2,$ and $E=0.06$. After t=5000, 
$t_{\rm NC}$ increases very rapidly, and thus we estimate $t_{\rm NC}$ from data obtained by simulations with $t_{\rm max}=5000$.
  }
  \label{RTF03E006k02}
\end{figure}

At finite temperatures, if we perform simulations of systems with a finite size for long times, the system should reach equilibrium. That is, the life time of metastable states is finite.
However, when we study the coercive force, the life time of metastable states is important.
For the estimation of the threshold fields of nucleation or domain wall depinning, we look for the parameter at which the relaxation time increases rapidly.
In realistic time scales, such as 1 second, the corresponding simulation time extremely large. However, it is fortunate that the relaxation time near the threshold increases very rapidly. 
In Fig.~\ref{RTF03E006k02} we show an example of the relaxation time for the case 
$F=0.3,$ $T=0.1,$ $K_1=0.2,$ $E=0.06$ which is found in the top left panel of Fig.~\ref{NucleationT}.
Simulations begin with the configuration $(+++)$, nucleation occurs in region II at a very small field $(h\simeq 0.15)$, and then the transition from $(+-+)$ to $(---)$ takes place at around $h\simeq 0.37$.

In the figure, we depict 
the average relaxation time $t_{\rm NC}$ obtained over 10 samples by performing long simulations which stopped when the relaxation from $(+-+)$ to $(---)$ occured.
We find a rapid increase of $t_{\rm NC}$ at around $h=0.37$, and thus in the present work, we decided to adopt 
$t_{\rm max}=5000$ to estimate $h_{\rm NC}(T)$. If we adopt a longer $t_{\rm max}$, then the threshold decreases slighty, but because of the rapid change we expect that the estimation here gives approximate information for the threshold fields.
This observation is also valid for the case in which the initial configuration is $(++-)$ and we study the threshold field for domain wall propagation.

%%%%%%%%%%%%%%%%%%%%%%%%%%%%%%%%%%%%%%%%%%%%%%%%%%%%%%%%%% C
\section{Bloch domain wall and Narrow domain wall}\label{AppB}
In the continuous limit the system is modeled by a one dimensional model
\beq
E=\int dx\left[ {A\over 2}\left({d\theta\over dx}\right)^2+K\sin^2\theta \right],
\eeq
where we put $M=1$.
% and 
%\beq
%A={J\over 2}.
%%, \quad K=D. 
%\eeq
The solution of domain wall type (Bloch wall) is given by
\beq
\theta(x)=2\tan^{-1}\left(e^{x/\xi}\right), \quad \xi=\sqrt{\frac{A}{2K}}.
\eeq

On the other hand, for the case of strong anisotropy, the discreteness of the lattice is relevant, and the model should be treated as a discrete lattice:\cite{narrowDW}
\beq
E=-A\sum_i\cos(\theta_i-\theta_{i+1})+K\sum_i\sin^2\theta_i.
\eeq
The minimum energy state is given by
\beq
{K\over A}\sin 2\theta_i+\sin(\theta_i-\theta_{i-1})+\sin(\theta_i-\theta_{i+1})=0, \quad {\rm for\ all}\ i.
\label{thetai}
\eeq
We assume a solution of domain wall type and set $\theta_{-\infty}=0$ and $\theta_{\infty}=\pi$. 
For the strong anisotropy case, we have $\theta_i\simeq 0$ for $i<0$, and we linearize the above relation.
\beq
{K\over A} 2\theta_i+(\theta_i-\theta_{i-1})+(\theta_i-\theta_{i+1})
=\left(2{K\over A}+2\right)\theta_i-\theta_{i+1}-\theta_{i-1}=0.
\eeq
This has the solution for $n<0$
\beq
\theta_n=\theta_{0}\lambda^{|n|},\quad {\rm with }\quad  \lambda=\rho-\sqrt{\rho^2-1},
\eeq
where $\rho=D/J+1$.
Assuming that  the center of the configuration is located at the middle of $i=0$ and 1, we set $\theta_{1}=\pi-\theta_0$.
The value of $\theta_0$ is determined by the relation (\ref{thetai}) at $i=0$
\beq
\left(1-{K\over A}\right)\sin 2\theta_0=\sin((1-\lambda)\theta_0).
\eeq
For $K>{2\over 3}A$, this relation only has the solution $\theta_0=0$, while $K<{2\over 3}A$ it has nonzero solution.

\end{document}